\documentclass[aip,jcp,amsmath,amssymb,floatfix,reprint,citeautoscript,noeprint,longbibliography]{revtex4-2}
\usepackage[utf8]{inputenc}
\usepackage{graphicx}
\usepackage{wrapfig}
\usepackage[colorlinks,allcolors=black,citecolor=blue,urlcolor=blue]{hyperref}
\usepackage[mathlines,displaymath]{lineno}
\usepackage{chemformula}
\usepackage{placeins}
\usepackage{booktabs}
\DeclareUnicodeCharacter{2212}{-}

\graphicspath{{figures/}}

\begin{document}
\def\mytitle{%
\ch{CO2} Hydration at the Air-Water Interface: \\
A Surface-Mediated ‘In and Out’ Mechanism}
\title{\mytitle}

\author{Samuel G.\ H.\ Brookes}
\affiliation{Yusuf Hamied Department of Chemistry, 
University of Cambridge, Lensfield Road, Cambridge, CB2 1EW, UK}
\affiliation{Lennard-Jones Centre, 
University of Cambridge, Trinity Ln, Cambridge, CB2 1TN, UK}
\affiliation{Cavendish Laboratory, Department of Physics, 
University of Cambridge, Cambridge, CB3 0HE, UK}

\author{Venkat Kapil}
\email{v.kapil@ucl.ac.uk}
\affiliation{Department of Physics and Astronomy, 
University College London, 17-19 Gordon Street, London WC1H 0AH, UK}
\affiliation{Thomas Young Centre and London Centre for Nanotechnology, 
19 Gordon Street, London WC1H 0AH, UK}
\affiliation{Yusuf Hamied Department of Chemistry, 
University of Cambridge, Lensfield Road, Cambridge, CB2 1EW, UK}
\affiliation{Lennard-Jones Centre, 
University of Cambridge, Trinity Ln, Cambridge, CB2 1TN, UK}

\author{Angelos Michaelides}
\email{am452@cam.ac.uk}
\affiliation{Yusuf Hamied Department of Chemistry, 
University of Cambridge, Lensfield Road, Cambridge, CB2 1EW, UK}
\affiliation{Lennard-Jones Centre, 
University of Cambridge, Trinity Ln, Cambridge, CB2 1TN, UK}

\author{Christoph Schran}
\email{cs2121@cam.ac.uk}
\affiliation{Cavendish Laboratory, Department of Physics, 
University of Cambridge, Cambridge, CB3 0HE, UK}
\affiliation{Lennard-Jones Centre, 
University of Cambridge, Trinity Ln, Cambridge, CB2 1TN, UK}

%

%
%

%
\keywords{Interfacial Reactivity $|$ Machine Learning Potentials $|$ Interfaces}

\begin{abstract}
An understanding of the \ch{CO2 + H2O} hydration reaction is crucial for modeling the effects of ocean acidification, for enabling novel carbon storage solutions, and as a model process in the geosciences. 
While the mechanism of this reaction has been investigated extensively in the condensed phase, its mechanism at the air-water interface remains elusive, leaving uncertain the contribution that surface-adsorbed \ch{CO2} makes to the overall acidification reaction. 
In this study, we employ machine-learned potentials 
trained to various levels of theory
to provide a molecular-level understanding of \ch{CO2} hydration at the air-water interface. 
We show that reaction at the interface follows a surface-mediated `In and Out' mechanism: \ch{CO2} diffuses into the aqueous surface layer, reacts to form carbonic acid, and is subsequently expelled from solution. 
We show that this surface layer provides a bulk-like solvation environment, engendering similar modes of reactivity and near-identical free energy profiles for the bulk and interfacial processes. 
Our study unveils a new, unconventional reaction mechanism that underscores the dynamic nature of the molecular reaction site at the air-water interface.
The similarity between bulk and interfacial profiles shows that \ch{CO2} hydration is equally as feasible under these two solvation environments and that acidification rates are likely enhanced by this additional surface contribution.  
\end{abstract}

\date{This manuscript was compiled on \today}

\maketitle

\section*{Significance Statement}
Reactions at interfaces are an important and ubiquitous type of process. 
Despite their prevalence in nature, obtaining a molecular-level understanding of these processes remains challenging due to difficulties associated with probing the interfacial regime. 
Using machine learning simulations
trained to various levels of theory,
we uncover new insights into how reactions proceed at the air-water interface. 
Specifically, we uncover a new type of reaction mechanism for \ch{CO2} hydration, one in which the position of the reaction site is coupled with the extent of reaction at the interface. 
This mechanism likely underpins a number of important surface reactions and forms an integral component of ocean acidification.  
Our work places a heightened importance on the contribution of surface-adsorbed \ch{CO2} to the overall acidification rate.

\section*{Introduction}
Over the past several decades, atmospheric \ch{CO2} levels have increased by over 100 ppm as a result of carbon-intensive anthropogenic activity \cite{Gattuso2015}. 
Alongside numerous other detrimental effects \cite{Dutton2015}, heightened \ch{CO2} emissions have led to a substantial acidification of Earth's oceans, which are now 30\,\% more acidic than preindustrial times \cite{Dupont2013}. 
Ocean acidification has already had a severe effect on marine wildlife, leading to the loss of marine biodiversity, the disruption of carbonate chemistry, and the unsettling of ecosystem stability, to name a few \cite{Orr2005,Kroeker2013,Gattuso2015,Foster2016}.

Important to understanding the acidification process is a knowledge of the underlying chemistry. 
Ocean acidification is caused by the reaction between \ch{CO2} and water to form carbonic acid (\ch{H2CO3}) or bicarbonate (\ch{HCO3-}).
This occurs via the following equilibria, 
\begin{equation}
\begin{gathered}
    \mathrm{CO_2 + 2H_2O \rightleftharpoons 
    \: H_2CO_3 + H_2O \rightleftharpoons \: HCO_3^- + H_3O^+,}
\end{gathered}
\end{equation}
where bicarbonate and carbonic acid inter-convert via proton exchange with the surrounding solvent. 
Alongside ocean acidification, this process underpins a number of key processes including carbon sequestration, mineralization, and the bicarbonate buffer system \cite{Leung2014,Snaebjornsdottir2020,Wang2021}. 
Realizing the implications of this process requires a detailed understanding of the \ch{CO2 + H2O} reaction at macroscopic and microscopic scales.
Concerning the macroscopic properties, experimental work employing spectrophotometric stop-flow measurements have revealed that, under ambient conditions and neutral pH, the \ch{CO2 + H2O} reaction is a thermodynamically unfavorable process in bulk solution \cite{Wang2010}.
The products of reaction, bicarbonate and carbonic acid, are destabilized by more than 6 kcal/mol relative to \ch{CO2}, and there exists a large free energy barrier separating the reactant and product states. 
Among the products, bicarbonate and carbonic acid interconvert via proton exchange with the surrounding water, with an associated pK$_\mathrm{a}$ of around 3.5 determined from infrared and fluorescence spectroscopy \cite{Adamczyk2009,Pines2016}.

\begin{figure*}[t!]
    \centering
    \includegraphics[scale=0.6]{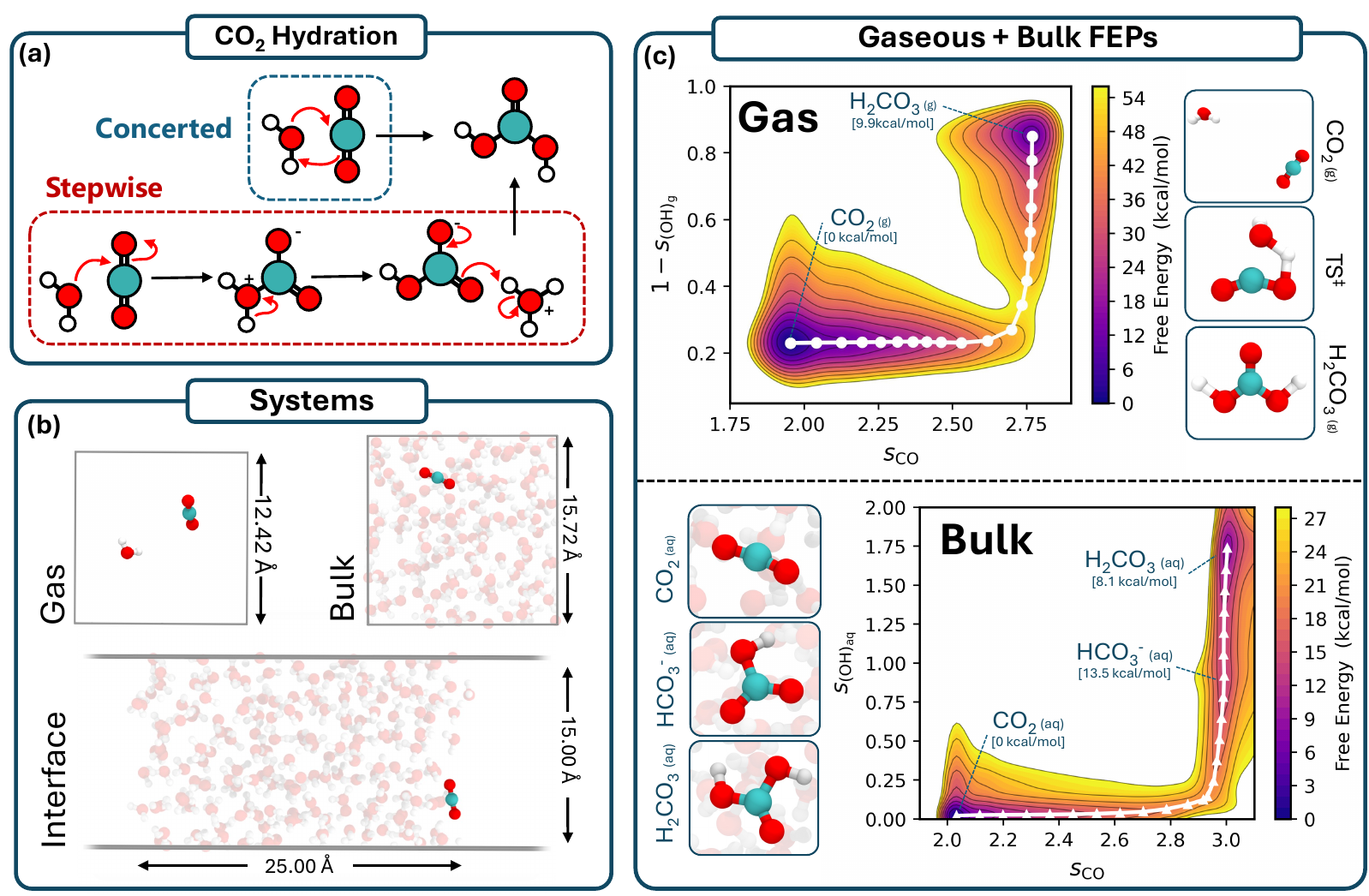}
    \caption{\label{fig:overview}
    Modeling the \ch{CO2 + H2O} reaction through enhanced sampling molecular simulations. 
    \textbf{(a)} Mechanisms detailing the reaction of \ch{CO2 + H2O} to form carbonic acid (\ch{H2CO3}) through both concerted and stepwise routes. 
    The concerted route is shown for a single reacting water molecule, though additional \ch{H2O} molecules may participate to form a proton transfer chain. 
    \textbf{(b)} System setups used to probe the \ch{CO2 + H2O} reaction under gaseous, bulk, and interfacial environments. 
    \textbf{(c)} Free energy profiles obtained from metadynamics simulations for the gaseous and bulk reactions. 
    Free energies are plotted as a function of both the C-O coordination number ($s_\mathrm{CO}$) as well as tailored protonation state collective variables for the gaseous ($s_\mathrm{(OH)_g}$) and aqueous ($s_\mathrm{(OH)_{aq}}$) phases. 
    Representative snapshots of the various species encountered during simulations are shown alongside the profiles. 
    }
\end{figure*}

In terms of the microscopic properties, much of our insight on the \ch{CO2 + H2O} reaction has come from computational work employing first-principles atomistic simulations \cite{Leung2007,Nguyen2008,Kumar2007,Kumar2009,Stirling2010,England2011,Wang_2013,Polino2020,Martirez2023,Bobell2024}.
Under partially solvated conditions, (e.g., in the gas-phase or in water clusters) the reaction proceeds in a concerted manner, with simultaneous C-O bond formation and proton transfer events. 
In the limit of a single reacting water molecule (i.e., n\ch{H2O} = 1), this proton is donated by the attacking water (see Figure \ref{fig:overview}a), and the overall process is characterized by a  large free energy barrier, $\Delta F^{\mathrm{\ddag}} \sim 50$ kcal/mol at 300 K. 
Increasing the number of explicit waters present can help reduce this barrier, enabling a proton-relay mechanism characterized by a cyclic transition-state structure. 
High-level correlated wave function studies have found that increasing n\ch{H2O} = 1 to n\ch{H2O} = 2 effectively halves the activation energy for reaction, whilst the presence of n\ch{H2O} = 3 reduces the barrier to $\Delta F^{\mathrm{\ddag}} \sim 22$ kcal/mol \cite{Nguyen2008}. 
Under these conditions, only carbonic acid forms as a viable product. 
However, by further increasing the number of explicit waters and moving towards fully solvated conditions, we also see the formation of bicarbonate ions, either through deprotonation of \ch{H2CO3} or as an intermediate in the step-wise conversion of \ch{CO2} to \ch{H2CO3} (see Figure \ref{fig:overview}a) \cite{Stirling2010}. 
\textit{Ab initio} molecular dynamics (AIMD) studies have been invaluable for investigating these reactive pathways. 
Studies have shown how the degree of hydrophobicity/hydrophilicity impacts the surrounding solvation environment \cite{Kumar2007}, how the choice of solvent alters the reaction energetics \cite{Polino2020}, and how pH can impact the outcome of reaction \cite{Martirez2023,Bobell2024}.

Despite the valuable insights acquired from these studies, they focus on either gaseous or bulk-solution modeling. 
Crucially, a detailed understanding of the \ch{CO2 + H2O} reaction at the air-water interface is lacking. 
This omission is significant; Earth's oceans span a vast surface area on the order of 350 million km$^2$, upon which are adsorbed large quantities of atmospheric \ch{CO2}.
The contribution that this surface-adsorbed and near-surface \ch{CO2} makes to the overall acidification process is currently unclear. 
Further, previous studies have identified that aqueous interfaces can display certain interesting properties and phenomena, including accelerated rates of reaction, altered product selectivities, and modified reaction mechanisms \cite{Baer2014,Lowe2015,nurev-physchem-121319-110654,Kusaka2021,Lee2024}.
Whether any of these effects are present for interfacial \ch{CO2} hydration also remains an open question.

To date, experimental and computational work have faced a number of challenges in modeling interfacial reactions and processes. 
Experimental work is hampered by the large volumetric ratio of bulk to interfacial environments, which makes the latter difficult to probe spectroscopically. 
Some work has been undertaken to identify \ch{CO2} and bicarbonate propensities for adsorption at the interface \cite{Tarbuck2006a,Devlin2023}, though such work is limited in the context of the hydration reaction and the various species involved. 
From a computational perspective, aqueous interfaces require both large length- and timescales to obtain converged results free from finite-size effects; this precludes the use of \textit{ab initio} methods, which are the go-to for performing reactivity analysis. 
As a result, there are major gaps in our understanding of this reaction at the air-water interface.

A potential solution to this problem arises in the form of machine-learned potentials (MLPs). 
Trained on first-principles data, these models offer an \textit{ab initio}-level of accuracy for a fraction of the computational cost. 
They have already been applied towards studying a variety of important reactions and processes at interfaces \cite{doi:10.1126/science.abd7716,DelaPuente2022,Kapil2024,Buttersack2024,DellagoPersp2024}.
We refer the reader to our recent introductory review paper for more details on MLPs~\cite{thiemann2024introduction} as well as other excellent reviews on the topic~\cite{Deringer2019,
Unke2021,
Behler2021,
Deringer2021}.
Building on recent developments in the field of MLPs, this paper aims to address the question of how exactly \ch{CO2} reacts at the air-water interface. 
Specifically, how does this reaction proceed? 
Where at the interface does the reaction occur? 
And how does this process differ from the equivalent reactions in bulk and in the gas phase?

To accomplish this, we train a multi atomic cluster expansion (MACE) potential capable of simulating the \ch{CO2 + H2O} reaction across a range of conditions. 
MACE combines a complete high-body order polynomial basis - known as the Atomic Cluster Expansion \cite{PhysRevB.99.014104} - with a message passing tensoral network to provide learnable representations of semi-local environments \cite{batatia2022mace}. 
This results is an efficient, high-accuracy potential that has excelled over other MLP variations and that is robust across a wide range of chemical systems, as discussed in Reference \citenum{Kovacs2023}.

In this work, we develop and validate 
MACE models at various level of theory,
for the study of the \ch{CO2 + H2O} reaction at the air-water interface. 
%
These models were%
trained on an extensive and representative collection of chemical structures, with subsequent validation of the 
models 
confirming the quality of 
their 
predictions.
%
Importantly, this includes predictions at beyond-DFT random phase approximation (RPA) level and benchmarks at the coupled-cluster level.
%
Using MACE, we performed well-tempered metadynamics \cite{Barducci2008} targeting the reaction in the gas phase, in bulk solution, and at the air-water interface - specifically, within 10 Å of the air-water interface. 
Our analysis uncovers an unconventional `In and Out' mechanism underpinning reaction at the air-water interface:
\ch{CO2} diffuses into the surface aqueous layer,
reacts to from carbonic acid,
and is subsequently expelled from solution. 
Characterization of this regime reveals that important solvation properties - specifically, the solute-solvent coordination number and hydrogen bond count - are bulk-like and uniform to within 1 \AA{} of the interface. 
The similarity between bulk and interfacial environments engenders near-identical chemical reactions in terms of the free energies, $\Delta F$, and free energy barriers, $\Delta F^{\mathrm{\ddag}}$.
Our work highlights the correlation between the extent of reaction and the location of reacting substituents at the air-water interface.
Further, it establishes the uniformity of solvation conditions for the near-interface aqueous regime. 
Together, these observations suggest that interfacial reactions can occur in very similar manner to their bulk counterparts, despite lower effective solvent densities. 
In the context of ocean acidification, our results place a heightened importance on the role of surface-adsorbed \ch{CO2} reactions and their contributions to the overall acidification process.

\section*{Results}
\subsection*{The interfacial reaction of \ch{CO2} is bulk-like}
To investigate the \ch{CO2 + H2O} reaction, we generated a series of free energy profiles mapping the conversion of \ch{CO2} to bicarbonate and carbonic acid. 
Profiles were generated for the gaseous, bulk, and interfacial reactions using the system setups in Figure \ref{fig:overview}b
with our MACE model trained against revPBE-D3 data; we refer to the supporting information for further analysis obtained at BLYP-D3 and RPA level confirm all conclusions shown here.
%
For each regime, an appropriate collective variable (CV) subspace was chosen to fully resolve the species encountered during reaction. 
This space was composed of a C-O coordination number, $s_\mathrm{CO}$, to monitor the initial attack of water on \ch{CO2} and a protonation state, $s_\mathrm{(OH)}$, to monitor the conversion between bicarbonate and carbonic acid. 
Tailored definitions of $s_\mathrm{(OH)}$ were used for the gas phase ($s_\mathrm{(OH)_{g}}$) and for solvated reactions ($s_\mathrm{(OH)_{aq.}}$)(see Methods). 
For each system, we performed well-tempered metadynamics simulations for a cumulative time of 50 ns. 
The extent of reaction was monitored by the tuple ($s_\mathrm{CO}$, $s_\mathrm{(OH)}$).

\begin{figure*}[t]
    \centering
    \includegraphics[scale=0.72]{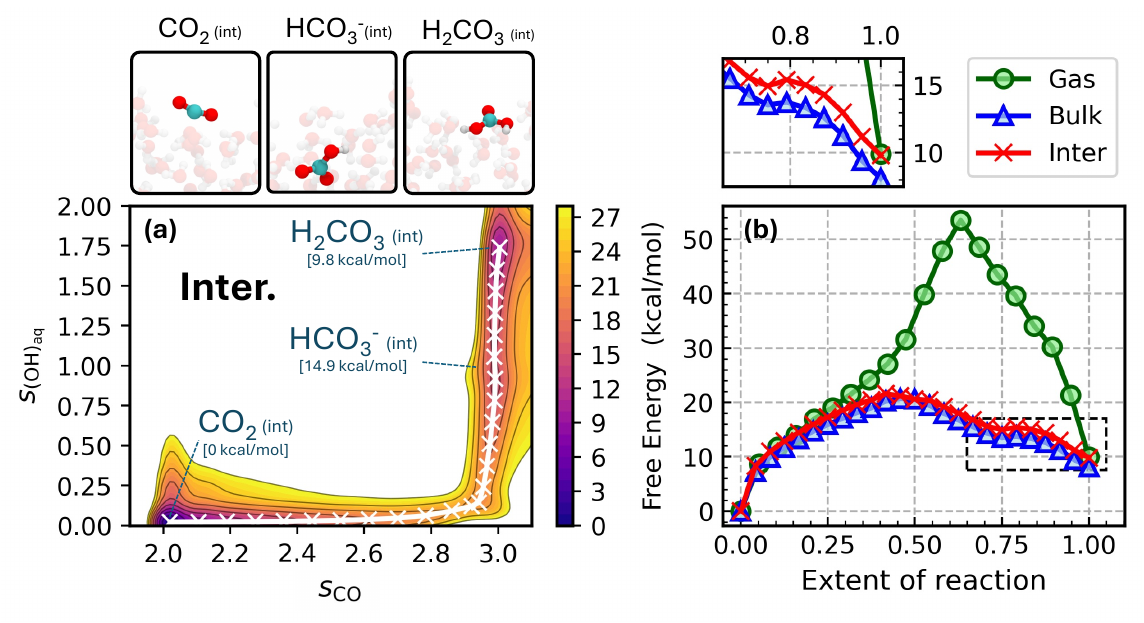}
    \caption{
    \label{fig:free_energy}
    Reaction free energies are almost identical for bulk and interfacial reactions. 
    \textbf{(a)} Free energy profile obtained from metadynamics simulations for the interfacial reaction. 
    Free energies are plotted as a function of both the C-O coordination number ($s_\mathrm{CO}$) as well as an aqueous-phase protonation state collective variable ($s_\mathrm{(OH)_{aq}}$). 
    Representative snapshots of the various species encountered during the simulation are shown alongside the profile.
    \textbf{(b)} Minimum free energy profiles (MFEPs) obtained for the \ch{CO2 + H2O} hydration reaction under gasous, bulk, and interfacial regimes.
    MFEPs are plotted as free energies as a function of the extent of reaction, where 0 corresponds to the reactants (i.e., \ch{CO2}) and 1.0 corresponds to the products (i.e., carbonic acid).
    The additional figure situated above the main plot highlights the `acidic' region of the graph, relating to the free energies of conversion between bicarbonate and carbonic acid. }
\end{figure*}

The free energy profiles obtained for the gaseous and bulk reactions are shown in Figure \ref{fig:overview}c. 
These reactions have been studied previously using correlated wavefunction and DFT modeling and thus serve as a good measure of the quality of our generated MACE potential \cite{Nguyen2008,Stirling2010,Polino2020}.
Looking at the profiles, we note several important differences between the gaseous and bulk reactions. 
For the gas-phase profile, there are only two local minima: a reactant basin at (2.0, 0.2) denoting \ch{CO2} and a product basin at (2.8, 0.8) denoting carbonic acid. 
Separating these minima is a large reaction barrier on the order of 50 kcal/mol. 
In contrast, the bulk reaction profile shows three distinct basins: a basin at (2.0, 0.0) denoting \ch{CO2}, a basin at (3.0, 1.0) corresponding to bicarbonate, and a basin at (3.0, 2.0) corresponding to carbonic acid. 
Separating the \ch{CO2} and \ch{HCO3-} basins is a reaction barrier on the order of $\mathrm{\Delta} F^{\mathrm{\ddag}}$ $\sim$ 20 kcal/mol, whilst a smaller barrier of $\mathrm{\Delta} F^{\mathrm{\ddag}}$ $<$ 1 kcal/mol connects \ch{H2CO3} to \ch{HCO3-} (see Figure S13 for more information).  
From these observations, we can make two key inferences: the bulk reaction occurs with enhanced kinetic favorability compared to the gas-phase and bicarbonate can only form under solvated conditions.

These results are consistent with previous studies of the \ch{CO2 + H2O} reaction. 
Our gas-phase predictions show excellent agreement with those of coupled cluster calculations  \cite{Nguyen2008}, where we find that we can reproduce $\mathrm{\Delta} F$ and $\mathrm{\Delta} F^{\mathrm{\ddag}}$ to within 1 kcal/mol.
We have also benchmarked our DFT setup with respect to coupled cluster calculations at the DLPNO-CCSD(T)-F12 level; we find good agreement for the minimum energy path of the gas-phase reaction, which can be seen in Figure S1 of the Supplementary Material.
Our bulk predictions also show close agreement with those of previous DFT-based studies employing the BLYP and revPBE-D3 functionals \cite{Stirling2010,Polino2020}.
Looking at the \ch{H2CO3/HCO3-} free energies, we derive a pK$_\mathrm{a}$ of 3.9 using $\Delta F = -RT\mathrm{ln}(K_\mathrm{a})$.
This result compares favorably with experimental pK$_\mathrm{a}$ values of 3.45 and 3.49 obtained from spectroscopic measurement \cite{Adamczyk2009,Pines2016}.
The similarity between these various results attest to the quality of our generated potential for investigating the \ch{CO2 + H2O} reaction under differing solvation conditions.

In Figure \ref{fig:free_energy}a, we show the free energy profile obtained using the interfacial setup. 
Similar to the bulk reaction, we observe three different basins corresponding to \ch{CO2}, bicarbonate, and carbonic acid.
We also observe free energies and reaction barriers that are remarkably similar to those of the bulk reaction, to within 1-2 kcal/mol. 
These similarities can be seen more clearly in Figure \ref{fig:free_energy}b, where we plot the minimum free energy profiles (MFEPs) extracted from each profile. 
Looking at these MFEPs, we see almost identical modes of reaction for the bulk and interfacial regimes in terms of the free energies, reaction barriers, and the overall reaction pathways. 
This may come somewhat as a surprise given the stark structural and compositional differences between the homogeneous bulk and the spatially anisotropic interface \cite{Bonn2015,Wolfhart2020}. 
Indeed, previous studies for other reactions have reported notable differences in the behavior of processes occurring at an interface versus in bulk. 
Work by de la Puente et al.\ has shown how certain acids can become more or less acidic depending on their location at the air-water interface \cite{DelaPuente2022}. 
Buttersack et al.\ demonstrated that sulfur species can undergo stabilization at the air-water interface, leading to enhanced dissociation relative to the bulk \cite{Buttersack2024}. 
For the \ch{CO2} hydration studied here, this is clearly not the case, with Figure \ref{fig:free_energy}b suggesting almost identical stabilization of each of the species under both types of environment. 
Likely this results from both the difference in reaction type being treated as well as our simulation setup, in which we allow species to react freely with no restraints imposed on the position relative to the interface.

\subsection*{\ch{CO2} reacts via surface-mediated `In and Out' Mechanism}
\begin{figure*}[t]
    \centering
    \includegraphics[width=0.95\textwidth]{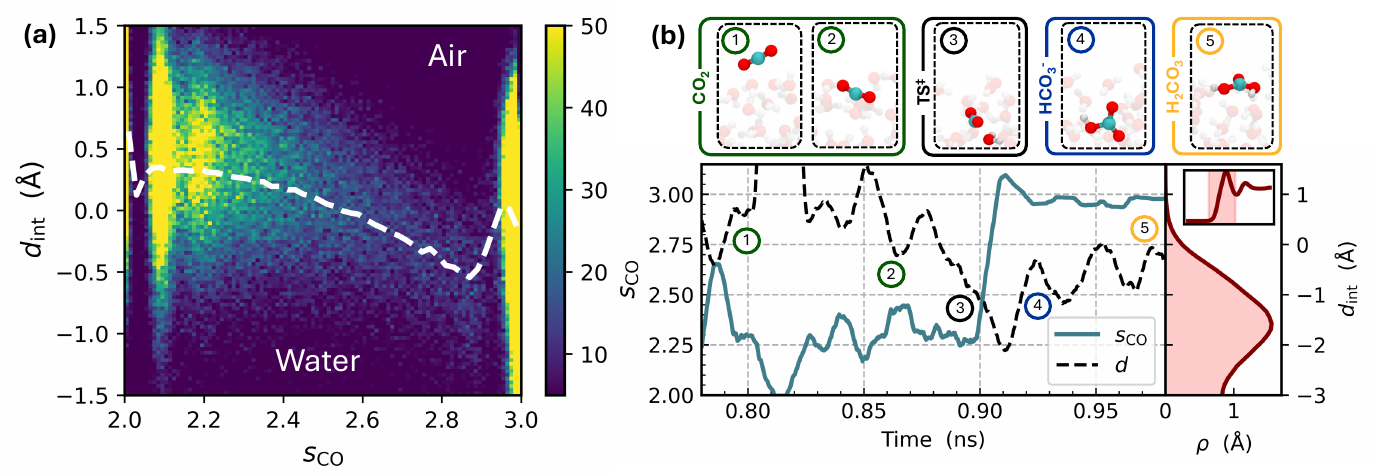}
    \caption{\label{fig:mechanism} 
     \ch{CO2} reacts with water via a surface-mediated `In and Out' mechanism. 
     \textbf{(a)} A 2D histogram showing the joint frequency distribution of the C-O coordination number ($s_\mathrm{CO}$) and the distance from the instantaneous interface ($d_\mathrm{int}$), calculated using the Willard-Chandler formalism \cite{Willard2010}. 
     The averaged $d_\mathrm{int}$ value, obtained for each $s_\mathrm{CO}$ within the limits $-1.5 \leq d_\mathrm{int} \leq +1.5$, is shown by the dashed curve. 
     The frequency colorbar is capped at 50 to prevent over-saturation of the figure due to reactant and product states. 
     \textbf{(b)} A representative reactive event, extracted from our metadynamics simulations, in which \ch{CO2} reacts within the first molecular layer of the aqueous phase during reaction.
     This event is followed in terms of both $s_\mathrm{CO}$ and $d_\mathrm{int}$, which are plotted as a function of the metadynamics walker time. 
     Reaction occurs at around 0.9 ns. 
     The density profile of water, plotted as $d_\mathrm{int}$ against density $\rho$, is shown for reference on the right-hand side of the figure. 
     Representative snapshots showing some of the species encountered during simulation are shown along the top of the figure. 
}
\end{figure*}

Figure \ref{fig:free_energy} demonstrates a likeness in the free energies of the bulk and interfacial \ch{CO2 + H2O} reactions.  
Understanding this similarity requires a more detailed investigation of the mechanisms underpinning reaction at the air-water interface. 
In Figure \ref{fig:mechanism}, we explore this mechanism in a statistical manner. 
Figure \ref{fig:mechanism}a shows a 2D histogram capturing the joint frequency distribution of the C-O coordination number, $s_\mathrm{CO}$, and the depth of the reacting species relative to the instantaneous air-water interface~\cite{Willard2010}, $d_\mathrm{int}$, across our multi-nanosecond simulation data. 
Frequency is encoded in the color bar, whilst the averaged interfacial distance $\langle d_\mathrm{int} \rangle$ at each $s_\mathrm{CO}$ is shown by the dashed line.

Looking at Figure \ref{fig:mechanism}a, we observe a clear correlation between the carbon-oxygen coordination number $s_\mathrm{CO}$ and the distance to the instantaneous interface $d_\mathrm{int}$.
As $s_\mathrm{CO}$ increases from left to right - representative of converting from the reactants (\ch{CO2 + 2 H2O}) to the products (\ch{H2CO3 + H2O}/\ch{HCO3- + H3O+}) - $\langle d_\mathrm{int} \rangle$ gradually decreases, switching from a positive value (air-side) to a negative value (water-side) at $s_\mathrm{CO} \sim 2.5$. 
A minimum in $\langle d_\mathrm{int} \rangle$ is observed at $s_\mathrm{CO} = 2.9$, which roughly coincides with the transition state of reaction. 
Beyond $s_\mathrm{CO} = 2.9$, $\langle d_\mathrm{int} \rangle$ starts to increase again and approache positive values.

Physically, these observations suggest an unconventional `In and Out' mechanism: \ch{CO2} initially starts on top of the air-water interface; \ch{CO2} then dives into the aqueous phase and reacts form carbonic acid; finally, carbonic acid is expelled from solution. 
A thorough investigation of the metadynamics trajectories supports this proposed mechanism.
In Figure \ref{fig:mechanism}b, we plot a representative reactive event - extracted from our metadynamics simulations - in terms of the $s_\mathrm{CO}$ and $d_\mathrm{int}$ against simulation time. 
We observe that, prior to reaction, \ch{CO2} resides on top of the air-water interface, with $d_\mathrm{int} \geq 0$.
Upon reaction - denoted by an increase in $s_\mathrm{CO}$ from 2 to 3 - the molecule dives into the first layer with $d_\mathrm{int}$ falling to a minimum value of $-2$ Å. 
Following the formation of carbonic acid and the completion of the reaction, these products return towards the surface.  
At the end of this simulation segment, $d_\mathrm{int}$ $\sim$ $-0.1$ \AA{} which corresponds to \ch{H2CO3} residing near the top of the first molecular layer of water.

Looking at Figure \ref{fig:mechanism}, we gain a number of valuable insights.
First, \ch{CO2} prefers to sit on top of the water-air interface, in accordance with previous studies~\cite{Brookes2024}.
Upon reaction, the surface-adsorbed \ch{CO2} will partially dissolve in the molecular surface layer of water. 
In doing this, \ch{CO2} moves from a solvent-deficient region to a solvent-rich one, thus stabilizing the emerging charge found in the  transition-state species. 
From this, we can infer that the \ch{CO2 + H2O} reaction is disfavored on top of the air-water interface and that \ch{CO2} must be immersed within the first contact layer to facilitate reaction. 
These insights suggest a correlation in the extent of reaction and the location of the reactive species at the interface. 
In the case of our proposed `In and Out' mechanism, in going from reactants to products, we observe $d_\mathrm{int}$ to initially decrease approaching the transition state, following which $d_\mathrm{int}$ starts to rise again. 
Validation of this proposed mechanism is provided in Figures S11-S12, where we show that similar results are observed for models trained at different levels of theory, specifically, BLYP-D3 and the random phase approximation (RPA). 
More details are provided in Section 3 of the Supplementary Material.

\subsection*{Solvation conditions are uniform approaching the air-water interface}
In Figure \ref{fig:mechanism}, we uncover a reaction mechanism that involves a positional change of the reacting species relative to the air-water interface.
We might expect such behavior to be born of the varying adsorption affinities that the species in Eq.\ 1 show for the air-water interface. 
To further explore this idea, we plot a series of time-averaged profiles - extracted from our metadynamics simulations - for \ch{CO2}, bicarbonate, carbonic acid, and transition-state-like species (which we denote generically by TS$^\mathrm{\ddag}$). 
For each species, we analyze the density profiles, solute-solvent coordination numbers, and hydrogen-bond counts per molecule as a function of the distance from the instantaneous interface, $d_\mathrm{int}$.

\begin{figure}[t]
    \centering
    \includegraphics[scale=0.75]{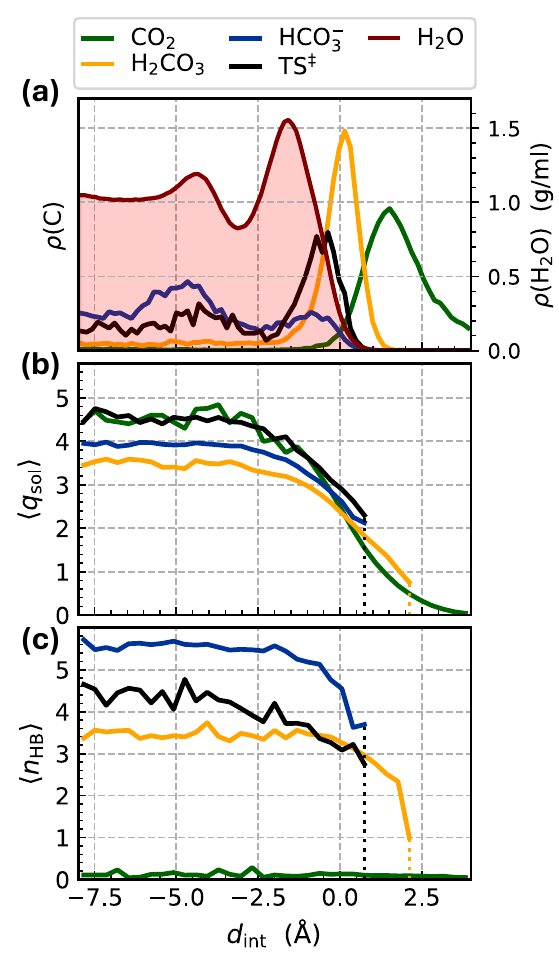}
    \caption{\label{fig:profiling} 
    Different carbon species reside at different distances from the instantaneous air-water interface. 
    \textbf{(a)} Normalised densities of \ch{CO2}, carbonic acid, bicarbonate, and transition-state-like structures plotted as a function of the distance from the instantaneous interface,  $d_\mathrm{int}$.
    The background density profile for water is shown for reference in red. 
    \textbf{(b)} The averaged solute-solvent coordination, $\langle  q_\mathrm{sol} \rangle$, plotted for each species as a function of $d_\mathrm{int}$. 
    $q_\mathrm{sol}$ is found by summing over all carbon-water coordinations, $n = (1 - (r/r_\mathrm{C})^p) / (1 - (r/r_\mathrm{C})^q)$, where $r$ gives the individual carbon-water distances, $r_\mathrm{C} = 3.5$ \AA{}, $p = 12$ and $q = 24$.
    Accordingly, $q_\mathrm{sol} = \Sigma n(r)$.
    \textbf{(c)} The averaged hydrogen-bond count, $\langle n_\mathrm{HB} \rangle$, for each species plotted as a function of $d_\mathrm{int}$.
    Hydrogen bond criteria: a donor-acceptor distance of less than 3.5 \AA{}, and an O-H-O angle of greater than 150$^{\circ}$.
    }
\end{figure}

Looking first at the density profiles shown in Figure \ref{fig:profiling}a, we see that different species adsorb at different distances relative to the air-water interface. 
\ch{CO2} resides mostly on top of the air-water interface (positive $d_\mathrm{int}$), with a peak in the distribution occurring at $d_\mathrm{int}$ = 1.6 Å. 
Carbonic acid also resides mostly on top of the interface, albeit with a smaller $d_\mathrm{int}$ equal to 0.1 Å at the distribution peak. 
Looking at the underlying water profile, this suggests that carbonic acid is partially hydrated by the first water layer when located at its equilibrium position. 
In contrast to both \ch{CO2} and carbonic acid, bicarbonate is able to freely move throughout the aqueous near-interface regime, with the equilibrium position residing at the second water layer, $d \sim -4.5$ Å. 
Finally, we observe that TS$^\mathrm{\ddag}$ structures can be located throughout the 10 Å regime. 
However, these species are found mostly in the topmost water layer, with $d \sim -1$ Å.
This serves as the most likely site of reaction for our interfacial setup.
Our profiles compare favorably with density profiles obtained from free molecular dynamics simulations, which are shown in Figure S15 of the Supplementary Material. 
The agreement between these profiles shows that we can recover equilibrium MD properties from our enhanced sampling MD.
It is also a testament to the converged nature of our metadynamics runs.

The density profiles shown in Figure \ref{fig:profiling}a indicate the preferential locations of each of the species in Eq.\ 1 relative to the air-water interface. 
There are a number of factors controlling the exact shapes of these distributions, and there is extensive discussion in literature as to why ions and molecules will adsorb to different extents depending on their chemical nature \cite{Jungwirth2006,Devlin2022,Ruiz-Lopez2020}. 
A high-level analysis of the profiles in Figure \ref{fig:profiling}a reveals that neutral species (\ch{CO2} and carbonic acid) reside on top of the interface (positive $d_\mathrm{int}$) whilst charged and partially charged species (bicarbonate, and the zwitterionic TS$^\mathrm{\ddag}$) are located within the aqueous regime (negative $d_\mathrm{int}$).
Theoretical analysis has shown that the surface adsorption of neutral molecules is principally driven by enthalpic considerations \cite{Hub2012}; the placement of hydrophobic \ch{CO2} on top of the air-water interface preserves the hydrogen bonding network of the near-interface regime, thereby maximizing solute-solute interactions. 
Comparing the positions of \ch{CO2} and carbonic acid, we observe that the latter molecule is more submerged within the topmost layer of water, a phenomenon attributed to the polarity of carbonic acid as well as its ability to partake in hydrogen bonding (see Figure \ref{fig:profiling}c). 
Concerning the `in-water' behavior of bicarbonate (and TS$^\mathrm{\ddag}$), the ionic nature of these species entails a large hydration enthalpy resulting from the stabilization of a permanent negative charge. 
The ability of bicarbonate to diffuse throughout the near-interface regime puts it in league with other relatively large, polarizable anions such as \ch{I-} and \ch{Br-}, which have also been shown to display a propensity for the air-water interface \cite{Jungwirth2006,nnurev.physchem.57.032905.104609}.
This contrasts with the behavior of harder or more highly charged anions, which are usually repelled from the interface \cite{Jungwirth2006}.
Our results suggest a clear tendency for bicarbonate to adsorb at the air-water interface \cite{Yan2018}.
This finding deviates from previous modeling using classical force-field MD \cite{Devlin2023}, likely due to our use of a more accurate underlying theory - one that better captures the solute-solvent interactions occurring in this complex regime. 
Additional surface stabilization of the interfacial reaction is due to the surface affinity of the hydronium ion \ch{H3O+}, which is central to the bicarbonate-carbonic acid equilibrium.  
Figure S16 of the Supplementary Material plots the density profile of \ch{H3O+} against distance from the instantaneous interface. 
Using $\Delta F = -RT \: \mathrm{ln}(\rho / {\rho}_0)$, we derive a free energy of adsorption of $-$ 1.3 kcal/mol, which matches exactly the $\Delta F$ value determined from experiment \cite{Das2020}.

Combining insights from Figure \ref{fig:mechanism} and Figure \ref{fig:profiling}, we deduce that reaction at the air-water interface predominantly occurs within the surface layer of water.
At a high level, we can rationalize this propensity by considering the charge of our reacting species. 
However, the question remains why reactions are concentrated in this specific region at the interface. 
To address this, in Figures \ref{fig:profiling}b and \ref{fig:profiling}c, averages of the solute-solvent coordination number, $\langle q_\mathrm{sol} \rangle$, and the solute hydrogen-bond count, $\langle n_\mathrm{HB} \rangle$, are plotted as a function of the distance from the interface. 
Looking at these properties, we observe that both are relatively uniform throughout the interfacial regime. 
For $\langle q_\mathrm{sol} \rangle$, values are relatively constant up to $d_\mathrm{int} = -2$ \AA{}, whilst for $\langle n_\mathrm{HB} \rangle$, values only start to reduce beyond $-1$ \AA{}. 
Looking at $d_\mathrm{int} \sim -1$, which we have determined to be the most favorable location for reaction to occur, both $\langle q_\mathrm{sol} \rangle$ and $\langle n_\mathrm{HB} \rangle$ are comparable to their values in bulk (i.e., towards $-8$ \AA{}) to within a couple of percent. 
This observation suggests that, despite lower effective solvent concentrations for this region, the surface layer of water provides bulk-like solvation conditions for reactive species. 
Therefore, reactions in this region can proceed with bulk-like free energies and reaction barriers in a way that minimizes disruption caused by hydrophobic \ch{CO2} to the hydrogen-bonded solvent structure deeper in the aqueous phase.

\section*{Summary and Outlook}
In this work, we have shown that the \ch{CO2 + H2O} hydration reaction proceeds via an `In and Out' mechanism at the air-water interface. 
This mechanism is characterized by a change in the position of the reacting species at the interface, a phenomenon that reflects evolving solute-solvent interactions in converting from reactants to products. 
Similarities between the bulk and interfacial solvation environments promote similar modes of reactivity and near-identical free energy profiles for the bulk and interfacial processes. 
These similarities suggest that the \ch{CO2 + H2O} reaction is equally as feasible at the interface as in bulk, therefore warranting greater attention for this region when considering the reactivity of complex, multi-component systems.

%
Importantly, the mechanistic insights reported here are consistent across three models employed, from revPBE-D3 to RPA-based machine-learned potentials, as detailed in the Supporting Information.
By incorporating beyond-DFT RPA reference level, our MLPs capture high-level electronic-structure accuracy while enabling statistically converged, multi-nanosecond sampling that would be out of scope for direct ab-initio molecular dynamics.
These extended simulations are essential for resolving the `In-and-Out’ mechanism of \ch{CO2} hydration at the air–water interface.
Despite their good description of interfacial properties, standard empirical force fields, being non-reactive, cannot be employed for such insight, underscoring the need for reactive MLPs.
The combination of high-accuracy reference data and enhanced sampling therefore establishes a new benchmark for exploring reactive events at interfaces and is readily transferable to other chemically complex systems.

In the context of ocean acidification, our results have revealed a viable path for carbonic acid formation at the air-water interface: surface-adsorbed \ch{CO2} diffuses into the aqueous surface layer, reacts to form carbonic acid, and is then ejected out of solution. 
Given the notable concentrations of \ch{CO2} adsorbed at the ocean's surface, our results suggest that \ch{H2CO3} and \ch{HCO3-} formation will occur within a couple of angstrom of the ocean's surface. 
The experimental determination of \ch{H2CO3}/\ch{HCO3-} concentrations using sum-frequency generation techniques would be ideal to test this hypothesis.  
Computationally, future work will explore the influence of ionic species, e.g., \ch{Na+} and \ch{Cl-}, on the course of \ch{CO2} hydration as well as potential clustering mechanisms for multiple carbon species \cite{Devlin2023}.
A recent geoscience study has shown that subtle temperature fluctuations at the ocean surface promote enhanced \ch{CO2} absorption, leading to a 7 \% increase in absorption rates compared to previous estimates \cite{Ford2024}.
Understanding this enhanced absorption, as well as the impacts on the ensuing acidification, will be important in helping to provide a more realistic picture of ocean acidification.

Beyond \ch{CO2 + H2O} systems, our work has important implications for more general chemistry at aqueous interfaces. 
Harnessing the accuracy and computational efficiency of MACE potentials, we have demonstrated that, for reactions involving some change in charge or polarity, there exists a correlation between the extent of reaction and molecular positioning at the aqueous interface. 
In simple terms, the depth of a reactive species relative to the air-water interface is determined by how far along the reactive process it is. 
The reactants, products, and transition state are all dynamic entities that adapt their positions in response to evolving solute-solvent interactions. 
Where previous studies considered only selected distances above or below an interface, our work considers the whole interfacial regime in evaluating reactivity.
We hypothesize that, for most `emerging-charge' reactions at the air-water interface, reaction occurs within the first molecular layer of water. 
This region offers a balance between charge stabilization and minimizing disruption to the hydrogen-bonded solvent structure. 
We expect that these insights may have broader implications for other types of chemical reactions at interfaces, for example, in phase-transfer catalysis, where the location of the reactive site is central to chemical transformation \cite{Ooi2007}.

In conclusion, our work addresses the nature of \ch{CO2 + H2O} hydration at the air-water interface and represents a step forward in the way we approach interfacial reactivity. 
We have shown that solvation conditions are bulk-like for the near-interface regime and that reactive processes are accompanied by positional change at the air-water interface. 
Given the ubiquity of interfaces in nature, we anticipate that such insights will be of direct importance to the study of other processes in nature, for example, `on-water' organic reactions, nanodroplet reactions, and phase-transfer catalysis.
Our work makes the case for the inclusion of translational considerations in the treatment and representation of chemical reactions, which we expect to lead to a re-evaluation of important chemical processes and surface phenomena.

\subsection*{Methods}
Our approach for this project can be separated into two main parts: first, the training and validation of a MACE model suitable for modeling the \ch{CO2 + H2O} reaction under differing solvation conditions; and second, the use of this model in performing well-tempered metadynamics simulations to extract free energies as well as structural and mechanistic data. 
Details on these procedures are provided below.

\subsubsection*{Training}
Our training procedure follows well-established protocols developed for training MACE models \cite{Kovacs2023}.
These procedures have been tested against a range of published datasets and system types, including aqueous systems, medicinal compounds, organic chemicals, and
functional materials.
%
A diverse selection of structures was collated for training our model. 
Initially, structures were generated using \textit{ab initio} molecular dynamics (AIMD). 
Subsequent iterations of our dataset were augmented with structures from MLP-MD, which allowed for for larger and more complex structure generation as well as structures from enhanced sampling and constrained MD runs. 
A variety of system types and compositions were incorporated into this training data, e.g., gaseous systems, periodic bulk and interfacial systems, pure water, pure \ch{CO2}, and systems of \ch{CO2} in water. 
In total, this initial dataset comprised some 150,000 datapoints. 
Following the procedure outlined in Reference \citenum{doi:10.1073/pnas.2110077118}, this dataset was optimized and its size reduced to $\sim$ 8000 representative structures. 
These structures are shown in Figure S2 of the Supplementary Material as a projected two-dimensional subspace using the UMAP transformation~\cite{UMAP2018,ASAP2020}.

Energies and atomistic forces were calculated for each structure at the DFT level of theory using the QUICKSTEP method implemented in \texttt{CP2K} \cite{VandeVondele2005,doi:10.1063/5.0007045}. 
We employed the revPBE functional augmented by Grimme's D3 corrections (zero damping) \cite{PhysRevLett.77.3865,doi:10.1063/1.3382344}.
%
%
%
revPBE-D3 benefits from a well-known error compensation: the neglect of nuclear quantum effects leads to overbinding, whilst revPBE-D3 is known to exhibit an underbinding effect. 
As a result, the use of revPBE-D3 for first-principles or machine-learned MD with classical nuclei often results in predictions consistent with those of experiment. 
This is true both for bulk and interfacial water, where revPBE-D3 has been shown to reproduce the structural, dynamical, and spectroscopic properties of liquid water \cite{Bankura2014,Soper2000,Skinner2013,Ohto2019}.
%
Further, previous metadynamics work have shown that revPBE-D3 can reproduce certain free energies and reaction barriers pertaining to both the \ch{CO2 + H2O} hydration and the bicarbonate/carbonic acid equilibrium to within 1 kcal/mol of experimental measurement \cite{Polino2020,Brookes2024}.
Goedecker–Teter–Hutter (GTH) pseudopotentials were used for the treatment of the core electrons, whilst the TZV2P-GTH basis sets were used for treating the valence electron density.
An auxiliary plane wave cutoff of 650 Ry was used.
We benchmark the predictions of this setup against those of DLPNO-CCSD(T)-F12 for simple nudged-elastic band calculations on the gas-phase reaction.
We find that DFT and coupled-cluster calculations are consistent to within several kcal/mol for transition-state and product energies. 
Coupled-cluster calculations were performed using \texttt{ORCA} \cite{ORCA,ORCA5}.
Further details can be found in the Supplementary Material.

With this optimised dataset, we trained a 2-layer MACE model with 128 equivariant messages and a maximal message equivariance of L = 1.
A cutoff of 5 Å was employed, which translates to an effective receptive field of 10 Å after message passing.  
Whilst long-range effects, important for modeling liquid-vapor interfaces \cite{Niblett2021}, are not explicitly included in our setup, we find that the large receptive field employed by our model is sufficient for recovering the interfacial properties of interest. 
Upon completion of the training, the model was validated against DFT energy and force predictions (RMSEs: 1.5 meV/atom and 28.8 meV/Å, respectively) as well as solvent and solute-solvent structural predictions.
%
A bulk water density of 0.991 ± 0.002 g/ml was attained by this model, consistent to within 3\% of previous revPBE-D3 estimates obtained using CP2K \cite{Galib2017} (see Table S1)
and also with the experimental value of 0.997 g/ml.
%
An interfacial tension (IFT) of 83 $\pm$ 1 mN/m was obtained for the air-water interface, also in good agreement with previous computational estimates and only 10 \% greater than the experimental value \cite{Nagata2016}.
Whilst classical force fields can achieve IFT predictions closer to experiment, it should be noted that these models are usually molecule-specific and generally unable to treat reactive events. 
Our MACE (revPBE-D3) model gives reasonably accurate interfacial predictions whilst simultaneously enabling treatment of bond-making and bond-breaking events. 
Given the remaining uncertainty in IFT, we tested the sensitivity of our main conclusions on the reaction mechanism using two additional MLPs trained to BLYP-D3 and RPA reference data.
Although a different treatment of the electronic structure is used, including beyond DFT methods in the case of RPA, the overall `In-and-Out' reaction mechanism remains unchanged as shown in detail in the supporting information, Figs. S13.
The fact that such results are observed across all three models - revPBE-D3, BLYP-D3, and RPA - gives us confidence that this mechanism is a general phenomenon occurring at air-water interfaces. 
%
We tested the interfacial predictions of our model further using targeted force error calculations and through the comparison of state energy predictions with coupled-cluster theory (see Figure S7). 
We find that, for the various states extracted along the interfacial reaction coordinate, our predicted energies agree with DLPNO-CCSD(T) values to within the combined uncertainties of both methods.
More details are provided in the Supporting Information.

\subsubsection*{System Setup}
Our work employed three system setups differing in size and composition. 
These are shown in Figure \ref{fig:overview}b. 
The gaseous setup consisted of a single \ch{CO2} molecule and a single \ch{H2O} molecule in a box of length 12.42 \AA{}.
The bulk system consisted of a single \ch{CO2} molecule and 127 \ch{H2O} molecules in a 15.72 \AA{} box, chosen to reproduce a bulk density of $\sim$ 1.00 g/ml.
And finally, the interfacial system consisted of a single \ch{CO2} molecule and 180 water molecules in a box of size 15 \AA{} $\times$ 15 \AA{} $\times$ 100 \AA{}. 
Starting configurations for the interfacial setup were initialized such that the interface resided parallel to the xy plane. 
The resulting water slab has a length approaching 25 \AA{}, which allows us to therefore probe the interfacial region within 10 \AA{} of each interface, recovering bulk-like water in the middle for the slab.

\subsubsection*{Metadynamics}
Obtaining meaningful results from enhanced sampling simulations requires careful selection of the underlying collective variables (CVs).
In our case, we made use of a C-O coordination number to track the principle attack of water on \ch{CO2} alongside a protonation state CV to differentiate between bicarbonate and carbonic acid products.  
We use the same underlying function, $s_\mathrm{CO}$, to describe the C-O coordination across all three systems, with only a minimal change in the CV parameters between gaseous and aqueous systems.  
For the protonation state, we define two different collective variables: a simple OH coordination CV, $s_\mathrm{(OH)_g}$ for probing the gas-phase reaction; and a radially dependent OH coordination number, $s_\mathrm{(OH)_{aq.}}$, to account for atom exchange with respect to the carbon-bound oxygens during the bulk and interfacial metadynamics runs. 
Exact details of these CVs are provided Supplementary Material.

Well-tempered metadynamics was performed by coupling \texttt{LAMMPS} with the multiple walker setup available in \texttt{PLUMED} \cite{LAMMPS,Tribello2014,Bonomi2019}. 
Three walkers were used to probe the gaseous reaction, whilst five walkers were used to look at bulk and interfacial reactions. 
Walkers were initiated from different starting configurations, generated from restrained MD simulations effecting the conversion of \ch{CO2} to carbonic acid. 
These configurations were chosen to represent the various species observed in the \ch{CO2 + H2O} reaction and to maximize the initial free energy surface coverage and exploration. 
Simulations were performed under the $NVT$ ensemble at 300 K, using the Nosé-Hoover thermostat with a time constant of 100 fs.
Deuterium masses were used for the hydrogen atoms along with a timestep of 1 fs.
Product free energies were monitored to gauge the profile convergence with time. 
For each system type, we achieve a cumulative simulation time of 50 ns across all walkers. 
Details on the metadynamics setup and the free-energy convergence are shown in the Supplementary Material (Figures S8-S10).

It is important to note that, for the interfacial metadynamics runs, no restraints were applied to the position of the reacting species. 
This means that the reactive species were free to diffuse throughout the interfacial system and adopt their preferred solvation environment. 
This differs from previous studies, which have used harmonic potentials to constrain the $z$ coordinate of the reacting species to a particular height relative to the interface. 
Our rationale for omitting such restraints lies in wanting to fully explore the positional dependence of molecules as a function of the extent of reaction.

\subsubsection*{Data availability}
All data required to reproduce the findings of this study will be made available upon acceptance of this manuscript.


\section*{Acknowledgments}
SGHB is supported by the Syntech CDT and funded by EPSRC (Grant No.\ EP/S024220/1).
V.K. acknowledges support from the Ernest Oppenheimer Early Career Fellowship and the Sydney Harvey Junior Research Fellowship, Churchill College, University of Cambridge. 
AM acknowledges support from the European Union under the “n-AQUA” European Research Council project (Grant No. 101071937).
CS acknowledges financial support from the Deutsche Forschungsgemeinschaft (DFG, German Research Foundation) project number 500244608, as well as from the Royal Society grant number RGS/R2/242614.
We are grateful for computational support and resources from the UK Materials and Molecular Modeling Hub which is partially funded by EPSRC (Grant Nos. EP/P020194/1 and EP/T022213/1).
We are also grateful for computational support and resources from the UK national high-performance computing service, Advanced Research Computing High End Resource (ARCHER2) and the Swiss National Supercomputing Centre under project s1209.
Access for both the UK Materials and Molecular Modeling Hub and ARCHER2 were obtained via the UK Car-Parrinello consortium, funded by EPSRC grant reference EP/P022561/1.
Access to CSD3 was obtained through a University of Cambridge EPSRC Core Equipment Award (EP/X034712/1).



%

%

\begin{thebibliography}{75}%
\makeatletter
\providecommand \@ifxundefined [1]{%
 \@ifx{#1\undefined}
}%
\providecommand \@ifnum [1]{%
 \ifnum #1\expandafter \@firstoftwo
 \else \expandafter \@secondoftwo
 \fi
}%
\providecommand \@ifx [1]{%
 \ifx #1\expandafter \@firstoftwo
 \else \expandafter \@secondoftwo
 \fi
}%
\providecommand \natexlab [1]{#1}%
\providecommand \enquote  [1]{``#1''}%
\providecommand \bibnamefont  [1]{#1}%
\providecommand \bibfnamefont [1]{#1}%
\providecommand \citenamefont [1]{#1}%
\providecommand \href@noop [0]{\@secondoftwo}%
\providecommand \href [0]{\begingroup \@sanitize@url \@href}%
\providecommand \@href[1]{\@@startlink{#1}\@@href}%
\providecommand \@@href[1]{\endgroup#1\@@endlink}%
\providecommand \@sanitize@url [0]{\catcode `\\12\catcode `\$12\catcode
  `\&12\catcode `\#12\catcode `\^12\catcode `\_12\catcode `\%12\relax}%
\providecommand \@@startlink[1]{}%
\providecommand \@@endlink[0]{}%
\providecommand \url  [0]{\begingroup\@sanitize@url \@url }%
\providecommand \@url [1]{\endgroup\@href {#1}{\urlprefix }}%
\providecommand \urlprefix  [0]{URL }%
\providecommand \Eprint [0]{\href }%
\providecommand \doibase [0]{https://doi.org/}%
\providecommand \selectlanguage [0]{\@gobble}%
\providecommand \bibinfo  [0]{\@secondoftwo}%
\providecommand \bibfield  [0]{\@secondoftwo}%
\providecommand \translation [1]{[#1]}%
\providecommand \BibitemOpen [0]{}%
\providecommand \bibitemStop [0]{}%
\providecommand \bibitemNoStop [0]{.\EOS\space}%
\providecommand \EOS [0]{\spacefactor3000\relax}%
\providecommand \BibitemShut  [1]{\csname bibitem#1\endcsname}%
\let\auto@bib@innerbib\@empty
\bibitem [{\citenamefont {Gattuso}\ \emph {et~al.}(2015)\citenamefont
  {Gattuso}, \citenamefont {Magnan}, \citenamefont {Billé}, \citenamefont
  {Cheung}, \citenamefont {Howes}, \citenamefont {Joos}, \citenamefont
  {Allemand}, \citenamefont {Bopp}, \citenamefont {Cooley}, \citenamefont
  {Eakin}, \citenamefont {Hoegh-Guldberg}, \citenamefont {Kelly}, \citenamefont
  {Pörtner}, \citenamefont {Rogers}, \citenamefont {Baxter}, \citenamefont
  {Laffoley}, \citenamefont {Osborn}, \citenamefont {Rankovic}, \citenamefont
  {Rochette}, \citenamefont {Sumaila}, \citenamefont {Treyer},\ and\
  \citenamefont {Turley}}]{Gattuso2015}%
  \BibitemOpen
  \bibfield  {author} {\bibinfo {author} {\bibfnamefont {J.-P.}\ \bibnamefont
  {Gattuso}}, \bibinfo {author} {\bibfnamefont {A.}~\bibnamefont {Magnan}},
  \bibinfo {author} {\bibfnamefont {R.}~\bibnamefont {Billé}}, \bibinfo
  {author} {\bibfnamefont {W.~W.~L.}\ \bibnamefont {Cheung}}, \bibinfo {author}
  {\bibfnamefont {E.~L.}\ \bibnamefont {Howes}}, \bibinfo {author}
  {\bibfnamefont {F.}~\bibnamefont {Joos}}, \bibinfo {author} {\bibfnamefont
  {D.}~\bibnamefont {Allemand}}, \bibinfo {author} {\bibfnamefont
  {L.}~\bibnamefont {Bopp}}, \bibinfo {author} {\bibfnamefont {S.~R.}\
  \bibnamefont {Cooley}}, \bibinfo {author} {\bibfnamefont {C.~M.}\
  \bibnamefont {Eakin}}, \bibinfo {author} {\bibfnamefont {O.}~\bibnamefont
  {Hoegh-Guldberg}}, \bibinfo {author} {\bibfnamefont {R.~P.}\ \bibnamefont
  {Kelly}}, \bibinfo {author} {\bibfnamefont {H.-O.}\ \bibnamefont {Pörtner}},
  \bibinfo {author} {\bibfnamefont {A.~D.}\ \bibnamefont {Rogers}}, \bibinfo
  {author} {\bibfnamefont {J.~M.}\ \bibnamefont {Baxter}}, \bibinfo {author}
  {\bibfnamefont {D.}~\bibnamefont {Laffoley}}, \bibinfo {author}
  {\bibfnamefont {D.}~\bibnamefont {Osborn}}, \bibinfo {author} {\bibfnamefont
  {A.}~\bibnamefont {Rankovic}}, \bibinfo {author} {\bibfnamefont
  {J.}~\bibnamefont {Rochette}}, \bibinfo {author} {\bibfnamefont {U.~R.}\
  \bibnamefont {Sumaila}}, \bibinfo {author} {\bibfnamefont {S.}~\bibnamefont
  {Treyer}},\ and\ \bibinfo {author} {\bibfnamefont {C.}~\bibnamefont
  {Turley}},\ }\bibfield  {title} {\enquote {\bibinfo {title} {Contrasting
  futures for ocean and society from different anthropogenic co2 emissions
  scenarios},}\ }\href {https://doi.org/10.1126/science.aac4722} {\bibfield
  {journal} {\bibinfo  {journal} {Science}\ }\textbf {\bibinfo {volume}
  {349}},\ \bibinfo {pages} {aac4722} (\bibinfo {year} {2015})}\BibitemShut
  {NoStop}%
\bibitem [{\citenamefont {Dutton}\ \emph {et~al.}(2015)\citenamefont {Dutton},
  \citenamefont {Carlson}, \citenamefont {Long}, \citenamefont {Milne},
  \citenamefont {Clark}, \citenamefont {DeConto}, \citenamefont {Horton},
  \citenamefont {Rahmstorf},\ and\ \citenamefont {Raymo}}]{Dutton2015}%
  \BibitemOpen
  \bibfield  {author} {\bibinfo {author} {\bibfnamefont {A.}~\bibnamefont
  {Dutton}}, \bibinfo {author} {\bibfnamefont {A.~E.}\ \bibnamefont {Carlson}},
  \bibinfo {author} {\bibfnamefont {A.~J.}\ \bibnamefont {Long}}, \bibinfo
  {author} {\bibfnamefont {G.~A.}\ \bibnamefont {Milne}}, \bibinfo {author}
  {\bibfnamefont {P.~U.}\ \bibnamefont {Clark}}, \bibinfo {author}
  {\bibfnamefont {R.}~\bibnamefont {DeConto}}, \bibinfo {author} {\bibfnamefont
  {B.~P.}\ \bibnamefont {Horton}}, \bibinfo {author} {\bibfnamefont
  {S.}~\bibnamefont {Rahmstorf}},\ and\ \bibinfo {author} {\bibfnamefont
  {M.~E.}\ \bibnamefont {Raymo}},\ }\bibfield  {title} {\enquote {\bibinfo
  {title} {Sea-level rise due to polar ice-sheet mass loss during past warm
  periods},}\ }\href {https://doi.org/10.1126/science.aaa4019} {\bibfield
  {journal} {\bibinfo  {journal} {Science}\ }\textbf {\bibinfo {volume}
  {349}},\ \bibinfo {pages} {aaa4019} (\bibinfo {year} {2015})}\BibitemShut
  {NoStop}%
\bibitem [{\citenamefont {Dupont}\ and\ \citenamefont
  {Pörtner}(2013)}]{Dupont2013}%
  \BibitemOpen
  \bibfield  {author} {\bibinfo {author} {\bibfnamefont {S.}~\bibnamefont
  {Dupont}}\ and\ \bibinfo {author} {\bibfnamefont {H.}~\bibnamefont
  {Pörtner}},\ }\bibfield  {title} {\enquote {\bibinfo {title} {Get ready for
  ocean acidification},}\ }\href {https://doi.org/10.1038/498429a} {\bibfield
  {journal} {\bibinfo  {journal} {Nature}\ }\textbf {\bibinfo {volume} {498}},\
  \bibinfo {pages} {429} (\bibinfo {year} {2013})}\BibitemShut {NoStop}%
\bibitem [{\citenamefont {Orr}\ \emph {et~al.}(2005)\citenamefont {Orr},
  \citenamefont {Fabry}, \citenamefont {Aumont}, \citenamefont {Bopp},
  \citenamefont {Doney}, \citenamefont {Feely}, \citenamefont {Gnanadesikan},
  \citenamefont {Gruber}, \citenamefont {Ishida}, \citenamefont {Joos},
  \citenamefont {Key}, \citenamefont {Lindsay}, \citenamefont {Maier-Reimer},
  \citenamefont {Matear}, \citenamefont {Monfray}, \citenamefont {Mouchet},
  \citenamefont {Najjar}, \citenamefont {Plattner}, \citenamefont {Rodgers},
  \citenamefont {Sabine}, \citenamefont {Sarmiento}, \citenamefont {Schlitzer},
  \citenamefont {Slater}, \citenamefont {Totterdell}, \citenamefont {Weirig},
  \citenamefont {Yamanaka},\ and\ \citenamefont {Yool}}]{Orr2005}%
  \BibitemOpen
  \bibfield  {author} {\bibinfo {author} {\bibfnamefont {J.~C.}\ \bibnamefont
  {Orr}}, \bibinfo {author} {\bibfnamefont {V.~J.}\ \bibnamefont {Fabry}},
  \bibinfo {author} {\bibfnamefont {O.}~\bibnamefont {Aumont}}, \bibinfo
  {author} {\bibfnamefont {L.}~\bibnamefont {Bopp}}, \bibinfo {author}
  {\bibfnamefont {S.~C.}\ \bibnamefont {Doney}}, \bibinfo {author}
  {\bibfnamefont {R.~A.}\ \bibnamefont {Feely}}, \bibinfo {author}
  {\bibfnamefont {A.}~\bibnamefont {Gnanadesikan}}, \bibinfo {author}
  {\bibfnamefont {N.}~\bibnamefont {Gruber}}, \bibinfo {author} {\bibfnamefont
  {A.}~\bibnamefont {Ishida}}, \bibinfo {author} {\bibfnamefont
  {F.}~\bibnamefont {Joos}}, \bibinfo {author} {\bibfnamefont {R.~M.}\
  \bibnamefont {Key}}, \bibinfo {author} {\bibfnamefont {K.}~\bibnamefont
  {Lindsay}}, \bibinfo {author} {\bibfnamefont {E.}~\bibnamefont
  {Maier-Reimer}}, \bibinfo {author} {\bibfnamefont {R.}~\bibnamefont
  {Matear}}, \bibinfo {author} {\bibfnamefont {P.}~\bibnamefont {Monfray}},
  \bibinfo {author} {\bibfnamefont {A.}~\bibnamefont {Mouchet}}, \bibinfo
  {author} {\bibfnamefont {R.~G.}\ \bibnamefont {Najjar}}, \bibinfo {author}
  {\bibfnamefont {G.-K.}\ \bibnamefont {Plattner}}, \bibinfo {author}
  {\bibfnamefont {K.~B.}\ \bibnamefont {Rodgers}}, \bibinfo {author}
  {\bibfnamefont {C.~L.}\ \bibnamefont {Sabine}}, \bibinfo {author}
  {\bibfnamefont {J.~L.}\ \bibnamefont {Sarmiento}}, \bibinfo {author}
  {\bibfnamefont {R.}~\bibnamefont {Schlitzer}}, \bibinfo {author}
  {\bibfnamefont {R.~D.}\ \bibnamefont {Slater}}, \bibinfo {author}
  {\bibfnamefont {I.~J.}\ \bibnamefont {Totterdell}}, \bibinfo {author}
  {\bibfnamefont {M.-F.}\ \bibnamefont {Weirig}}, \bibinfo {author}
  {\bibfnamefont {Y.}~\bibnamefont {Yamanaka}},\ and\ \bibinfo {author}
  {\bibfnamefont {A.}~\bibnamefont {Yool}},\ }\bibfield  {title} {\enquote
  {\bibinfo {title} {Anthropogenic ocean acidification over the twenty-first
  century and its impact on calcifying organisms},}\ }\href
  {https://doi.org/10.1038/nature04095} {\bibfield  {journal} {\bibinfo
  {journal} {Nature}\ }\textbf {\bibinfo {volume} {437}},\ \bibinfo {pages}
  {681--686} (\bibinfo {year} {2005})}\BibitemShut {NoStop}%
\bibitem [{\citenamefont {Kroeker}, \citenamefont {Micheli},\ and\
  \citenamefont {Gambi}(2013)}]{Kroeker2013}%
  \BibitemOpen
  \bibfield  {author} {\bibinfo {author} {\bibfnamefont {K.~J.}\ \bibnamefont
  {Kroeker}}, \bibinfo {author} {\bibfnamefont {F.}~\bibnamefont {Micheli}},\
  and\ \bibinfo {author} {\bibfnamefont {M.~C.}\ \bibnamefont {Gambi}},\
  }\bibfield  {title} {\enquote {\bibinfo {title} {Ocean acidification causes
  ecosystem shifts via altered competitive interactions},}\ }\href
  {https://doi.org/10.1038/nclimate1680} {\bibfield  {journal} {\bibinfo
  {journal} {Nature Climate Change}\ }\textbf {\bibinfo {volume} {3}},\
  \bibinfo {pages} {156--159} (\bibinfo {year} {2013})}\BibitemShut {NoStop}%
\bibitem [{\citenamefont {Foster}\ \emph {et~al.}(2016)\citenamefont {Foster},
  \citenamefont {Falter}, \citenamefont {McCulloch},\ and\ \citenamefont
  {Clode}}]{Foster2016}%
  \BibitemOpen
  \bibfield  {author} {\bibinfo {author} {\bibfnamefont {T.}~\bibnamefont
  {Foster}}, \bibinfo {author} {\bibfnamefont {J.~L.}\ \bibnamefont {Falter}},
  \bibinfo {author} {\bibfnamefont {M.~T.}\ \bibnamefont {McCulloch}},\ and\
  \bibinfo {author} {\bibfnamefont {P.~L.}\ \bibnamefont {Clode}},\ }\bibfield
  {title} {\enquote {\bibinfo {title} {Ocean acidification causes structural
  deformities in juvenile coral skeletons},}\ }\href
  {https://doi.org/10.1126/sciadv.1501130} {\bibfield  {journal} {\bibinfo
  {journal} {Science Advances}\ }\textbf {\bibinfo {volume} {2}},\ \bibinfo
  {pages} {e1501130} (\bibinfo {year} {2016})}\BibitemShut {NoStop}%
\bibitem [{\citenamefont {Leung}, \citenamefont {Caramanna},\ and\
  \citenamefont {Maroto-Valer}(2014)}]{Leung2014}%
  \BibitemOpen
  \bibfield  {author} {\bibinfo {author} {\bibfnamefont {D.~Y.~C.}\
  \bibnamefont {Leung}}, \bibinfo {author} {\bibfnamefont {G.}~\bibnamefont
  {Caramanna}},\ and\ \bibinfo {author} {\bibfnamefont {M.~M.}\ \bibnamefont
  {Maroto-Valer}},\ }\bibfield  {title} {\enquote {\bibinfo {title} {An
  overview of current status of carbon dioxide capture and storage
  technologies},}\ }\href
  {https://doi.org/https://doi.org/10.1016/j.rser.2014.07.093} {\bibfield
  {journal} {\bibinfo  {journal} {Renewable and Sustainable Energy Reviews}\
  }\textbf {\bibinfo {volume} {39}},\ \bibinfo {pages} {426--443} (\bibinfo
  {year} {2014})}\BibitemShut {NoStop}%
\bibitem [{\citenamefont {Sn\{\ae\}björnsdóttir}\ \emph
  {et~al.}(2020)\citenamefont {Sn\{\ae\}björnsdóttir}, \citenamefont
  {Sigfússon}, \citenamefont {Marieni}, \citenamefont {Goldberg},
  \citenamefont {Gislason},\ and\ \citenamefont
  {Oelkers}}]{Snaebjornsdottir2020}%
  \BibitemOpen
  \bibfield  {author} {\bibinfo {author} {\bibfnamefont {S.~O.}\ \bibnamefont
  {Sn\{\ae\}björnsdóttir}}, \bibinfo {author} {\bibfnamefont
  {B.}~\bibnamefont {Sigfússon}}, \bibinfo {author} {\bibfnamefont
  {C.}~\bibnamefont {Marieni}}, \bibinfo {author} {\bibfnamefont
  {D.}~\bibnamefont {Goldberg}}, \bibinfo {author} {\bibfnamefont {S.~R.}\
  \bibnamefont {Gislason}},\ and\ \bibinfo {author} {\bibfnamefont {E.~H.}\
  \bibnamefont {Oelkers}},\ }\bibfield  {title} {\enquote {\bibinfo {title}
  {Carbon dioxide storage through mineral carbonation},}\ }\href
  {https://doi.org/10.1038/s43017-019-0011-8} {\bibfield  {journal} {\bibinfo
  {journal} {Nature Reviews Earth \& Environment}\ }\textbf {\bibinfo {volume}
  {1}},\ \bibinfo {pages} {90--102} (\bibinfo {year} {2020})}\BibitemShut
  {NoStop}%
\bibitem [{\citenamefont {Wang}\ \emph {et~al.}(2021)\citenamefont {Wang},
  \citenamefont {Zhang}, \citenamefont {Li}, \citenamefont {Zhang},\ and\
  \citenamefont {Deng}}]{Wang2021}%
  \BibitemOpen
  \bibfield  {author} {\bibinfo {author} {\bibfnamefont {X.}~\bibnamefont
  {Wang}}, \bibinfo {author} {\bibfnamefont {F.}~\bibnamefont {Zhang}},
  \bibinfo {author} {\bibfnamefont {L.}~\bibnamefont {Li}}, \bibinfo {author}
  {\bibfnamefont {H.}~\bibnamefont {Zhang}},\ and\ \bibinfo {author}
  {\bibfnamefont {S.}~\bibnamefont {Deng}},\ }\bibfield  {title} {\enquote
  {\bibinfo {title} {Carbon dioxide capture},}\ }\href
  {https://doi.org/10.1016/bs.ache.2021.10.005} {\bibfield  {journal} {\bibinfo
   {journal} {Advances in Chemical Engineering}\ }\textbf {\bibinfo {volume}
  {58}},\ \bibinfo {pages} {297--348} (\bibinfo {year} {2021})}\BibitemShut
  {NoStop}%
\bibitem [{\citenamefont {Wang}\ \emph {et~al.}(2010)\citenamefont {Wang},
  \citenamefont {Conway}, \citenamefont {Burns}, \citenamefont {McCann},\ and\
  \citenamefont {Maeder}}]{Wang2010}%
  \BibitemOpen
  \bibfield  {author} {\bibinfo {author} {\bibfnamefont {X.}~\bibnamefont
  {Wang}}, \bibinfo {author} {\bibfnamefont {W.}~\bibnamefont {Conway}},
  \bibinfo {author} {\bibfnamefont {R.}~\bibnamefont {Burns}}, \bibinfo
  {author} {\bibfnamefont {N.}~\bibnamefont {McCann}},\ and\ \bibinfo {author}
  {\bibfnamefont {M.}~\bibnamefont {Maeder}},\ }\bibfield  {title} {\enquote
  {\bibinfo {title} {Comprehensive study of the hydration and dehydration
  reactions of carbon dioxide in aqueous solution},}\ }\href
  {https://doi.org/10.1021/jp909019u} {\bibfield  {journal} {\bibinfo
  {journal} {The Journal of Physical Chemistry A}\ }\textbf {\bibinfo {volume}
  {114}},\ \bibinfo {pages} {1734--1740} (\bibinfo {year} {2010})}\BibitemShut
  {NoStop}%
\bibitem [{\citenamefont {Adamczyk}\ \emph {et~al.}(2009)\citenamefont
  {Adamczyk}, \citenamefont {Prémont-Schwarz}, \citenamefont {Pines},
  \citenamefont {Pines},\ and\ \citenamefont {Nibbering}}]{Adamczyk2009}%
  \BibitemOpen
  \bibfield  {author} {\bibinfo {author} {\bibfnamefont {K.}~\bibnamefont
  {Adamczyk}}, \bibinfo {author} {\bibfnamefont {M.}~\bibnamefont
  {Prémont-Schwarz}}, \bibinfo {author} {\bibfnamefont {D.}~\bibnamefont
  {Pines}}, \bibinfo {author} {\bibfnamefont {E.}~\bibnamefont {Pines}},\ and\
  \bibinfo {author} {\bibfnamefont {E.~T.~J.}\ \bibnamefont {Nibbering}},\
  }\bibfield  {title} {\enquote {\bibinfo {title} {Real-time observation of
  carbonic acid formation in aqueous solution},}\ }\href
  {https://doi.org/10.1126/science.1180060} {\bibfield  {journal} {\bibinfo
  {journal} {Science}\ }\textbf {\bibinfo {volume} {326}},\ \bibinfo {pages}
  {1690--1694} (\bibinfo {year} {2009})}\BibitemShut {NoStop}%
\bibitem [{\citenamefont {Pines}\ \emph {et~al.}(2016)\citenamefont {Pines},
  \citenamefont {Ditkovich}, \citenamefont {Mukra}, \citenamefont {Miller},
  \citenamefont {Kiefer}, \citenamefont {Daschakraborty}, \citenamefont
  {Hynes},\ and\ \citenamefont {Pines}}]{Pines2016}%
  \BibitemOpen
  \bibfield  {author} {\bibinfo {author} {\bibfnamefont {D.}~\bibnamefont
  {Pines}}, \bibinfo {author} {\bibfnamefont {J.}~\bibnamefont {Ditkovich}},
  \bibinfo {author} {\bibfnamefont {T.}~\bibnamefont {Mukra}}, \bibinfo
  {author} {\bibfnamefont {Y.}~\bibnamefont {Miller}}, \bibinfo {author}
  {\bibfnamefont {P.~M.}\ \bibnamefont {Kiefer}}, \bibinfo {author}
  {\bibfnamefont {S.}~\bibnamefont {Daschakraborty}}, \bibinfo {author}
  {\bibfnamefont {J.~T.}\ \bibnamefont {Hynes}},\ and\ \bibinfo {author}
  {\bibfnamefont {E.}~\bibnamefont {Pines}},\ }\bibfield  {title} {\enquote
  {\bibinfo {title} {How acidic is carbonic acid?}}\ }\href
  {https://doi.org/10.1021/acs.jpcb.5b12428} {\bibfield  {journal} {\bibinfo
  {journal} {The Journal of Physical Chemistry B}\ }\textbf {\bibinfo {volume}
  {120}},\ \bibinfo {pages} {2440--2451} (\bibinfo {year} {2016})}\BibitemShut
  {NoStop}%
\bibitem [{\citenamefont {Leung}, \citenamefont {Nielsen},\ and\ \citenamefont
  {Kurtz}(2007)}]{Leung2007}%
  \BibitemOpen
  \bibfield  {author} {\bibinfo {author} {\bibfnamefont {K.}~\bibnamefont
  {Leung}}, \bibinfo {author} {\bibfnamefont {I.~M.~B.}\ \bibnamefont
  {Nielsen}},\ and\ \bibinfo {author} {\bibfnamefont {I.}~\bibnamefont
  {Kurtz}},\ }\bibfield  {title} {\enquote {\bibinfo {title} {Ab initio
  molecular dynamics study of carbon dioxide and bicarbonate hydration and the
  nucleophilic attack of hydroxide on co2},}\ }\href
  {https://doi.org/10.1021/jp068475l} {\bibfield  {journal} {\bibinfo
  {journal} {The Journal of Physical Chemistry B}\ }\textbf {\bibinfo {volume}
  {111}},\ \bibinfo {pages} {4453--4459} (\bibinfo {year} {2007})}\BibitemShut
  {NoStop}%
\bibitem [{\citenamefont {Nguyen}\ \emph {et~al.}(2008)\citenamefont {Nguyen},
  \citenamefont {Matus}, \citenamefont {Jackson}, \citenamefont {Ngan},
  \citenamefont {Rustad},\ and\ \citenamefont {Dixon}}]{Nguyen2008}%
  \BibitemOpen
  \bibfield  {author} {\bibinfo {author} {\bibfnamefont {M.~T.}\ \bibnamefont
  {Nguyen}}, \bibinfo {author} {\bibfnamefont {M.~H.}\ \bibnamefont {Matus}},
  \bibinfo {author} {\bibfnamefont {V.~E.}\ \bibnamefont {Jackson}}, \bibinfo
  {author} {\bibfnamefont {V.~T.}\ \bibnamefont {Ngan}}, \bibinfo {author}
  {\bibfnamefont {J.~R.}\ \bibnamefont {Rustad}},\ and\ \bibinfo {author}
  {\bibfnamefont {D.~A.}\ \bibnamefont {Dixon}},\ }\bibfield  {title} {\enquote
  {\bibinfo {title} {Mechanism of the hydration of carbon dioxide: Direct
  participation of h2o versus microsolvation},}\ }\href
  {https://doi.org/10.1021/jp804715j} {\bibfield  {journal} {\bibinfo
  {journal} {The Journal of Physical Chemistry A}\ }\textbf {\bibinfo {volume}
  {112}},\ \bibinfo {pages} {10386--10398} (\bibinfo {year}
  {2008})}\BibitemShut {NoStop}%
\bibitem [{\citenamefont {Kumar}, \citenamefont {Kalinichev},\ and\
  \citenamefont {Kirkpatrick}(2007)}]{Kumar2007}%
  \BibitemOpen
  \bibfield  {author} {\bibinfo {author} {\bibfnamefont {P.~P.}\ \bibnamefont
  {Kumar}}, \bibinfo {author} {\bibfnamefont {A.~G.}\ \bibnamefont
  {Kalinichev}},\ and\ \bibinfo {author} {\bibfnamefont {R.~J.}\ \bibnamefont
  {Kirkpatrick}},\ }\bibfield  {title} {\enquote {\bibinfo {title}
  {Dissociation of carbonic acid: Gas phase energetics and mechanism from ab
  initio metadynamics simulations},}\ }\href
  {https://doi.org/10.1063/1.2741552} {\bibfield  {journal} {\bibinfo
  {journal} {The Journal of Chemical Physics}\ }\textbf {\bibinfo {volume}
  {126}},\ \bibinfo {pages} {204315} (\bibinfo {year} {2007})}\BibitemShut
  {NoStop}%
\bibitem [{\citenamefont {Kumar}, \citenamefont {Kalinichev},\ and\
  \citenamefont {Kirkpatrick}(2009)}]{Kumar2009}%
  \BibitemOpen
  \bibfield  {author} {\bibinfo {author} {\bibfnamefont {P.~P.}\ \bibnamefont
  {Kumar}}, \bibinfo {author} {\bibfnamefont {A.~G.}\ \bibnamefont
  {Kalinichev}},\ and\ \bibinfo {author} {\bibfnamefont {R.~J.}\ \bibnamefont
  {Kirkpatrick}},\ }\bibfield  {title} {\enquote {\bibinfo {title}
  {Hydrogen-bonding structure and dynamics of aqueous carbonate species from
  car−parrinello molecular dynamics simulations},}\ }\href
  {https://doi.org/10.1021/jp809069g} {\bibfield  {journal} {\bibinfo
  {journal} {The Journal of Physical Chemistry B}\ }\textbf {\bibinfo {volume}
  {113}},\ \bibinfo {pages} {794--802} (\bibinfo {year} {2009})}\BibitemShut
  {NoStop}%
\bibitem [{\citenamefont {Stirling}\ and\ \citenamefont
  {Pápai}(2010)}]{Stirling2010}%
  \BibitemOpen
  \bibfield  {author} {\bibinfo {author} {\bibfnamefont {A.}~\bibnamefont
  {Stirling}}\ and\ \bibinfo {author} {\bibfnamefont {I.}~\bibnamefont
  {Pápai}},\ }\bibfield  {title} {\enquote {\bibinfo {title} {H2co3 forms via
  hco3− in water},}\ }\href {https://doi.org/10.1021/jp1099909} {\bibfield
  {journal} {\bibinfo  {journal} {The Journal of Physical Chemistry B}\
  }\textbf {\bibinfo {volume} {114}},\ \bibinfo {pages} {16854--16859}
  (\bibinfo {year} {2010})}\BibitemShut {NoStop}%
\bibitem [{\citenamefont {England}\ \emph {et~al.}(2011)\citenamefont
  {England}, \citenamefont {Duffin}, \citenamefont {Schwartz}, \citenamefont
  {Uejio}, \citenamefont {Prendergast},\ and\ \citenamefont
  {Saykally}}]{England2011}%
  \BibitemOpen
  \bibfield  {author} {\bibinfo {author} {\bibfnamefont {A.~H.}\ \bibnamefont
  {England}}, \bibinfo {author} {\bibfnamefont {A.~M.}\ \bibnamefont {Duffin}},
  \bibinfo {author} {\bibfnamefont {C.~P.}\ \bibnamefont {Schwartz}}, \bibinfo
  {author} {\bibfnamefont {J.~S.}\ \bibnamefont {Uejio}}, \bibinfo {author}
  {\bibfnamefont {D.}~\bibnamefont {Prendergast}},\ and\ \bibinfo {author}
  {\bibfnamefont {R.~J.}\ \bibnamefont {Saykally}},\ }\bibfield  {title}
  {\enquote {\bibinfo {title} {On the hydration and hydrolysis of carbon
  dioxide},}\ }\href
  {https://doi.org/https://doi.org/10.1016/j.cplett.2011.08.063} {\bibfield
  {journal} {\bibinfo  {journal} {Chemical Physics Letters}\ }\textbf {\bibinfo
  {volume} {514}},\ \bibinfo {pages} {187--195} (\bibinfo {year}
  {2011})}\BibitemShut {NoStop}%
\bibitem [{\citenamefont {Wang}\ and\ \citenamefont {Cao}(2013)}]{Wang_2013}%
  \BibitemOpen
  \bibfield  {author} {\bibinfo {author} {\bibfnamefont {B.}~\bibnamefont
  {Wang}}\ and\ \bibinfo {author} {\bibfnamefont {Z.}~\bibnamefont {Cao}},\
  }\bibfield  {title} {\enquote {\bibinfo {title} {How water molecules modulate
  the hydration of co2 in water solution: Insight from the cluster-continuum
  model calculations},}\ }\href
  {https://doi.org/https://doi.org/10.1002/jcc.23144} {\bibfield  {journal}
  {\bibinfo  {journal} {Journal of Computational Chemistry}\ }\textbf {\bibinfo
  {volume} {34}},\ \bibinfo {pages} {372--378} (\bibinfo {year}
  {2013})}\BibitemShut {NoStop}%
\bibitem [{\citenamefont {Polino}\ \emph {et~al.}(2020)\citenamefont {Polino},
  \citenamefont {Grifoni}, \citenamefont {Rousseau}, \citenamefont
  {Parrinello},\ and\ \citenamefont {Glezakou}}]{Polino2020}%
  \BibitemOpen
  \bibfield  {author} {\bibinfo {author} {\bibfnamefont {D.}~\bibnamefont
  {Polino}}, \bibinfo {author} {\bibfnamefont {E.}~\bibnamefont {Grifoni}},
  \bibinfo {author} {\bibfnamefont {R.}~\bibnamefont {Rousseau}}, \bibinfo
  {author} {\bibfnamefont {M.}~\bibnamefont {Parrinello}},\ and\ \bibinfo
  {author} {\bibfnamefont {V.-A.}\ \bibnamefont {Glezakou}},\ }\bibfield
  {title} {\enquote {\bibinfo {title} {How collective phenomena impact co2
  reactivity and speciation in different media},}\ }\href
  {https://doi.org/10.1021/acs.jpca.9b11744} {\bibfield  {journal} {\bibinfo
  {journal} {The Journal of Physical Chemistry A}\ }\textbf {\bibinfo {volume}
  {124}},\ \bibinfo {pages} {3963--3975} (\bibinfo {year} {2020})}\BibitemShut
  {NoStop}%
\bibitem [{\citenamefont {Martirez}\ and\ \citenamefont
  {Carter}(2023)}]{Martirez2023}%
  \BibitemOpen
  \bibfield  {author} {\bibinfo {author} {\bibfnamefont {J.~M.~P.}\
  \bibnamefont {Martirez}}\ and\ \bibinfo {author} {\bibfnamefont {E.~A.}\
  \bibnamefont {Carter}},\ }\bibfield  {title} {\enquote {\bibinfo {title}
  {Solvent dynamics are critical to understanding carbon dioxide dissolution
  and hydration in water},}\ }\href {https://doi.org/10.1021/jacs.3c01283}
  {\bibfield  {journal} {\bibinfo  {journal} {Journal of the American Chemical
  Society}\ }\textbf {\bibinfo {volume} {145}},\ \bibinfo {pages}
  {12561--12575} (\bibinfo {year} {2023})}\BibitemShut {NoStop}%
\bibitem [{\citenamefont {Bobell}\ \emph {et~al.}(2024)\citenamefont {Bobell},
  \citenamefont {Boyn}, \citenamefont {Martirez},\ and\ \citenamefont
  {Carter}}]{Bobell2024}%
  \BibitemOpen
  \bibfield  {author} {\bibinfo {author} {\bibfnamefont {B.}~\bibnamefont
  {Bobell}}, \bibinfo {author} {\bibfnamefont {J.-N.}\ \bibnamefont {Boyn}},
  \bibinfo {author} {\bibfnamefont {J.~M.~P.}\ \bibnamefont {Martirez}},\ and\
  \bibinfo {author} {\bibfnamefont {E.~A.}\ \bibnamefont {Carter}},\ }\bibfield
   {title} {\enquote {\bibinfo {title} {Modeling bicarbonate formation in an
  alkaline solution with multi-level quantum mechanics/molecular dynamics
  simulations},}\ }\href {https://doi.org/10.1080/00268976.2024.2375370}
  {\bibfield  {journal} {\bibinfo  {journal} {Molecular Physics}\ ,\ \bibinfo
  {pages} {e2375370}} (\bibinfo {year} {2024})}\BibitemShut {NoStop}%
\bibitem [{\citenamefont {Baer}, \citenamefont {Tobias},\ and\ \citenamefont
  {Mundy}(2014)}]{Baer2014}%
  \BibitemOpen
  \bibfield  {author} {\bibinfo {author} {\bibfnamefont {M.~D.}\ \bibnamefont
  {Baer}}, \bibinfo {author} {\bibfnamefont {D.~J.}\ \bibnamefont {Tobias}},\
  and\ \bibinfo {author} {\bibfnamefont {C.~J.}\ \bibnamefont {Mundy}},\
  }\bibfield  {title} {\enquote {\bibinfo {title} {Investigation of interfacial
  and bulk dissociation of hbr, hcl, and hno3 using density functional
  theory-based molecular dynamics simulations},}\ }\href
  {https://doi.org/10.1021/jp5062896} {\bibfield  {journal} {\bibinfo
  {journal} {The Journal of Physical Chemistry C}\ }\textbf {\bibinfo {volume}
  {118}},\ \bibinfo {pages} {29412--29420} (\bibinfo {year}
  {2014})}\BibitemShut {NoStop}%
\bibitem [{\citenamefont {Lowe}, \citenamefont {Skylaris},\ and\ \citenamefont
  {Green}(2015)}]{Lowe2015}%
  \BibitemOpen
  \bibfield  {author} {\bibinfo {author} {\bibfnamefont {B.~M.}\ \bibnamefont
  {Lowe}}, \bibinfo {author} {\bibfnamefont {C.-K.}\ \bibnamefont {Skylaris}},\
  and\ \bibinfo {author} {\bibfnamefont {N.~G.}\ \bibnamefont {Green}},\
  }\bibfield  {title} {\enquote {\bibinfo {title} {Acid-base dissociation
  mechanisms and energetics at the silica–water interface: An activationless
  process},}\ }\href
  {https://doi.org/https://doi.org/10.1016/j.jcis.2015.01.094} {\bibfield
  {journal} {\bibinfo  {journal} {Journal of Colloid and Interface Science}\
  }\textbf {\bibinfo {volume} {451}},\ \bibinfo {pages} {231--244} (\bibinfo
  {year} {2015})}\BibitemShut {NoStop}%
\bibitem [{\citenamefont {Wei}\ \emph {et~al.}(2020)\citenamefont {Wei},
  \citenamefont {Li}, \citenamefont {Cooks},\ and\ \citenamefont
  {Yan}}]{nurev-physchem-121319-110654}%
  \BibitemOpen
  \bibfield  {author} {\bibinfo {author} {\bibfnamefont {Z.}~\bibnamefont
  {Wei}}, \bibinfo {author} {\bibfnamefont {Y.}~\bibnamefont {Li}}, \bibinfo
  {author} {\bibfnamefont {R.~G.}\ \bibnamefont {Cooks}},\ and\ \bibinfo
  {author} {\bibfnamefont {X.}~\bibnamefont {Yan}},\ }\bibfield  {title}
  {\enquote {\bibinfo {title} {Accelerated reaction kinetics in microdroplets:
  Overview and recent developments},}\ }\href
  {https://doi.org/https://doi.org/10.1146/annurev-physchem-121319-110654}
  {\bibfield  {journal} {\bibinfo  {journal} {Annual Review of Physical
  Chemistry}\ }\textbf {\bibinfo {volume} {71}},\ \bibinfo {pages} {31--51}
  (\bibinfo {year} {2020})}\BibitemShut {NoStop}%
\bibitem [{\citenamefont {Kusaka}, \citenamefont {Nihonyanagi},\ and\
  \citenamefont {Tahara}(2021)}]{Kusaka2021}%
  \BibitemOpen
  \bibfield  {author} {\bibinfo {author} {\bibfnamefont {R.}~\bibnamefont
  {Kusaka}}, \bibinfo {author} {\bibfnamefont {S.}~\bibnamefont
  {Nihonyanagi}},\ and\ \bibinfo {author} {\bibfnamefont {T.}~\bibnamefont
  {Tahara}},\ }\bibfield  {title} {\enquote {\bibinfo {title} {The
  photochemical reaction of phenol becomes ultrafast at the air–water
  interface},}\ }\href {https://doi.org/10.1038/s41557-020-00619-5} {\bibfield
  {journal} {\bibinfo  {journal} {Nature Chemistry}\ }\textbf {\bibinfo
  {volume} {13}},\ \bibinfo {pages} {306--311} (\bibinfo {year}
  {2021})}\BibitemShut {NoStop}%
\bibitem [{\citenamefont {Lee}\ \emph {et~al.}(2024)\citenamefont {Lee},
  \citenamefont {Cho}, \citenamefont {Kim}, \citenamefont {Choi}, \citenamefont
  {Kim}, \citenamefont {Lee}, \citenamefont {Li}, \citenamefont {Kwak},\ and\
  \citenamefont {Choi}}]{Lee2024}%
  \BibitemOpen
  \bibfield  {author} {\bibinfo {author} {\bibfnamefont {K.}~\bibnamefont
  {Lee}}, \bibinfo {author} {\bibfnamefont {Y.}~\bibnamefont {Cho}}, \bibinfo
  {author} {\bibfnamefont {J.~C.}\ \bibnamefont {Kim}}, \bibinfo {author}
  {\bibfnamefont {C.}~\bibnamefont {Choi}}, \bibinfo {author} {\bibfnamefont
  {J.}~\bibnamefont {Kim}}, \bibinfo {author} {\bibfnamefont {J.~K.}\
  \bibnamefont {Lee}}, \bibinfo {author} {\bibfnamefont {S.}~\bibnamefont
  {Li}}, \bibinfo {author} {\bibfnamefont {S.~K.}\ \bibnamefont {Kwak}},\ and\
  \bibinfo {author} {\bibfnamefont {S.~Q.}\ \bibnamefont {Choi}},\ }\bibfield
  {title} {\enquote {\bibinfo {title} {Catalyst-free selective oxidation of
  c(sp3)-h bonds in toluene on water},}\ }\href
  {https://doi.org/10.1038/s41467-024-50352-7} {\bibfield  {journal} {\bibinfo
  {journal} {Nature Communications}\ }\textbf {\bibinfo {volume} {15}},\
  \bibinfo {pages} {6127} (\bibinfo {year} {2024})}\BibitemShut {NoStop}%
\bibitem [{\citenamefont {Tarbuck}\ and\ \citenamefont
  {Richmond}(2006)}]{Tarbuck2006a}%
  \BibitemOpen
  \bibfield  {author} {\bibinfo {author} {\bibfnamefont {T.~L.}\ \bibnamefont
  {Tarbuck}}\ and\ \bibinfo {author} {\bibfnamefont {G.~L.}\ \bibnamefont
  {Richmond}},\ }\bibfield  {title} {\enquote {\bibinfo {title} {Adsorption and
  reaction of co2 and so2 at a water surface},}\ }\href
  {https://doi.org/10.1021/ja057375a} {\bibfield  {journal} {\bibinfo
  {journal} {Journal of the American Chemical Society}\ }\textbf {\bibinfo
  {volume} {128}},\ \bibinfo {pages} {3256--3267} (\bibinfo {year}
  {2006})}\BibitemShut {NoStop}%
\bibitem [{\citenamefont {Devlin}\ \emph {et~al.}(2023)\citenamefont {Devlin},
  \citenamefont {Jamnuch}, \citenamefont {Xu}, \citenamefont {Chen},
  \citenamefont {Qian}, \citenamefont {Pascal},\ and\ \citenamefont
  {Saykally}}]{Devlin2023}%
  \BibitemOpen
  \bibfield  {author} {\bibinfo {author} {\bibfnamefont {S.~W.}\ \bibnamefont
  {Devlin}}, \bibinfo {author} {\bibfnamefont {S.}~\bibnamefont {Jamnuch}},
  \bibinfo {author} {\bibfnamefont {Q.}~\bibnamefont {Xu}}, \bibinfo {author}
  {\bibfnamefont {A.~A.}\ \bibnamefont {Chen}}, \bibinfo {author}
  {\bibfnamefont {J.}~\bibnamefont {Qian}}, \bibinfo {author} {\bibfnamefont
  {T.~A.}\ \bibnamefont {Pascal}},\ and\ \bibinfo {author} {\bibfnamefont
  {R.~J.}\ \bibnamefont {Saykally}},\ }\bibfield  {title} {\enquote {\bibinfo
  {title} {1},}\ }\href {https://doi.org/10.1021/jacs.3c05093} {\bibfield
  {journal} {\bibinfo  {journal} {Journal of the American Chemical Society}\
  }\textbf {\bibinfo {volume} {145}},\ \bibinfo {pages} {22384--22393}
  (\bibinfo {year} {2023})}\BibitemShut {NoStop}%
\bibitem [{\citenamefont {Galib}\ and\ \citenamefont
  {Limmer}(2021)}]{doi:10.1126/science.abd7716}%
  \BibitemOpen
  \bibfield  {author} {\bibinfo {author} {\bibfnamefont {M.}~\bibnamefont
  {Galib}}\ and\ \bibinfo {author} {\bibfnamefont {D.~T.}\ \bibnamefont
  {Limmer}},\ }\bibfield  {title} {\enquote {\bibinfo {title} {Reactive uptake
  of n2o5 by atmospheric aerosol is dominated by interfacial processes},}\
  }\href {https://doi.org/10.1126/science.abd7716} {\bibfield  {journal}
  {\bibinfo  {journal} {Science}\ }\textbf {\bibinfo {volume} {371}},\ \bibinfo
  {pages} {921--925} (\bibinfo {year} {2021})}\BibitemShut {NoStop}%
\bibitem [{\citenamefont {de~la Puente}\ \emph {et~al.}(2022)\citenamefont
  {de~la Puente}, \citenamefont {David}, \citenamefont {Gomez},\ and\
  \citenamefont {Laage}}]{DelaPuente2022}%
  \BibitemOpen
  \bibfield  {author} {\bibinfo {author} {\bibfnamefont {M.}~\bibnamefont
  {de~la Puente}}, \bibinfo {author} {\bibfnamefont {R.}~\bibnamefont {David}},
  \bibinfo {author} {\bibfnamefont {A.}~\bibnamefont {Gomez}},\ and\ \bibinfo
  {author} {\bibfnamefont {D.}~\bibnamefont {Laage}},\ }\bibfield  {title}
  {\enquote {\bibinfo {title} {Acids at the edge: Why nitric and formic acid
  dissociations at air–water interfaces depend on depth and on interface
  specific area},}\ }\href {https://doi.org/10.1021/jacs.2c03099} {\bibfield
  {journal} {\bibinfo  {journal} {Journal of the American Chemical Society}\
  }\textbf {\bibinfo {volume} {144}},\ \bibinfo {pages} {10524--10529}
  (\bibinfo {year} {2022})}\BibitemShut {NoStop}%
\bibitem [{\citenamefont {Kapil}\ \emph {et~al.}(2024)\citenamefont {Kapil},
  \citenamefont {Kovács}, \citenamefont {Csányi},\ and\ \citenamefont
  {Michaelides}}]{Kapil2024}%
  \BibitemOpen
  \bibfield  {author} {\bibinfo {author} {\bibfnamefont {V.}~\bibnamefont
  {Kapil}}, \bibinfo {author} {\bibfnamefont {D.~P.}\ \bibnamefont {Kovács}},
  \bibinfo {author} {\bibfnamefont {G.}~\bibnamefont {Csányi}},\ and\ \bibinfo
  {author} {\bibfnamefont {A.}~\bibnamefont {Michaelides}},\ }\bibfield
  {title} {\enquote {\bibinfo {title} {First-principles spectroscopy of aqueous
  interfaces using machine-learned electronic and quantum nuclear effects},}\
  }\href {https://doi.org/10.1039/D3FD00113J} {\bibfield  {journal} {\bibinfo
  {journal} {Faraday Discussions}\ }\textbf {\bibinfo {volume} {249}},\
  \bibinfo {pages} {50--68} (\bibinfo {year} {2024})}\BibitemShut {NoStop}%
\bibitem [{\citenamefont {Buttersack}\ \emph {et~al.}(2024)\citenamefont
  {Buttersack}, \citenamefont {Gladich}, \citenamefont {Gholami}, \citenamefont
  {Richter}, \citenamefont {Dupuy}, \citenamefont {Nicolas}, \citenamefont
  {Trinter}, \citenamefont {Trunschke}, \citenamefont {Delgado}, \citenamefont
  {Arroyo}, \citenamefont {Parmentier}, \citenamefont {Winter}, \citenamefont
  {Iezzi}, \citenamefont {Roose}, \citenamefont {Boucly}, \citenamefont
  {Artiglia}, \citenamefont {Ammann}, \citenamefont {Signorell},\ and\
  \citenamefont {Bluhm}}]{Buttersack2024}%
  \BibitemOpen
  \bibfield  {author} {\bibinfo {author} {\bibfnamefont {T.}~\bibnamefont
  {Buttersack}}, \bibinfo {author} {\bibfnamefont {I.}~\bibnamefont {Gladich}},
  \bibinfo {author} {\bibfnamefont {S.}~\bibnamefont {Gholami}}, \bibinfo
  {author} {\bibfnamefont {C.}~\bibnamefont {Richter}}, \bibinfo {author}
  {\bibfnamefont {R.}~\bibnamefont {Dupuy}}, \bibinfo {author} {\bibfnamefont
  {C.}~\bibnamefont {Nicolas}}, \bibinfo {author} {\bibfnamefont
  {F.}~\bibnamefont {Trinter}}, \bibinfo {author} {\bibfnamefont
  {A.}~\bibnamefont {Trunschke}}, \bibinfo {author} {\bibfnamefont
  {D.}~\bibnamefont {Delgado}}, \bibinfo {author} {\bibfnamefont {P.~C.}\
  \bibnamefont {Arroyo}}, \bibinfo {author} {\bibfnamefont {E.~A.}\
  \bibnamefont {Parmentier}}, \bibinfo {author} {\bibfnamefont
  {B.}~\bibnamefont {Winter}}, \bibinfo {author} {\bibfnamefont
  {L.}~\bibnamefont {Iezzi}}, \bibinfo {author} {\bibfnamefont
  {A.}~\bibnamefont {Roose}}, \bibinfo {author} {\bibfnamefont
  {A.}~\bibnamefont {Boucly}}, \bibinfo {author} {\bibfnamefont
  {L.}~\bibnamefont {Artiglia}}, \bibinfo {author} {\bibfnamefont
  {M.}~\bibnamefont {Ammann}}, \bibinfo {author} {\bibfnamefont
  {R.}~\bibnamefont {Signorell}},\ and\ \bibinfo {author} {\bibfnamefont
  {H.}~\bibnamefont {Bluhm}},\ }\bibfield  {title} {\enquote {\bibinfo {title}
  {Direct observation of the complex s(iv) equilibria at the liquid-vapor
  interface},}\ }\href {https://doi.org/10.1038/s41467-024-53186-5} {\bibfield
  {journal} {\bibinfo  {journal} {Nature Communications}\ }\textbf {\bibinfo
  {volume} {15}},\ \bibinfo {pages} {8987} (\bibinfo {year}
  {2024})}\BibitemShut {NoStop}%
\bibitem [{\citenamefont {Omranpour}\ \emph {et~al.}(2024)\citenamefont
  {Omranpour}, \citenamefont {Hijes}, \citenamefont {Behler},\ and\
  \citenamefont {Dellago}}]{DellagoPersp2024}%
  \BibitemOpen
  \bibfield  {author} {\bibinfo {author} {\bibfnamefont {A.}~\bibnamefont
  {Omranpour}}, \bibinfo {author} {\bibfnamefont {P.~M.~D.}\ \bibnamefont
  {Hijes}}, \bibinfo {author} {\bibfnamefont {J.}~\bibnamefont {Behler}},\ and\
  \bibinfo {author} {\bibfnamefont {C.}~\bibnamefont {Dellago}},\ }\bibfield
  {title} {\enquote {\bibinfo {title} {Perspective: Atomistic simulations of
  water and aqueous systems with machine learning potentials},}\ }\href
  {https://doi.org/10.1063/5.0201241} {\bibfield  {journal} {\bibinfo
  {journal} {Journal of Chemical Physics}\ }\textbf {\bibinfo {volume} {160}}
  (\bibinfo {year} {2024}),\ 10.1063/5.0201241}\BibitemShut {NoStop}%
\bibitem [{\citenamefont {Thiemann}\ \emph {et~al.}(2024)\citenamefont
  {Thiemann}, \citenamefont {O’neill}, \citenamefont {Kapil}, \citenamefont
  {Michaelides},\ and\ \citenamefont {Schran}}]{thiemann2024introduction}%
  \BibitemOpen
  \bibfield  {author} {\bibinfo {author} {\bibfnamefont {F.~L.}\ \bibnamefont
  {Thiemann}}, \bibinfo {author} {\bibfnamefont {N.}~\bibnamefont {O’neill}},
  \bibinfo {author} {\bibfnamefont {V.}~\bibnamefont {Kapil}}, \bibinfo
  {author} {\bibfnamefont {A.}~\bibnamefont {Michaelides}},\ and\ \bibinfo
  {author} {\bibfnamefont {C.}~\bibnamefont {Schran}},\ }\bibfield  {title}
  {\enquote {\bibinfo {title} {Introduction to machine learning potentials for
  atomistic simulations},}\ }\href {https://doi.org/10.1088/1361-648X/ad9657}
  {\bibfield  {journal} {\bibinfo  {journal} {Journal of Physics: Condensed
  Matter}\ }\textbf {\bibinfo {volume} {37}},\ \bibinfo {pages} {73002}
  (\bibinfo {year} {2024})}\BibitemShut {NoStop}%
\bibitem [{\citenamefont {Deringer}, \citenamefont {Caro},\ and\ \citenamefont
  {Csányi}(2019)}]{Deringer2019}%
  \BibitemOpen
  \bibfield  {author} {\bibinfo {author} {\bibfnamefont {V.~L.}\ \bibnamefont
  {Deringer}}, \bibinfo {author} {\bibfnamefont {M.~A.}\ \bibnamefont {Caro}},\
  and\ \bibinfo {author} {\bibfnamefont {G.}~\bibnamefont {Csányi}},\
  }\bibfield  {title} {\enquote {\bibinfo {title} {Machine learning interatomic
  potentials as emerging tools for materials science},}\ }\href
  {https://doi.org/https://doi.org/10.1002/adma.201902765} {\bibfield
  {journal} {\bibinfo  {journal} {Advanced Materials}\ }\textbf {\bibinfo
  {volume} {31}},\ \bibinfo {pages} {1902765} (\bibinfo {year}
  {2019})}\BibitemShut {NoStop}%
\bibitem [{\citenamefont {Unke}\ \emph {et~al.}(2021)\citenamefont {Unke},
  \citenamefont {Chmiela}, \citenamefont {Sauceda}, \citenamefont {Gastegger},
  \citenamefont {Poltavsky}, \citenamefont {Schütt}, \citenamefont
  {Tkatchenko},\ and\ \citenamefont {Müller}}]{Unke2021}%
  \BibitemOpen
  \bibfield  {author} {\bibinfo {author} {\bibfnamefont {O.~T.}\ \bibnamefont
  {Unke}}, \bibinfo {author} {\bibfnamefont {S.}~\bibnamefont {Chmiela}},
  \bibinfo {author} {\bibfnamefont {H.~E.}\ \bibnamefont {Sauceda}}, \bibinfo
  {author} {\bibfnamefont {M.}~\bibnamefont {Gastegger}}, \bibinfo {author}
  {\bibfnamefont {I.}~\bibnamefont {Poltavsky}}, \bibinfo {author}
  {\bibfnamefont {K.~T.}\ \bibnamefont {Schütt}}, \bibinfo {author}
  {\bibfnamefont {A.}~\bibnamefont {Tkatchenko}},\ and\ \bibinfo {author}
  {\bibfnamefont {K.-R.}\ \bibnamefont {Müller}},\ }\bibfield  {title}
  {\enquote {\bibinfo {title} {Machine learning force fields},}\ }\href
  {https://doi.org/10.1021/acs.chemrev.0c01111} {\bibfield  {journal} {\bibinfo
   {journal} {Chemical Reviews}\ }\textbf {\bibinfo {volume} {121}},\ \bibinfo
  {pages} {10142--10186} (\bibinfo {year} {2021})}\BibitemShut {NoStop}%
\bibitem [{\citenamefont {Behler}(2021)}]{Behler2021}%
  \BibitemOpen
  \bibfield  {author} {\bibinfo {author} {\bibfnamefont {J.}~\bibnamefont
  {Behler}},\ }\bibfield  {title} {\enquote {\bibinfo {title} {Four generations
  of high-dimensional neural network potentials},}\ }\href
  {https://doi.org/10.1021/acs.chemrev.0c00868} {\bibfield  {journal} {\bibinfo
   {journal} {Chemical Reviews}\ }\textbf {\bibinfo {volume} {121}},\ \bibinfo
  {pages} {10037--10072} (\bibinfo {year} {2021})}\BibitemShut {NoStop}%
\bibitem [{\citenamefont {Deringer}\ \emph {et~al.}(2021)\citenamefont
  {Deringer}, \citenamefont {Bartók}, \citenamefont {Bernstein}, \citenamefont
  {Wilkins}, \citenamefont {Ceriotti},\ and\ \citenamefont
  {Csányi}}]{Deringer2021}%
  \BibitemOpen
  \bibfield  {author} {\bibinfo {author} {\bibfnamefont {V.~L.}\ \bibnamefont
  {Deringer}}, \bibinfo {author} {\bibfnamefont {A.~P.}\ \bibnamefont
  {Bartók}}, \bibinfo {author} {\bibfnamefont {N.}~\bibnamefont {Bernstein}},
  \bibinfo {author} {\bibfnamefont {D.~M.}\ \bibnamefont {Wilkins}}, \bibinfo
  {author} {\bibfnamefont {M.}~\bibnamefont {Ceriotti}},\ and\ \bibinfo
  {author} {\bibfnamefont {G.}~\bibnamefont {Csányi}},\ }\bibfield  {title}
  {\enquote {\bibinfo {title} {Gaussian process regression for materials and
  molecules},}\ }\href {https://doi.org/10.1021/acs.chemrev.1c00022} {\bibfield
   {journal} {\bibinfo  {journal} {Chemical Reviews}\ }\textbf {\bibinfo
  {volume} {121}},\ \bibinfo {pages} {10073--10141} (\bibinfo {year}
  {2021})}\BibitemShut {NoStop}%
\bibitem [{\citenamefont {Drautz}(2019)}]{PhysRevB.99.014104}%
  \BibitemOpen
  \bibfield  {author} {\bibinfo {author} {\bibfnamefont {R.}~\bibnamefont
  {Drautz}},\ }\bibfield  {title} {\enquote {\bibinfo {title} {Atomic cluster
  expansion for accurate and transferable interatomic potentials},}\ }\href
  {https://doi.org/10.1103/PhysRevB.99.014104} {\bibfield  {journal} {\bibinfo
  {journal} {Phys. Rev. B}\ }\textbf {\bibinfo {volume} {99}},\ \bibinfo
  {pages} {14104} (\bibinfo {year} {2019})}\BibitemShut {NoStop}%
\bibitem [{\citenamefont {Batatia}\ \emph {et~al.}(2022)\citenamefont
  {Batatia}, \citenamefont {Kovacs}, \citenamefont {Simm}, \citenamefont
  {Ortner},\ and\ \citenamefont {Csányi}}]{batatia2022mace}%
  \BibitemOpen
  \bibfield  {author} {\bibinfo {author} {\bibfnamefont {I.}~\bibnamefont
  {Batatia}}, \bibinfo {author} {\bibfnamefont {D.~P.}\ \bibnamefont {Kovacs}},
  \bibinfo {author} {\bibfnamefont {G.}~\bibnamefont {Simm}}, \bibinfo {author}
  {\bibfnamefont {C.}~\bibnamefont {Ortner}},\ and\ \bibinfo {author}
  {\bibfnamefont {G.}~\bibnamefont {Csányi}},\ }\bibfield  {title} {\enquote
  {\bibinfo {title} {Mace: Higher order equivariant message passing neural
  networks for fast and accurate force fields},}\ }\href@noop {} {\bibfield
  {journal} {\bibinfo  {journal} {Advances in Neural Information Processing
  Systems}\ }\textbf {\bibinfo {volume} {35}},\ \bibinfo {pages} {11423--11436}
  (\bibinfo {year} {2022})}\BibitemShut {NoStop}%
\bibitem [{\citenamefont {Kovács}\ \emph {et~al.}(2023)\citenamefont
  {Kovács}, \citenamefont {Batatia}, \citenamefont {Arany},\ and\
  \citenamefont {Csányi}}]{Kovacs2023}%
  \BibitemOpen
  \bibfield  {author} {\bibinfo {author} {\bibfnamefont {D.~P.}\ \bibnamefont
  {Kovács}}, \bibinfo {author} {\bibfnamefont {I.}~\bibnamefont {Batatia}},
  \bibinfo {author} {\bibfnamefont {E.~S.}\ \bibnamefont {Arany}},\ and\
  \bibinfo {author} {\bibfnamefont {G.}~\bibnamefont {Csányi}},\ }\bibfield
  {title} {\enquote {\bibinfo {title} {Evaluation of the mace force field
  architecture: From medicinal chemistry to materials science},}\ }\href
  {https://doi.org/10.1063/5.0155322} {\bibfield  {journal} {\bibinfo
  {journal} {The Journal of Chemical Physics}\ }\textbf {\bibinfo {volume}
  {159}},\ \bibinfo {pages} {44118} (\bibinfo {year} {2023})}\BibitemShut
  {NoStop}%
\bibitem [{\citenamefont {Barducci}, \citenamefont {Bussi},\ and\ \citenamefont
  {Parrinello}(2008)}]{Barducci2008}%
  \BibitemOpen
  \bibfield  {author} {\bibinfo {author} {\bibfnamefont {A.}~\bibnamefont
  {Barducci}}, \bibinfo {author} {\bibfnamefont {G.}~\bibnamefont {Bussi}},\
  and\ \bibinfo {author} {\bibfnamefont {M.}~\bibnamefont {Parrinello}},\
  }\bibfield  {title} {\enquote {\bibinfo {title} {Well-tempered metadynamics:
  A smoothly converging and tunable free-energy method},}\ }\href
  {https://doi.org/10.1103/PhysRevLett.100.020603} {\bibfield  {journal}
  {\bibinfo  {journal} {Physical Review Letters}\ }\textbf {\bibinfo {volume}
  {100}},\ \bibinfo {pages} {20603} (\bibinfo {year} {2008})}\BibitemShut
  {NoStop}%
\bibitem [{\citenamefont {Bonn}, \citenamefont {Nagata},\ and\ \citenamefont
  {Backus}(2015)}]{Bonn2015}%
  \BibitemOpen
  \bibfield  {author} {\bibinfo {author} {\bibfnamefont {M.}~\bibnamefont
  {Bonn}}, \bibinfo {author} {\bibfnamefont {Y.}~\bibnamefont {Nagata}},\ and\
  \bibinfo {author} {\bibfnamefont {E.~H.~G.}\ \bibnamefont {Backus}},\
  }\bibfield  {title} {\enquote {\bibinfo {title} {Molecular structure and
  dynamics of water at the water–air interface studied with surface-specific
  vibrational spectroscopy},}\ }\href
  {https://doi.org/https://doi.org/10.1002/anie.201411188} {\bibfield
  {journal} {\bibinfo  {journal} {Angewandte Chemie International Edition}\
  }\textbf {\bibinfo {volume} {54}},\ \bibinfo {pages} {5560--5576} (\bibinfo
  {year} {2015})}\BibitemShut {NoStop}%
\bibitem [{\citenamefont {Wohlfahrt}, \citenamefont {Dellago},\ and\
  \citenamefont {Sega}(2020)}]{Wolfhart2020}%
  \BibitemOpen
  \bibfield  {author} {\bibinfo {author} {\bibfnamefont {O.}~\bibnamefont
  {Wohlfahrt}}, \bibinfo {author} {\bibfnamefont {C.}~\bibnamefont {Dellago}},\
  and\ \bibinfo {author} {\bibfnamefont {M.}~\bibnamefont {Sega}},\ }\bibfield
  {title} {\enquote {\bibinfo {title} {Ab initio structure and thermodynamics
  of the rpbe-d3 water/vapor interface by neural-network molecular dynamics},}\
  }\href {https://doi.org/10.1063/5.0021852} {\bibfield  {journal} {\bibinfo
  {journal} {The Journal of Chemical Physics}\ }\textbf {\bibinfo {volume}
  {153}},\ \bibinfo {pages} {144710} (\bibinfo {year} {2020})}\BibitemShut
  {NoStop}%
\bibitem [{\citenamefont {Willard}\ and\ \citenamefont
  {Chandler}(2010)}]{Willard2010}%
  \BibitemOpen
  \bibfield  {author} {\bibinfo {author} {\bibfnamefont {A.~P.}\ \bibnamefont
  {Willard}}\ and\ \bibinfo {author} {\bibfnamefont {D.}~\bibnamefont
  {Chandler}},\ }\bibfield  {title} {\enquote {\bibinfo {title} {Instantaneous
  liquid interfaces},}\ }\href {https://doi.org/10.1021/jp909219k} {\bibfield
  {journal} {\bibinfo  {journal} {The Journal of Physical Chemistry B}\
  }\textbf {\bibinfo {volume} {114}},\ \bibinfo {pages} {1954--1958} (\bibinfo
  {year} {2010})}\BibitemShut {NoStop}%
\bibitem [{\citenamefont {Brookes}\ \emph {et~al.}(2024)\citenamefont
  {Brookes}, \citenamefont {Kapil}, \citenamefont {Schran},\ and\ \citenamefont
  {Michaelides}}]{Brookes2024}%
  \BibitemOpen
  \bibfield  {author} {\bibinfo {author} {\bibfnamefont {S.~G.~H.}\
  \bibnamefont {Brookes}}, \bibinfo {author} {\bibfnamefont {V.}~\bibnamefont
  {Kapil}}, \bibinfo {author} {\bibfnamefont {C.}~\bibnamefont {Schran}},\ and\
  \bibinfo {author} {\bibfnamefont {A.}~\bibnamefont {Michaelides}},\
  }\bibfield  {title} {\enquote {\bibinfo {title} {The wetting of h2o by
  co2},}\ }\href {https://doi.org/10.1063/5.0224230} {\bibfield  {journal}
  {\bibinfo  {journal} {The Journal of Chemical Physics}\ }\textbf {\bibinfo
  {volume} {161}},\ \bibinfo {pages} {84711} (\bibinfo {year}
  {2024})}\BibitemShut {NoStop}%
\bibitem [{\citenamefont {Jungwirth}\ and\ \citenamefont
  {Tobias}(2006)}]{Jungwirth2006}%
  \BibitemOpen
  \bibfield  {author} {\bibinfo {author} {\bibfnamefont {P.}~\bibnamefont
  {Jungwirth}}\ and\ \bibinfo {author} {\bibfnamefont {D.~J.}\ \bibnamefont
  {Tobias}},\ }\bibfield  {title} {\enquote {\bibinfo {title} {Specific ion
  effects at the air/water interface},}\ }\href
  {https://doi.org/10.1021/cr0403741} {\bibfield  {journal} {\bibinfo
  {journal} {Chemical Reviews}\ }\textbf {\bibinfo {volume} {106}},\ \bibinfo
  {pages} {1259--1281} (\bibinfo {year} {2006})}\BibitemShut {NoStop}%
\bibitem [{\citenamefont {Devlin}, \citenamefont {Benjamin},\ and\
  \citenamefont {Saykally}(2022)}]{Devlin2022}%
  \BibitemOpen
  \bibfield  {author} {\bibinfo {author} {\bibfnamefont {S.~W.}\ \bibnamefont
  {Devlin}}, \bibinfo {author} {\bibfnamefont {I.}~\bibnamefont {Benjamin}},\
  and\ \bibinfo {author} {\bibfnamefont {R.~J.}\ \bibnamefont {Saykally}},\
  }\bibfield  {title} {\enquote {\bibinfo {title} {On the mechanisms of ion
  adsorption to aqueous interfaces: air-water vs. oil-water},}\ }\href
  {https://doi.org/10.1073/pnas.2210857119} {\bibfield  {journal} {\bibinfo
  {journal} {Proceedings of the National Academy of Sciences}\ }\textbf
  {\bibinfo {volume} {119}},\ \bibinfo {pages} {e2210857119} (\bibinfo {year}
  {2022})}\BibitemShut {NoStop}%
\bibitem [{\citenamefont {Ruiz-Lopez}\ \emph {et~al.}(2020)\citenamefont
  {Ruiz-Lopez}, \citenamefont {Francisco}, \citenamefont {Martins-Costa},\ and\
  \citenamefont {Anglada}}]{Ruiz-Lopez2020}%
  \BibitemOpen
  \bibfield  {author} {\bibinfo {author} {\bibfnamefont {M.~F.}\ \bibnamefont
  {Ruiz-Lopez}}, \bibinfo {author} {\bibfnamefont {J.~S.}\ \bibnamefont
  {Francisco}}, \bibinfo {author} {\bibfnamefont {M.~T.~C.}\ \bibnamefont
  {Martins-Costa}},\ and\ \bibinfo {author} {\bibfnamefont {J.~M.}\
  \bibnamefont {Anglada}},\ }\bibfield  {title} {\enquote {\bibinfo {title}
  {Molecular reactions at aqueous interfaces},}\ }\href
  {https://doi.org/10.1038/s41570-020-0203-2} {\bibfield  {journal} {\bibinfo
  {journal} {Nature Reviews Chemistry}\ }\textbf {\bibinfo {volume} {4}},\
  \bibinfo {pages} {459--475} (\bibinfo {year} {2020})}\BibitemShut {NoStop}%
\bibitem [{\citenamefont {Hub}, \citenamefont {Caleman},\ and\ \citenamefont
  {van~der Spoel}(2012)}]{Hub2012}%
  \BibitemOpen
  \bibfield  {author} {\bibinfo {author} {\bibfnamefont {J.~S.}\ \bibnamefont
  {Hub}}, \bibinfo {author} {\bibfnamefont {C.}~\bibnamefont {Caleman}},\ and\
  \bibinfo {author} {\bibfnamefont {D.}~\bibnamefont {van~der Spoel}},\
  }\bibfield  {title} {\enquote {\bibinfo {title} {Organic molecules on the
  surface of water droplets – an energetic perspective},}\ }\href
  {https://doi.org/10.1039/C2CP40483D} {\bibfield  {journal} {\bibinfo
  {journal} {Physical Chemistry Chemical Physics}\ }\textbf {\bibinfo {volume}
  {14}},\ \bibinfo {pages} {9537--9545} (\bibinfo {year} {2012})}\BibitemShut
  {NoStop}%
\bibitem [{\citenamefont {Petersen}\ and\ \citenamefont
  {Saykally}(2006)}]{nnurev.physchem.57.032905.104609}%
  \BibitemOpen
  \bibfield  {author} {\bibinfo {author} {\bibfnamefont {P.~B.}\ \bibnamefont
  {Petersen}}\ and\ \bibinfo {author} {\bibfnamefont {R.~J.}\ \bibnamefont
  {Saykally}},\ }\bibfield  {title} {\enquote {\bibinfo {title} {On the nature
  of ions at the liquid water surface},}\ }\href
  {https://doi.org/https://doi.org/10.1146/annurev.physchem.57.032905.104609}
  {\bibfield  {journal} {\bibinfo  {journal} {Annual Review of Physical
  Chemistry}\ }\textbf {\bibinfo {volume} {57}},\ \bibinfo {pages} {333--364}
  (\bibinfo {year} {2006})}\BibitemShut {NoStop}%
\bibitem [{\citenamefont {Yan}\ \emph {et~al.}(2018)\citenamefont {Yan},
  \citenamefont {Delgado}, \citenamefont {Aubry}, \citenamefont {Gribelin},
  \citenamefont {Stocco}, \citenamefont {Cruz}, \citenamefont {Bernard},\ and\
  \citenamefont {Ganachaud}}]{Yan2018}%
  \BibitemOpen
  \bibfield  {author} {\bibinfo {author} {\bibfnamefont {X.}~\bibnamefont
  {Yan}}, \bibinfo {author} {\bibfnamefont {M.}~\bibnamefont {Delgado}},
  \bibinfo {author} {\bibfnamefont {J.}~\bibnamefont {Aubry}}, \bibinfo
  {author} {\bibfnamefont {O.}~\bibnamefont {Gribelin}}, \bibinfo {author}
  {\bibfnamefont {A.}~\bibnamefont {Stocco}}, \bibinfo {author} {\bibfnamefont
  {F.~B.-D.}\ \bibnamefont {Cruz}}, \bibinfo {author} {\bibfnamefont
  {J.}~\bibnamefont {Bernard}},\ and\ \bibinfo {author} {\bibfnamefont
  {F.}~\bibnamefont {Ganachaud}},\ }\bibfield  {title} {\enquote {\bibinfo
  {title} {Central role of bicarbonate anions in charging water/hydrophobic
  interfaces},}\ }\href {https://doi.org/10.1021/acs.jpclett.7b02993}
  {\bibfield  {journal} {\bibinfo  {journal} {The Journal of Physical Chemistry
  Letters}\ }\textbf {\bibinfo {volume} {9}},\ \bibinfo {pages} {96--103}
  (\bibinfo {year} {2018})}\BibitemShut {NoStop}%
\bibitem [{\citenamefont {Das}\ \emph {et~al.}(2020)\citenamefont {Das},
  \citenamefont {Imoto}, \citenamefont {Sun}, \citenamefont {Nagata},
  \citenamefont {Backus},\ and\ \citenamefont {Bonn}}]{Das2020}%
  \BibitemOpen
  \bibfield  {author} {\bibinfo {author} {\bibfnamefont {S.}~\bibnamefont
  {Das}}, \bibinfo {author} {\bibfnamefont {S.}~\bibnamefont {Imoto}}, \bibinfo
  {author} {\bibfnamefont {S.}~\bibnamefont {Sun}}, \bibinfo {author}
  {\bibfnamefont {Y.}~\bibnamefont {Nagata}}, \bibinfo {author} {\bibfnamefont
  {E.~H.~G.}\ \bibnamefont {Backus}},\ and\ \bibinfo {author} {\bibfnamefont
  {M.}~\bibnamefont {Bonn}},\ }\bibfield  {title} {\enquote {\bibinfo {title}
  {Nature of excess hydrated proton at the water–air interface},}\ }\href
  {https://doi.org/10.1021/jacs.9b10807} {\bibfield  {journal} {\bibinfo
  {journal} {Journal of the American Chemical Society}\ }\textbf {\bibinfo
  {volume} {142}},\ \bibinfo {pages} {945--952} (\bibinfo {year}
  {2020})}\BibitemShut {NoStop}%
\bibitem [{\citenamefont {Ford}\ \emph {et~al.}(2024)\citenamefont {Ford},
  \citenamefont {Shutler}, \citenamefont {Blanco-Sacristán}, \citenamefont
  {Corrigan}, \citenamefont {Bell}, \citenamefont {Yang}, \citenamefont
  {Kitidis}, \citenamefont {Nightingale}, \citenamefont {Brown}, \citenamefont
  {Wimmer}, \citenamefont {Woolf}, \citenamefont {Casal}, \citenamefont
  {Donlon}, \citenamefont {Tilstone},\ and\ \citenamefont {Ashton}}]{Ford2024}%
  \BibitemOpen
  \bibfield  {author} {\bibinfo {author} {\bibfnamefont {D.~J.}\ \bibnamefont
  {Ford}}, \bibinfo {author} {\bibfnamefont {J.~D.}\ \bibnamefont {Shutler}},
  \bibinfo {author} {\bibfnamefont {J.}~\bibnamefont {Blanco-Sacristán}},
  \bibinfo {author} {\bibfnamefont {S.}~\bibnamefont {Corrigan}}, \bibinfo
  {author} {\bibfnamefont {T.~G.}\ \bibnamefont {Bell}}, \bibinfo {author}
  {\bibfnamefont {M.}~\bibnamefont {Yang}}, \bibinfo {author} {\bibfnamefont
  {V.}~\bibnamefont {Kitidis}}, \bibinfo {author} {\bibfnamefont {P.~D.}\
  \bibnamefont {Nightingale}}, \bibinfo {author} {\bibfnamefont
  {I.}~\bibnamefont {Brown}}, \bibinfo {author} {\bibfnamefont
  {W.}~\bibnamefont {Wimmer}}, \bibinfo {author} {\bibfnamefont {D.~K.}\
  \bibnamefont {Woolf}}, \bibinfo {author} {\bibfnamefont {T.}~\bibnamefont
  {Casal}}, \bibinfo {author} {\bibfnamefont {C.}~\bibnamefont {Donlon}},
  \bibinfo {author} {\bibfnamefont {G.~H.}\ \bibnamefont {Tilstone}},\ and\
  \bibinfo {author} {\bibfnamefont {I.}~\bibnamefont {Ashton}},\ }\bibfield
  {title} {\enquote {\bibinfo {title} {Enhanced ocean co2 uptake due to
  near-surface temperature gradients},}\ }\href
  {https://doi.org/10.1038/s41561-024-01570-7} {\bibfield  {journal} {\bibinfo
  {journal} {Nature Geoscience}\ }\textbf {\bibinfo {volume} {17}},\ \bibinfo
  {pages} {1135--1140} (\bibinfo {year} {2024})}\BibitemShut {NoStop}%
\bibitem [{\citenamefont {Ooi}\ and\ \citenamefont {Maruoka}(2007)}]{Ooi2007}%
  \BibitemOpen
  \bibfield  {author} {\bibinfo {author} {\bibfnamefont {T.}~\bibnamefont
  {Ooi}}\ and\ \bibinfo {author} {\bibfnamefont {K.}~\bibnamefont {Maruoka}},\
  }\bibfield  {title} {\enquote {\bibinfo {title} {Recent advances in
  asymmetric phase-transfer catalysis},}\ }\href
  {https://doi.org/https://doi.org/10.1002/anie.200601737} {\bibfield
  {journal} {\bibinfo  {journal} {Angewandte Chemie International Edition}\
  }\textbf {\bibinfo {volume} {46}},\ \bibinfo {pages} {4222--4266} (\bibinfo
  {year} {2007})}\BibitemShut {NoStop}%
\bibitem [{\citenamefont {Schran}\ \emph {et~al.}(2021)\citenamefont {Schran},
  \citenamefont {Thiemann}, \citenamefont {Rowe}, \citenamefont {Müller},
  \citenamefont {Marsalek},\ and\ \citenamefont
  {Michaelides}}]{doi:10.1073/pnas.2110077118}%
  \BibitemOpen
  \bibfield  {author} {\bibinfo {author} {\bibfnamefont {C.}~\bibnamefont
  {Schran}}, \bibinfo {author} {\bibfnamefont {F.~L.}\ \bibnamefont
  {Thiemann}}, \bibinfo {author} {\bibfnamefont {P.}~\bibnamefont {Rowe}},
  \bibinfo {author} {\bibfnamefont {E.~A.}\ \bibnamefont {Müller}}, \bibinfo
  {author} {\bibfnamefont {O.}~\bibnamefont {Marsalek}},\ and\ \bibinfo
  {author} {\bibfnamefont {A.}~\bibnamefont {Michaelides}},\ }\bibfield
  {title} {\enquote {\bibinfo {title} {Machine learning potentials for complex
  aqueous systems made simple},}\ }\href
  {https://doi.org/10.1073/pnas.2110077118} {\bibfield  {journal} {\bibinfo
  {journal} {Proceedings of the National Academy of Sciences}\ }\textbf
  {\bibinfo {volume} {118}},\ \bibinfo {pages} {e2110077118} (\bibinfo {year}
  {2021})}\BibitemShut {NoStop}%
\bibitem [{\citenamefont {McInnes}, \citenamefont {Healy},\ and\ \citenamefont
  {Melville}(2018)}]{UMAP2018}%
  \BibitemOpen
  \bibfield  {author} {\bibinfo {author} {\bibfnamefont {L.}~\bibnamefont
  {McInnes}}, \bibinfo {author} {\bibfnamefont {J.}~\bibnamefont {Healy}},\
  and\ \bibinfo {author} {\bibfnamefont {J.}~\bibnamefont {Melville}},\
  }\bibfield  {title} {\enquote {\bibinfo {title} {Umap: Uniform manifold
  approximation and projection for dimension reduction},}\ }\href@noop {}
  {\bibfield  {journal} {\bibinfo  {journal} {arXiv preprint arXiv:1802.03426}\
  } (\bibinfo {year} {2018})}\BibitemShut {NoStop}%
\bibitem [{\citenamefont {Cheng}\ \emph {et~al.}(2020)\citenamefont {Cheng},
  \citenamefont {Griffiths}, \citenamefont {Wengert}, \citenamefont {Kunkel},
  \citenamefont {Stenczel}, \citenamefont {Zhu}, \citenamefont {Deringer},
  \citenamefont {Bernstein}, \citenamefont {Margraf}, \citenamefont {Reuter},\
  and\ \citenamefont {Csanyi}}]{ASAP2020}%
  \BibitemOpen
  \bibfield  {author} {\bibinfo {author} {\bibfnamefont {B.}~\bibnamefont
  {Cheng}}, \bibinfo {author} {\bibfnamefont {R.-R.}\ \bibnamefont
  {Griffiths}}, \bibinfo {author} {\bibfnamefont {S.}~\bibnamefont {Wengert}},
  \bibinfo {author} {\bibfnamefont {C.}~\bibnamefont {Kunkel}}, \bibinfo
  {author} {\bibfnamefont {T.}~\bibnamefont {Stenczel}}, \bibinfo {author}
  {\bibfnamefont {B.}~\bibnamefont {Zhu}}, \bibinfo {author} {\bibfnamefont
  {V.~L.}\ \bibnamefont {Deringer}}, \bibinfo {author} {\bibfnamefont
  {N.}~\bibnamefont {Bernstein}}, \bibinfo {author} {\bibfnamefont {J.~T.}\
  \bibnamefont {Margraf}}, \bibinfo {author} {\bibfnamefont {K.}~\bibnamefont
  {Reuter}},\ and\ \bibinfo {author} {\bibfnamefont {G.}~\bibnamefont
  {Csanyi}},\ }\bibfield  {title} {\enquote {\bibinfo {title} {Mapping
  materials and molecules},}\ }\href
  {https://doi.org/10.1021/acs.accounts.0c00403} {\bibfield  {journal}
  {\bibinfo  {journal} {Accounts of Chemical Research}\ }\textbf {\bibinfo
  {volume} {53}},\ \bibinfo {pages} {1981--1991} (\bibinfo {year}
  {2020})}\BibitemShut {NoStop}%
\bibitem [{\citenamefont {VandeVondele}\ \emph {et~al.}(2005)\citenamefont
  {VandeVondele}, \citenamefont {Krack}, \citenamefont {Mohamed}, \citenamefont
  {Parrinello}, \citenamefont {Chassaing},\ and\ \citenamefont
  {Hutter}}]{VandeVondele2005}%
  \BibitemOpen
  \bibfield  {author} {\bibinfo {author} {\bibfnamefont {J.}~\bibnamefont
  {VandeVondele}}, \bibinfo {author} {\bibfnamefont {M.}~\bibnamefont {Krack}},
  \bibinfo {author} {\bibfnamefont {F.}~\bibnamefont {Mohamed}}, \bibinfo
  {author} {\bibfnamefont {M.}~\bibnamefont {Parrinello}}, \bibinfo {author}
  {\bibfnamefont {T.}~\bibnamefont {Chassaing}},\ and\ \bibinfo {author}
  {\bibfnamefont {J.}~\bibnamefont {Hutter}},\ }\bibfield  {title} {\enquote
  {\bibinfo {title} {Quickstep: Fast and accurate density functional
  calculations using a mixed gaussian and plane waves approach},}\ }\href
  {https://doi.org/https://doi.org/10.1016/j.cpc.2004.12.014} {\bibfield
  {journal} {\bibinfo  {journal} {Computer Physics Communications}\ }\textbf
  {\bibinfo {volume} {167}},\ \bibinfo {pages} {103--128} (\bibinfo {year}
  {2005})}\BibitemShut {NoStop}%
\bibitem [{\citenamefont {Kühne}\ \emph {et~al.}(2020)\citenamefont {Kühne},
  \citenamefont {Iannuzzi}, \citenamefont {Ben}, \citenamefont {Rybkin},
  \citenamefont {Seewald}, \citenamefont {Stein}, \citenamefont {Laino},
  \citenamefont {Khaliullin}, \citenamefont {Schütt}, \citenamefont
  {Schiffmann}, \citenamefont {Golze}, \citenamefont {Wilhelm}, \citenamefont
  {Chulkov}, \citenamefont {Bani-Hashemian}, \citenamefont {Weber},
  \citenamefont {Borštnik}, \citenamefont {Taillefumier}, \citenamefont
  {Jakobovits}, \citenamefont {Lazzaro}, \citenamefont {Pabst}, \citenamefont
  {Müller}, \citenamefont {Schade}, \citenamefont {Guidon}, \citenamefont
  {Andermatt}, \citenamefont {Holmberg}, \citenamefont {Schenter},
  \citenamefont {Hehn}, \citenamefont {Bussy}, \citenamefont {Belleflamme},
  \citenamefont {Tabacchi}, \citenamefont {Glöß}, \citenamefont {Lass},
  \citenamefont {Bethune}, \citenamefont {Mundy}, \citenamefont {Plessl},
  \citenamefont {Watkins}, \citenamefont {VandeVondele}, \citenamefont
  {Krack},\ and\ \citenamefont {Hutter}}]{doi:10.1063/5.0007045}%
  \BibitemOpen
  \bibfield  {author} {\bibinfo {author} {\bibfnamefont {T.~D.}\ \bibnamefont
  {Kühne}}, \bibinfo {author} {\bibfnamefont {M.}~\bibnamefont {Iannuzzi}},
  \bibinfo {author} {\bibfnamefont {M.~D.}\ \bibnamefont {Ben}}, \bibinfo
  {author} {\bibfnamefont {V.~V.}\ \bibnamefont {Rybkin}}, \bibinfo {author}
  {\bibfnamefont {P.}~\bibnamefont {Seewald}}, \bibinfo {author} {\bibfnamefont
  {F.}~\bibnamefont {Stein}}, \bibinfo {author} {\bibfnamefont
  {T.}~\bibnamefont {Laino}}, \bibinfo {author} {\bibfnamefont {R.~Z.}\
  \bibnamefont {Khaliullin}}, \bibinfo {author} {\bibfnamefont
  {O.}~\bibnamefont {Schütt}}, \bibinfo {author} {\bibfnamefont
  {F.}~\bibnamefont {Schiffmann}}, \bibinfo {author} {\bibfnamefont
  {D.}~\bibnamefont {Golze}}, \bibinfo {author} {\bibfnamefont
  {J.}~\bibnamefont {Wilhelm}}, \bibinfo {author} {\bibfnamefont
  {S.}~\bibnamefont {Chulkov}}, \bibinfo {author} {\bibfnamefont {M.~H.}\
  \bibnamefont {Bani-Hashemian}}, \bibinfo {author} {\bibfnamefont
  {V.}~\bibnamefont {Weber}}, \bibinfo {author} {\bibfnamefont
  {U.}~\bibnamefont {Borštnik}}, \bibinfo {author} {\bibfnamefont
  {M.}~\bibnamefont {Taillefumier}}, \bibinfo {author} {\bibfnamefont {A.~S.}\
  \bibnamefont {Jakobovits}}, \bibinfo {author} {\bibfnamefont
  {A.}~\bibnamefont {Lazzaro}}, \bibinfo {author} {\bibfnamefont
  {H.}~\bibnamefont {Pabst}}, \bibinfo {author} {\bibfnamefont
  {T.}~\bibnamefont {Müller}}, \bibinfo {author} {\bibfnamefont
  {R.}~\bibnamefont {Schade}}, \bibinfo {author} {\bibfnamefont
  {M.}~\bibnamefont {Guidon}}, \bibinfo {author} {\bibfnamefont
  {S.}~\bibnamefont {Andermatt}}, \bibinfo {author} {\bibfnamefont
  {N.}~\bibnamefont {Holmberg}}, \bibinfo {author} {\bibfnamefont {G.~K.}\
  \bibnamefont {Schenter}}, \bibinfo {author} {\bibfnamefont {A.}~\bibnamefont
  {Hehn}}, \bibinfo {author} {\bibfnamefont {A.}~\bibnamefont {Bussy}},
  \bibinfo {author} {\bibfnamefont {F.}~\bibnamefont {Belleflamme}}, \bibinfo
  {author} {\bibfnamefont {G.}~\bibnamefont {Tabacchi}}, \bibinfo {author}
  {\bibfnamefont {A.}~\bibnamefont {Glöß}}, \bibinfo {author} {\bibfnamefont
  {M.}~\bibnamefont {Lass}}, \bibinfo {author} {\bibfnamefont {I.}~\bibnamefont
  {Bethune}}, \bibinfo {author} {\bibfnamefont {C.~J.}\ \bibnamefont {Mundy}},
  \bibinfo {author} {\bibfnamefont {C.}~\bibnamefont {Plessl}}, \bibinfo
  {author} {\bibfnamefont {M.}~\bibnamefont {Watkins}}, \bibinfo {author}
  {\bibfnamefont {J.}~\bibnamefont {VandeVondele}}, \bibinfo {author}
  {\bibfnamefont {M.}~\bibnamefont {Krack}},\ and\ \bibinfo {author}
  {\bibfnamefont {J.}~\bibnamefont {Hutter}},\ }\bibfield  {title} {\enquote
  {\bibinfo {title} {Cp2k: An electronic structure and molecular dynamics
  software package - quickstep: Efficient and accurate electronic structure
  calculations},}\ }\href {https://doi.org/10.1063/5.0007045} {\bibfield
  {journal} {\bibinfo  {journal} {The Journal of Chemical Physics}\ }\textbf
  {\bibinfo {volume} {152}},\ \bibinfo {pages} {194103} (\bibinfo {year}
  {2020})}\BibitemShut {NoStop}%
\bibitem [{\citenamefont {Perdew}, \citenamefont {Burke},\ and\ \citenamefont
  {Ernzerhof}(1996)}]{PhysRevLett.77.3865}%
  \BibitemOpen
  \bibfield  {author} {\bibinfo {author} {\bibfnamefont {J.~P.}\ \bibnamefont
  {Perdew}}, \bibinfo {author} {\bibfnamefont {K.}~\bibnamefont {Burke}},\ and\
  \bibinfo {author} {\bibfnamefont {M.}~\bibnamefont {Ernzerhof}},\ }\bibfield
  {title} {\enquote {\bibinfo {title} {Generalized gradient approximation made
  simple},}\ }\href {https://doi.org/10.1103/PhysRevLett.77.3865} {\bibfield
  {journal} {\bibinfo  {journal} {Phys. Rev. Lett.}\ }\textbf {\bibinfo
  {volume} {77}},\ \bibinfo {pages} {3865--3868} (\bibinfo {year}
  {1996})}\BibitemShut {NoStop}%
\bibitem [{\citenamefont {Grimme}\ \emph {et~al.}(2010)\citenamefont {Grimme},
  \citenamefont {Antony}, \citenamefont {Ehrlich},\ and\ \citenamefont
  {Krieg}}]{doi:10.1063/1.3382344}%
  \BibitemOpen
  \bibfield  {author} {\bibinfo {author} {\bibfnamefont {S.}~\bibnamefont
  {Grimme}}, \bibinfo {author} {\bibfnamefont {J.}~\bibnamefont {Antony}},
  \bibinfo {author} {\bibfnamefont {S.}~\bibnamefont {Ehrlich}},\ and\ \bibinfo
  {author} {\bibfnamefont {H.}~\bibnamefont {Krieg}},\ }\bibfield  {title}
  {\enquote {\bibinfo {title} {A consistent and accurate ab initio
  parametrization of density functional dispersion correction (dft-d) for the
  94 elements h-pu},}\ }\href {https://doi.org/10.1063/1.3382344} {\bibfield
  {journal} {\bibinfo  {journal} {The Journal of Chemical Physics}\ }\textbf
  {\bibinfo {volume} {132}},\ \bibinfo {pages} {154104} (\bibinfo {year}
  {2010})}\BibitemShut {NoStop}%
\bibitem [{\citenamefont {Bankura}\ \emph {et~al.}(2014)\citenamefont
  {Bankura}, \citenamefont {Karmakar}, \citenamefont {Carnevale}, \citenamefont
  {Chandra},\ and\ \citenamefont {Klein}}]{Bankura2014}%
  \BibitemOpen
  \bibfield  {author} {\bibinfo {author} {\bibfnamefont {A.}~\bibnamefont
  {Bankura}}, \bibinfo {author} {\bibfnamefont {A.}~\bibnamefont {Karmakar}},
  \bibinfo {author} {\bibfnamefont {V.}~\bibnamefont {Carnevale}}, \bibinfo
  {author} {\bibfnamefont {A.}~\bibnamefont {Chandra}},\ and\ \bibinfo {author}
  {\bibfnamefont {M.~L.}\ \bibnamefont {Klein}},\ }\bibfield  {title} {\enquote
  {\bibinfo {title} {Structure, dynamics, and spectral diffusion of water from
  first-principles molecular dynamics},}\ }\href
  {https://doi.org/10.1021/jp506120t} {\bibfield  {journal} {\bibinfo
  {journal} {The Journal of Physical Chemistry C}\ }\textbf {\bibinfo {volume}
  {118}},\ \bibinfo {pages} {29401--29411} (\bibinfo {year}
  {2014})}\BibitemShut {NoStop}%
\bibitem [{\citenamefont {Soper}(2000)}]{Soper2000}%
  \BibitemOpen
  \bibfield  {author} {\bibinfo {author} {\bibfnamefont {A.~K.}\ \bibnamefont
  {Soper}},\ }\bibfield  {title} {\enquote {\bibinfo {title} {The radial
  distribution functions of water and ice from 220 to 673 k and at pressures up
  to 400 mpa},}\ }\href
  {https://doi.org/https://doi.org/10.1016/S0301-0104(00)00179-8} {\bibfield
  {journal} {\bibinfo  {journal} {Chemical Physics}\ }\textbf {\bibinfo
  {volume} {258}},\ \bibinfo {pages} {121--137} (\bibinfo {year}
  {2000})}\BibitemShut {NoStop}%
\bibitem [{\citenamefont {Skinner}\ \emph {et~al.}(2013)\citenamefont
  {Skinner}, \citenamefont {Huang}, \citenamefont {Schlesinger}, \citenamefont
  {Pettersson}, \citenamefont {Nilsson},\ and\ \citenamefont
  {Benmore}}]{Skinner2013}%
  \BibitemOpen
  \bibfield  {author} {\bibinfo {author} {\bibfnamefont {L.~B.}\ \bibnamefont
  {Skinner}}, \bibinfo {author} {\bibfnamefont {C.}~\bibnamefont {Huang}},
  \bibinfo {author} {\bibfnamefont {D.}~\bibnamefont {Schlesinger}}, \bibinfo
  {author} {\bibfnamefont {L.~G.~M.}\ \bibnamefont {Pettersson}}, \bibinfo
  {author} {\bibfnamefont {A.}~\bibnamefont {Nilsson}},\ and\ \bibinfo {author}
  {\bibfnamefont {C.~J.}\ \bibnamefont {Benmore}},\ }\bibfield  {title}
  {\enquote {\bibinfo {title} {Benchmark oxygen-oxygen pair-distribution
  function of ambient water from x-ray diffraction measurements with a wide
  q-range},}\ }\href {https://doi.org/10.1063/1.4790861} {\bibfield  {journal}
  {\bibinfo  {journal} {The Journal of Chemical Physics}\ }\textbf {\bibinfo
  {volume} {138}},\ \bibinfo {pages} {74506} (\bibinfo {year}
  {2013})}\BibitemShut {NoStop}%
\bibitem [{\citenamefont {Ohto}\ \emph {et~al.}(2019)\citenamefont {Ohto},
  \citenamefont {Dodia}, \citenamefont {Xu}, \citenamefont {Imoto},
  \citenamefont {Tang}, \citenamefont {Zysk}, \citenamefont {Kühne},
  \citenamefont {Shigeta}, \citenamefont {Bonn}, \citenamefont {Wu},\ and\
  \citenamefont {Nagata}}]{Ohto2019}%
  \BibitemOpen
  \bibfield  {author} {\bibinfo {author} {\bibfnamefont {T.}~\bibnamefont
  {Ohto}}, \bibinfo {author} {\bibfnamefont {M.}~\bibnamefont {Dodia}},
  \bibinfo {author} {\bibfnamefont {J.}~\bibnamefont {Xu}}, \bibinfo {author}
  {\bibfnamefont {S.}~\bibnamefont {Imoto}}, \bibinfo {author} {\bibfnamefont
  {F.}~\bibnamefont {Tang}}, \bibinfo {author} {\bibfnamefont {F.}~\bibnamefont
  {Zysk}}, \bibinfo {author} {\bibfnamefont {T.~D.}\ \bibnamefont {Kühne}},
  \bibinfo {author} {\bibfnamefont {Y.}~\bibnamefont {Shigeta}}, \bibinfo
  {author} {\bibfnamefont {M.}~\bibnamefont {Bonn}}, \bibinfo {author}
  {\bibfnamefont {X.}~\bibnamefont {Wu}},\ and\ \bibinfo {author}
  {\bibfnamefont {Y.}~\bibnamefont {Nagata}},\ }\bibfield  {title} {\enquote
  {\bibinfo {title} {Accessing the accuracy of density functional theory
  through structure and dynamics of the water–air interface},}\ }\href
  {https://doi.org/10.1021/acs.jpclett.9b01983} {\bibfield  {journal} {\bibinfo
   {journal} {The Journal of Physical Chemistry Letters}\ }\textbf {\bibinfo
  {volume} {10}},\ \bibinfo {pages} {4914--4919} (\bibinfo {year}
  {2019})}\BibitemShut {NoStop}%
\bibitem [{\citenamefont {Neese}(2012)}]{ORCA}%
  \BibitemOpen
  \bibfield  {author} {\bibinfo {author} {\bibfnamefont {F.}~\bibnamefont
  {Neese}},\ }\bibfield  {title} {\enquote {\bibinfo {title} {The orca program
  system},}\ }\href {https://doi.org/10.1002/wcms.81} {\bibfield  {journal}
  {\bibinfo  {journal} {WIRES Comput. Molec. Sci.}\ }\textbf {\bibinfo {volume}
  {2}},\ \bibinfo {pages} {73--78} (\bibinfo {year} {2012})}\BibitemShut
  {NoStop}%
\bibitem [{\citenamefont {Neese}(2022)}]{ORCA5}%
  \BibitemOpen
  \bibfield  {author} {\bibinfo {author} {\bibfnamefont {F.}~\bibnamefont
  {Neese}},\ }\bibfield  {title} {\enquote {\bibinfo {title} {Software update:
  the orca program system, version 5.0},}\ }\href
  {https://doi.org/10.1002/wcms.1606} {\bibfield  {journal} {\bibinfo
  {journal} {WIRES Comput. Molec. Sci.}\ }\textbf {\bibinfo {volume} {12}},\
  \bibinfo {pages} {e1606} (\bibinfo {year} {2022})}\BibitemShut {NoStop}%
\bibitem [{\citenamefont {Niblett}, \citenamefont {Galib},\ and\ \citenamefont
  {Limmer}(2021)}]{Niblett2021}%
  \BibitemOpen
  \bibfield  {author} {\bibinfo {author} {\bibfnamefont {S.~P.}\ \bibnamefont
  {Niblett}}, \bibinfo {author} {\bibfnamefont {M.}~\bibnamefont {Galib}},\
  and\ \bibinfo {author} {\bibfnamefont {D.~T.}\ \bibnamefont {Limmer}},\
  }\bibfield  {title} {\enquote {\bibinfo {title} {Learning intermolecular
  forces at liquid–vapor interfaces},}\ }\href
  {https://doi.org/10.1063/5.0067565} {\bibfield  {journal} {\bibinfo
  {journal} {The Journal of Chemical Physics}\ }\textbf {\bibinfo {volume}
  {155}},\ \bibinfo {pages} {164101} (\bibinfo {year} {2021})}\BibitemShut
  {NoStop}%
\bibitem [{\citenamefont {Galib}\ \emph {et~al.}(2017)\citenamefont {Galib},
  \citenamefont {Duignan}, \citenamefont {Misteli}, \citenamefont {Baer},
  \citenamefont {Schenter}, \citenamefont {Hutter},\ and\ \citenamefont
  {Mundy}}]{Galib2017}%
  \BibitemOpen
  \bibfield  {author} {\bibinfo {author} {\bibfnamefont {M.}~\bibnamefont
  {Galib}}, \bibinfo {author} {\bibfnamefont {T.~T.}\ \bibnamefont {Duignan}},
  \bibinfo {author} {\bibfnamefont {Y.}~\bibnamefont {Misteli}}, \bibinfo
  {author} {\bibfnamefont {M.~D.}\ \bibnamefont {Baer}}, \bibinfo {author}
  {\bibfnamefont {G.~K.}\ \bibnamefont {Schenter}}, \bibinfo {author}
  {\bibfnamefont {J.}~\bibnamefont {Hutter}},\ and\ \bibinfo {author}
  {\bibfnamefont {C.~J.}\ \bibnamefont {Mundy}},\ }\bibfield  {title} {\enquote
  {\bibinfo {title} {Mass density fluctuations in quantum and classical
  descriptions of liquid water},}\ }\href {https://doi.org/10.1063/1.4986284}
  {\bibfield  {journal} {\bibinfo  {journal} {The Journal of Chemical Physics}\
  }\textbf {\bibinfo {volume} {146}},\ \bibinfo {pages} {244501} (\bibinfo
  {year} {2017})}\BibitemShut {NoStop}%
\bibitem [{\citenamefont {Nagata}\ \emph {et~al.}(2016)\citenamefont {Nagata},
  \citenamefont {Ohto}, \citenamefont {Bonn},\ and\ \citenamefont
  {Kühne}}]{Nagata2016}%
  \BibitemOpen
  \bibfield  {author} {\bibinfo {author} {\bibfnamefont {Y.}~\bibnamefont
  {Nagata}}, \bibinfo {author} {\bibfnamefont {T.}~\bibnamefont {Ohto}},
  \bibinfo {author} {\bibfnamefont {M.}~\bibnamefont {Bonn}},\ and\ \bibinfo
  {author} {\bibfnamefont {T.~D.}\ \bibnamefont {Kühne}},\ }\bibfield  {title}
  {\enquote {\bibinfo {title} {Surface tension of ab initio liquid water at the
  water-air interface},}\ }\href {https://doi.org/10.1063/1.4951710} {\bibfield
   {journal} {\bibinfo  {journal} {The Journal of Chemical Physics}\ }\textbf
  {\bibinfo {volume} {144}},\ \bibinfo {pages} {204705} (\bibinfo {year}
  {2016})}\BibitemShut {NoStop}%
\bibitem [{\citenamefont {Plimpton}(1995)}]{LAMMPS}%
  \BibitemOpen
  \bibfield  {author} {\bibinfo {author} {\bibfnamefont {S.}~\bibnamefont
  {Plimpton}},\ }\bibfield  {title} {\enquote {\bibinfo {title} {Fast parallel
  algorithms for short-range molecular dynamics},}\ }\href
  {https://doi.org/https://doi.org/10.1006/jcph.1995.1039} {\bibfield
  {journal} {\bibinfo  {journal} {Journal of Computational Physics}\ }\textbf
  {\bibinfo {volume} {117}},\ \bibinfo {pages} {1--19} (\bibinfo {year}
  {1995})}\BibitemShut {NoStop}%
\bibitem [{\citenamefont {Tribello}\ \emph {et~al.}(2014)\citenamefont
  {Tribello}, \citenamefont {Bonomi}, \citenamefont {Branduardi}, \citenamefont
  {Camilloni},\ and\ \citenamefont {Bussi}}]{Tribello2014}%
  \BibitemOpen
  \bibfield  {author} {\bibinfo {author} {\bibfnamefont {G.~A.}\ \bibnamefont
  {Tribello}}, \bibinfo {author} {\bibfnamefont {M.}~\bibnamefont {Bonomi}},
  \bibinfo {author} {\bibfnamefont {D.}~\bibnamefont {Branduardi}}, \bibinfo
  {author} {\bibfnamefont {C.}~\bibnamefont {Camilloni}},\ and\ \bibinfo
  {author} {\bibfnamefont {G.}~\bibnamefont {Bussi}},\ }\bibfield  {title}
  {\enquote {\bibinfo {title} {Plumed 2: New feathers for an old bird},}\
  }\href {https://doi.org/https://doi.org/10.1016/j.cpc.2013.09.018} {\bibfield
   {journal} {\bibinfo  {journal} {Computer Physics Communications}\ }\textbf
  {\bibinfo {volume} {185}},\ \bibinfo {pages} {604--613} (\bibinfo {year}
  {2014})}\BibitemShut {NoStop}%
\bibitem [{\citenamefont {Bonomi}\ \emph {et~al.}(2019)\citenamefont {Bonomi},
  \citenamefont {Bussi}, \citenamefont {Camilloni}, \citenamefont {Tribello},
  \citenamefont {Banáš}, \citenamefont {Barducci}, \citenamefont {Bernetti},
  \citenamefont {Bolhuis}, \citenamefont {Bottaro}, \citenamefont {Branduardi},
  \citenamefont {Capelli}, \citenamefont {Carloni}, \citenamefont {Ceriotti},
  \citenamefont {Cesari}, \citenamefont {Chen}, \citenamefont {Chen},
  \citenamefont {Colizzi}, \citenamefont {De}, \citenamefont {Pierre},
  \citenamefont {Donadio}, \citenamefont {Drobot}, \citenamefont {Ensing},
  \citenamefont {Ferguson}, \citenamefont {Filizola}, \citenamefont {Fraser},
  \citenamefont {Fu}, \citenamefont {Gasparotto}, \citenamefont {Gervasio},
  \citenamefont {Giberti}, \citenamefont {Gil-Ley}, \citenamefont {Giorgino},
  \citenamefont {Heller}, \citenamefont {Hocky}, \citenamefont {Iannuzzi},
  \citenamefont {Invernizzi}, \citenamefont {Jelfs}, \citenamefont {Jussupow},
  \citenamefont {Kirilin}, \citenamefont {Laio}, \citenamefont {Limongelli},
  \citenamefont {Lindorff-Larsen}, \citenamefont {Löhr}, \citenamefont
  {Marinelli}, \citenamefont {Martin-Samos}, \citenamefont {Masetti},
  \citenamefont {Meyer}, \citenamefont {Michaelides}, \citenamefont {Molteni},
  \citenamefont {Morishita}, \citenamefont {Nava}, \citenamefont {Paissoni},
  \citenamefont {Papaleo}, \citenamefont {Parrinello}, \citenamefont
  {Pfaendtner}, \citenamefont {Piaggi}, \citenamefont {Piccini}, \citenamefont
  {Pietropaolo}, \citenamefont {Pietrucci}, \citenamefont {Pipolo},
  \citenamefont {Provasi}, \citenamefont {Quigley}, \citenamefont {Raiteri},
  \citenamefont {Raniolo}, \citenamefont {Rydzewski}, \citenamefont
  {Salvalaglio}, \citenamefont {Sosso}, \citenamefont {Spiwok}, \citenamefont
  {Šponer}, \citenamefont {Swenson}, \citenamefont {Tiwary}, \citenamefont
  {Valsson}, \citenamefont {Vendruscolo}, \citenamefont {Voth}, \citenamefont
  {White},\ and\ \citenamefont {consortium}}]{Bonomi2019}%
  \BibitemOpen
  \bibfield  {author} {\bibinfo {author} {\bibfnamefont {M.}~\bibnamefont
  {Bonomi}}, \bibinfo {author} {\bibfnamefont {G.}~\bibnamefont {Bussi}},
  \bibinfo {author} {\bibfnamefont {C.}~\bibnamefont {Camilloni}}, \bibinfo
  {author} {\bibfnamefont {G.~A.}\ \bibnamefont {Tribello}}, \bibinfo {author}
  {\bibfnamefont {P.}~\bibnamefont {Banáš}}, \bibinfo {author} {\bibfnamefont
  {A.}~\bibnamefont {Barducci}}, \bibinfo {author} {\bibfnamefont
  {M.}~\bibnamefont {Bernetti}}, \bibinfo {author} {\bibfnamefont {P.~G.}\
  \bibnamefont {Bolhuis}}, \bibinfo {author} {\bibfnamefont {S.}~\bibnamefont
  {Bottaro}}, \bibinfo {author} {\bibfnamefont {D.}~\bibnamefont {Branduardi}},
  \bibinfo {author} {\bibfnamefont {R.}~\bibnamefont {Capelli}}, \bibinfo
  {author} {\bibfnamefont {P.}~\bibnamefont {Carloni}}, \bibinfo {author}
  {\bibfnamefont {M.}~\bibnamefont {Ceriotti}}, \bibinfo {author}
  {\bibfnamefont {A.}~\bibnamefont {Cesari}}, \bibinfo {author} {\bibfnamefont
  {H.}~\bibnamefont {Chen}}, \bibinfo {author} {\bibfnamefont {W.}~\bibnamefont
  {Chen}}, \bibinfo {author} {\bibfnamefont {F.}~\bibnamefont {Colizzi}},
  \bibinfo {author} {\bibfnamefont {S.}~\bibnamefont {De}}, \bibinfo {author}
  {\bibfnamefont {M.~D.~L.}\ \bibnamefont {Pierre}}, \bibinfo {author}
  {\bibfnamefont {D.}~\bibnamefont {Donadio}}, \bibinfo {author} {\bibfnamefont
  {V.}~\bibnamefont {Drobot}}, \bibinfo {author} {\bibfnamefont
  {B.}~\bibnamefont {Ensing}}, \bibinfo {author} {\bibfnamefont {A.~L.}\
  \bibnamefont {Ferguson}}, \bibinfo {author} {\bibfnamefont {M.}~\bibnamefont
  {Filizola}}, \bibinfo {author} {\bibfnamefont {J.~S.}\ \bibnamefont
  {Fraser}}, \bibinfo {author} {\bibfnamefont {H.}~\bibnamefont {Fu}}, \bibinfo
  {author} {\bibfnamefont {P.}~\bibnamefont {Gasparotto}}, \bibinfo {author}
  {\bibfnamefont {F.~L.}\ \bibnamefont {Gervasio}}, \bibinfo {author}
  {\bibfnamefont {F.}~\bibnamefont {Giberti}}, \bibinfo {author} {\bibfnamefont
  {A.}~\bibnamefont {Gil-Ley}}, \bibinfo {author} {\bibfnamefont
  {T.}~\bibnamefont {Giorgino}}, \bibinfo {author} {\bibfnamefont {G.~T.}\
  \bibnamefont {Heller}}, \bibinfo {author} {\bibfnamefont {G.~M.}\
  \bibnamefont {Hocky}}, \bibinfo {author} {\bibfnamefont {M.}~\bibnamefont
  {Iannuzzi}}, \bibinfo {author} {\bibfnamefont {M.}~\bibnamefont
  {Invernizzi}}, \bibinfo {author} {\bibfnamefont {K.~E.}\ \bibnamefont
  {Jelfs}}, \bibinfo {author} {\bibfnamefont {A.}~\bibnamefont {Jussupow}},
  \bibinfo {author} {\bibfnamefont {E.}~\bibnamefont {Kirilin}}, \bibinfo
  {author} {\bibfnamefont {A.}~\bibnamefont {Laio}}, \bibinfo {author}
  {\bibfnamefont {V.}~\bibnamefont {Limongelli}}, \bibinfo {author}
  {\bibfnamefont {K.}~\bibnamefont {Lindorff-Larsen}}, \bibinfo {author}
  {\bibfnamefont {T.}~\bibnamefont {Löhr}}, \bibinfo {author} {\bibfnamefont
  {F.}~\bibnamefont {Marinelli}}, \bibinfo {author} {\bibfnamefont
  {L.}~\bibnamefont {Martin-Samos}}, \bibinfo {author} {\bibfnamefont
  {M.}~\bibnamefont {Masetti}}, \bibinfo {author} {\bibfnamefont
  {R.}~\bibnamefont {Meyer}}, \bibinfo {author} {\bibfnamefont
  {A.}~\bibnamefont {Michaelides}}, \bibinfo {author} {\bibfnamefont
  {C.}~\bibnamefont {Molteni}}, \bibinfo {author} {\bibfnamefont
  {T.}~\bibnamefont {Morishita}}, \bibinfo {author} {\bibfnamefont
  {M.}~\bibnamefont {Nava}}, \bibinfo {author} {\bibfnamefont {C.}~\bibnamefont
  {Paissoni}}, \bibinfo {author} {\bibfnamefont {E.}~\bibnamefont {Papaleo}},
  \bibinfo {author} {\bibfnamefont {M.}~\bibnamefont {Parrinello}}, \bibinfo
  {author} {\bibfnamefont {J.}~\bibnamefont {Pfaendtner}}, \bibinfo {author}
  {\bibfnamefont {P.}~\bibnamefont {Piaggi}}, \bibinfo {author} {\bibfnamefont
  {G.}~\bibnamefont {Piccini}}, \bibinfo {author} {\bibfnamefont
  {A.}~\bibnamefont {Pietropaolo}}, \bibinfo {author} {\bibfnamefont
  {F.}~\bibnamefont {Pietrucci}}, \bibinfo {author} {\bibfnamefont
  {S.}~\bibnamefont {Pipolo}}, \bibinfo {author} {\bibfnamefont
  {D.}~\bibnamefont {Provasi}}, \bibinfo {author} {\bibfnamefont
  {D.}~\bibnamefont {Quigley}}, \bibinfo {author} {\bibfnamefont
  {P.}~\bibnamefont {Raiteri}}, \bibinfo {author} {\bibfnamefont
  {S.}~\bibnamefont {Raniolo}}, \bibinfo {author} {\bibfnamefont
  {J.}~\bibnamefont {Rydzewski}}, \bibinfo {author} {\bibfnamefont
  {M.}~\bibnamefont {Salvalaglio}}, \bibinfo {author} {\bibfnamefont {G.~C.}\
  \bibnamefont {Sosso}}, \bibinfo {author} {\bibfnamefont {V.}~\bibnamefont
  {Spiwok}}, \bibinfo {author} {\bibfnamefont {J.}~\bibnamefont {Šponer}},
  \bibinfo {author} {\bibfnamefont {D.~W.~H.}\ \bibnamefont {Swenson}},
  \bibinfo {author} {\bibfnamefont {P.}~\bibnamefont {Tiwary}}, \bibinfo
  {author} {\bibfnamefont {O.}~\bibnamefont {Valsson}}, \bibinfo {author}
  {\bibfnamefont {M.}~\bibnamefont {Vendruscolo}}, \bibinfo {author}
  {\bibfnamefont {G.~A.}\ \bibnamefont {Voth}}, \bibinfo {author}
  {\bibfnamefont {A.}~\bibnamefont {White}},\ and\ \bibinfo {author}
  {\bibfnamefont {T.~P.}\ \bibnamefont {consortium}},\ }\bibfield  {title}
  {\enquote {\bibinfo {title} {Promoting transparency and reproducibility in
  enhanced molecular simulations},}\ }\href
  {https://doi.org/10.1038/s41592-019-0506-8} {\bibfield  {journal} {\bibinfo
  {journal} {Nature Methods}\ }\textbf {\bibinfo {volume} {16}},\ \bibinfo
  {pages} {670--673} (\bibinfo {year} {2019})}\BibitemShut {NoStop}%
\end{thebibliography}
%

\end{document}


\def\mytitle{%
\ch{CO2} Hydration at the Air-Water Interface: \\
A Surface-Mediated ‘In and Out’ Mechanism}
\title{Supporting Information for: \mytitle}

\author{Samuel G.\ H.\ Brookes}
\affiliation{Yusuf Hamied Department of Chemistry, 
University of Cambridge, Lensfield Road, Cambridge, CB2 1EW, UK}
\affiliation{Lennard-Jones Centre, 
University of Cambridge, Trinity Ln, Cambridge, CB2 1TN, UK}
\affiliation{Cavendish Laboratory, Department of Physics, 
University of Cambridge, Cambridge, CB3 0HE, UK}

\author{Venkat Kapil}
\email{v.kapil@ucl.ac.uk}
\affiliation{Department of Physics and Astronomy, 
University College London, 17-19 Gordon Street, London WC1H 0AH, UK}
\affiliation{Thomas Young Centre and London Centre for Nanotechnology, 
19 Gordon Street, London WC1H 0AH, UK}
\affiliation{Yusuf Hamied Department of Chemistry, 
University of Cambridge, Lensfield Road, Cambridge, CB2 1EW, UK}
\affiliation{Lennard-Jones Centre, 
University of Cambridge, Trinity Ln, Cambridge, CB2 1TN, UK}

\author{Angelos Michaelides}
\email{am452@cam.ac.uk}
\affiliation{Yusuf Hamied Department of Chemistry, 
University of Cambridge, Lensfield Road, Cambridge, CB2 1EW, UK}
\affiliation{Lennard-Jones Centre, 
University of Cambridge, Trinity Ln, Cambridge, CB2 1TN, UK}

\author{Christoph Schran}
\email{cs2121@cam.ac.uk}
\affiliation{Cavendish Laboratory, Department of Physics, 
University of Cambridge, Cambridge, CB3 0HE, UK}
\affiliation{Lennard-Jones Centre, 
University of Cambridge, Trinity Ln, Cambridge, CB2 1TN, UK}

{\maketitle}

\tableofcontents

\onecolumngrid

\section{Model Development}
\subsection*{Reference method validation}
\begin{figure}[p]
    \centering
    \includegraphics[scale=0.7]{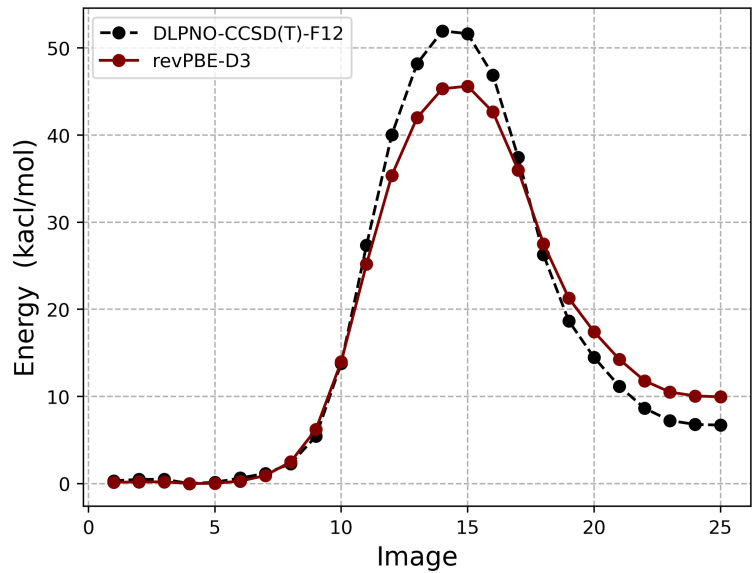}
    \caption{\label{fig:neb}
    Potential energy curve obtained from NEB simulations for the gaseous \ch{CO2 + H2O} reaction. 
    Results are plotted for both revPBE-D3 and DLPNO-CCSD(T)-F12 (def2-QZVPPD basis set) levels of theory. }
\end{figure}

The following section contains details on our model and the validation tests used to assess its quality. 
In Figure \ref{fig:neb}, we compare the predictions of our chosen functional, revPBE-D3 - used to generate the training forces and energies - with those of DLPNO-CCSD(T)-F12 (def2-QZVPPD basis set).
We obtained an optimized gas-phase reaction path using NEB simulations at the revPBE-D3 level of theory. 
Predicted energies were compared with DLPNO-CCSD(T)-F12 energies obtained using Orca for the same structures \cite{ORCA,ORCA5}. 
Overall, we find satisfactory agreement between the two methods of both for $\mathrm{\Delta} F$ and $\mathrm{\Delta} F^{\mathrm{\ddag}}$ to within 1.5 kcal/mol.

\subsection*{Model training and validation}
\begin{figure}[p]
    \centering
    \includegraphics[scale=0.7]{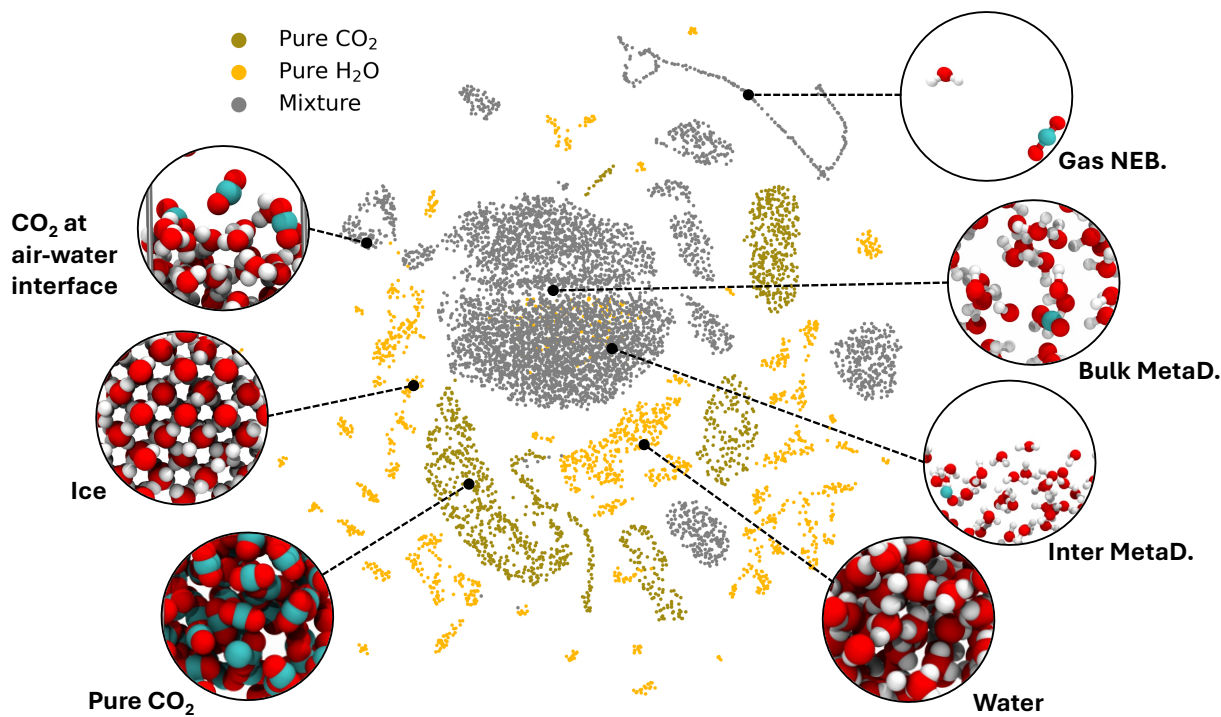}
    \caption{\label{fig:umap}
     2D UMAP projection of the structural dataset used to train our MACE model, shown alongside snapshots of representative structures.
     A total of some 8000 structures were used for training. 
     Structures were extracted from free MD simulations, from simulations employing some a restraint on some distance or coordination CV, and from preliminary metadynamics runs. 
     The most appropriate structures for training were identified using a `Query-by-Committee' procedure, which has been described in Ref.\ \cite{doi:10.1073/pnas.2110077118}.
     }
\end{figure}

\begin{figure}[p]
    \centering
    \includegraphics[scale=0.7]{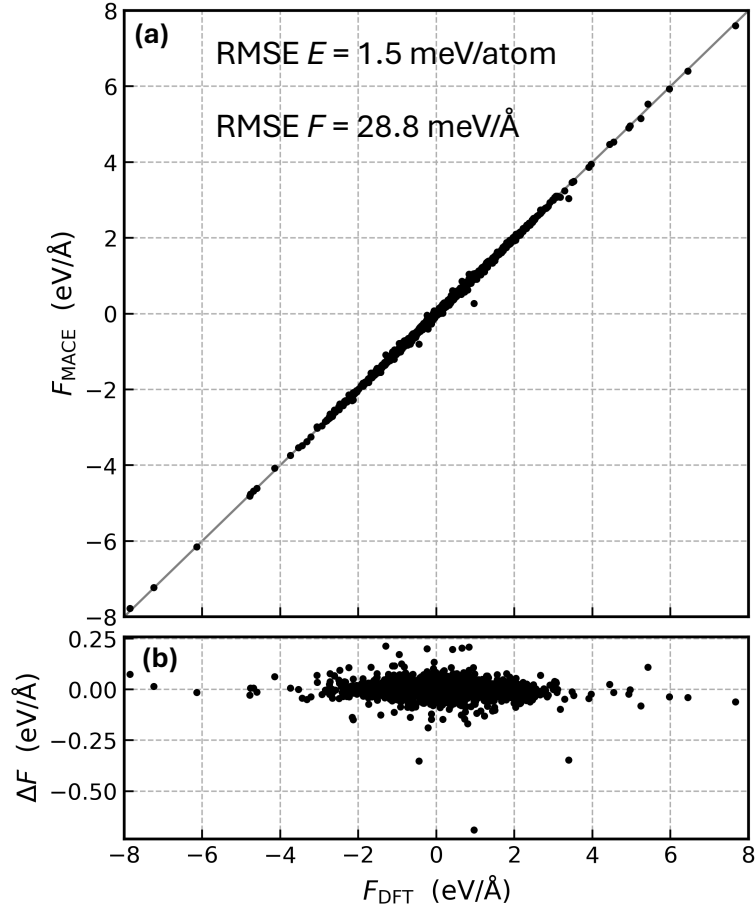}
    \caption{\label{fig:frc_nrg}
    Comparing MACE and DFT force predictions. 
    (\textbf{a}) Plot of the forces predicted by MACE against those of DFT. 
    Forces are calculated for a random selection of test 500 structures. 
    Total energy and force RMSEs are shown at the top of this figure. 
    (\textbf{b}) Plot of the difference in MACE and DFT forces against the reference DFT forces.}
\end{figure}

In total, some 8000 structures - labelled with revPBE-D3 forces and energies - were used in the training of our MACE model. 
These structures are shown as a 2D projection in Figure \ref{fig:umap} and consist of a mixture of pure \ch{CO2}, pure \ch{H2O}, and combined \ch{CO2/H2O} systems. 
The ability of our model to reproduce the underlying revPBE-D3 forces and energies for these structures is shown in Figure \ref{fig:frc_nrg}. 
Here, we plot MACE predictions on the forces against those of DFT for a random subset of test-set structures. 
Additionally, we compare MACE and DFT structural predictions for pure water, \ch{CO2} in water, and carbonic acid in bulk water in Figures \ref{fig:rdf_h2o}, \ref{fig:rdf_co2}, and \ref{fig:rdf_ca}, respectively. 
For both force/energy and structural predictions, we find a satisfactory agreement between the predictions of our model and those of reference revPBE-D3 calculations.

To test the interfacial predictions of our model, we extracted a compact subset of 50 structures corresponding to the five different molecular states found at the air-water interface: 
\ch{CO2} on top of the interface (\ch{CO2}($\uparrow$)); 
\ch{CO2} within the air-water interface (\ch{CO2}($\downarrow$)); 
the transition state (TS$^{\ddag}$); bicarbonate (BiC); and carbonic acid (CA). 
For each state, the force predictions of our MACE model were compared with those of revPBE-D3 for increasingly large radial cutoffs about the central, reactive carbon molecule. 
Results of this analysis are displayed in Figure \ref{fig:inter_states}a, where we can see consistently low errors across all five interfacial states.
The more prominent force deviations seen for TS$^\mathrm{\ddag}$ and BiC likely stem from having fewer of these types of structures in our dataset; nonetheless, the overall errors are lower relative to the total dataset error (28.8 meV/\AA{}), suggesting a good ability of our model to reproduce the reaction of \ch{CO2} at the air-water interface.

We assessed the ability of our revPBE-D3 model to accurately describe the chemistry of the air-water interface by comparing its energy predictions with those of coupled-cluster theory. 
Taking the 50-structure interfacial subset discussed above, we extracted carved-out clusters of the reactive carbon species and the closest 9 or 10 water molecules, keeping each cluster at 33 atoms in total. 
Single-point potential energies were calculated at the DLPNO-CCSD(T) (def2-TZVPPD
basis set) level of theory using ORCA and then compared to our MACE predictions. 
The results of this analysis are shown in Figure \ref{fig:inter_states}b, which plots the potential energy against the state identity. 
We see a close agreement between CCSD(T) and MACE energies, which agree within the uncertainty of their error bars.
Assuming that the most prominent contributions to the potential energy of interfacial states arise from local contributions, these results suggest we can be confident in our model's reactivity predictions for interfacial structures and processes.

The ambient nature of our 300\,K simulations is confirmed through solid-liquid coexistence simulations.
In this analysis, we run $NPT$ simulations using equilibrated ice-water biphasic systems. 
In Figure~\ref{fig:melting}, we show how the fraction of ice structures gradually reduces with time before disappearing completely at around 550 ps. 
These observations confirm that our MACE model correctly predicts water to be the most stable phase under ambient conditions. 

\section{Metadynamics}
Performing metadynamics simulations requires an appropriate selection of collective variables (CVs). 
For the \ch{CO2 + H2O} reaction, we use two main CVs: a coordination CV to probe the number of oxygens attached to carbon, thereby allowing us to distinguish between \ch{CO2} and the product species; and a protonation-state CV to count the number of hydrogen attached to carbon-bound oxygens. 
The nature of these CVs varies with system type.

\subsection*{Gas-phase Collective Variables}
We define the C-O coordination CV as follows:
\begin{equation}\label{eqn:co_coord}
    s_{\mathrm{CO}} = \sum_{i \in O} \frac{1 - \left(\dfrac{r_{\mathrm{Ci}}}{r_0}\right)^{6}}{{1 - \left(\dfrac{r_{\mathrm{Ci}}}{r_0}\right)^{12}}}
\end{equation}
where $r_{\mathrm{ij}}$ gives the pairwise carbon-oxygen distances, and $r_0 = 2.0$ Å.
Under this representation, $s_{\mathrm{CO}}$ $\sim$ 2 when the system is in the reactant state (i.e., CO$_2$) and $s_{\mathrm{CO}}$ $\sim$ 3 in the product state (i.e., \ch{HCO3-} or \ch{H2CO3}).

We define the gaseous protonation state CV as follows:
\begin{equation}\label{eqn:gas_coord_1}
    C^{\mathrm{gas}}_{\mathrm{iH}} = \sum_{j \in H} \frac{1 - \left(\dfrac{r_{\mathrm{ij}}}{r_0}\right)^{6}}{{1 - \left(\dfrac{r_{\mathrm{ij}}}{r_0}\right)^{12}}}
\end{equation}
\begin{equation}\label{eqn:gas_prot_state}
    s_\mathrm{(OH)_{g}} =  \sum_{i \in O} 1 - \frac{1 - \left(\dfrac{C^{\mathrm{gas}}_{\mathrm{iH}}}{C^0_{\mathrm{OH}}}\right)^{6}}{{1 - \left(\dfrac{C^{\mathrm{gas}}_{\mathrm{iH}}}{C^0_{\mathrm{OH}}}\right)^{12}}}
\end{equation}
where $r_{\mathrm{ij}}$ gives the pairwise carbon-oxygen distances, $r_0 = 1.5$ Å, and $C^0_{\mathrm{OH}}$ = 1.5.
This CV amounts to taking the number of hydrogen atoms attached to each oxygen (Eq.\ \ref{eqn:gas_coord_1}) and calculating how many of these oxygens have more than 1.5 hydrogens attached (Eq.\ \ref{eqn:gas_prot_state}). 
Under the formalism used in Figure 1 (i.e., $1- s_\mathrm{(OH)_{g}}$), we obtain values of $\sim$ 0.2 for \ch{CO2} and $\sim$ 0.8 for carbonic acid.

\subsection*{Bulk and Interfacial Collective Variables}
Collective variables used for the gas-phase \ch{CO2 + H2O} are simplified by the fact that there are only three oxygen atoms present in the system. 
However, in bulk solution and at the interface, we need to be able to differentiate between oxygens bound to the carbon and those that belong to free water molecules. 
Accordingly, we require a new set of collective variables to reflect this.

To describe C-O bond formation, we use the C-O coordination function described in Eq.\ \ref{eqn:co_coord},
\begin{equation}\label{eqn:co_coord_2}
    s_{\mathrm{CO}} = \sum_{i \in O} \frac{1 - \left(\dfrac{r_{\mathrm{Ci}}}{r_0}\right)^{12}}{{1 - \left(\dfrac{r_{\mathrm{Ci}}}{r_0}\right)^{24}}},
\end{equation}
albeit with stricter cutoff exponents in the numerator and denominator (i.e., 12/24 instead of 6/12). 
To describe the protonation state of our carbon species, we use the following CV:
\begin{equation}\label{eqn:sol_prot_state}
    s_\mathrm{(OH)_{aq}} = \sum_{\mathrm{j \in H}}
    \frac{ \sum_{\mathrm{i \in O}} C^{\mathrm{sol}}_{\mathrm{ij}} \sigma_{\mathrm{i}}}
    {\sum_{\mathrm{i \in O}} \sigma_{\mathrm{i}}}       
\end{equation}
%
In this expression, $C^{\mathrm{sol}}_{\mathrm{ij}}$ is the coordination between the ith oxygen and the jth hydrogen and is defined as:
\begin{equation}\label{eqn:sol_coord_1}
    C^{\mathrm{sol}}_{\mathrm{ij}} =  
    \frac{1 - \left(\dfrac{r_{\mathrm{ij}}}{r_0}\right)^{12}}
    {{1 - \left(\dfrac{r_{\mathrm{ij}}}{r_0}\right)^{24}}}
\end{equation}
%
where $r_{\mathrm{ij}}$ is the distance between these species, and $r_0$ is set to 1.25 Å.
$\sigma_{\mathrm{i}}$ is a switching function that acts on the distance between the calculated coordination numbers $C^{\mathrm{sol}}_{\mathrm{ij}}$ and carbon, $r_{\mathrm{Ci}}$, according to 
\begin{equation}\label{eqn:sol_dist}
    \sigma_{\mathrm{i}} = 
    \frac{1 - \left(\dfrac{r_{\mathrm{Ci}}}{r_\mathrm{C}}\right)^{25}}
    {{1 - \left(\dfrac{r_{\mathrm{Ci}}}{r_\mathrm{C}}\right)^{50}}}
\end{equation}
%
where $r_\mathrm{C}$ is set to 1.5 Å. 
Accordingly, the collective variable defined in Eq.\ \ref{eqn:sol_prot_state} amounts to calculating all O-H coordinations and summing only those that reside within 1.5 Å of the reacting molecule. 
When $s_\mathrm{(OH)_{aq}} \sim 1.0$, we have bicarbonate; when $s_\mathrm{(OH)_{aq}} \sim 2.0$, we have carbonic acid.

\subsection*{Free energy convergence}
For each metadynamics simulation, we obtain more than 50 ns worth of structural and thermodynamic data. 
To check the free energy convergence, we plot $\Delta F$ values for the product(s) of reaction as a function of individual walker time. 
These are shown in Figures \ref{fig:conv_gas}-\ref{fig:conv_inter} for the gaseous, bulk, and interfacial reactions.

\section{Additional Analysis}
\subsection*{BLYP-D3 and RPA Models}

The bulk and interfacial properties of condensed-phase systems are highly dependent upon the choice of underlying theory. 
Whilst our decision to train a MACE model on revPBE-D3 is grounded on its well-documented performance for aqueous systems,\cite{Bankura2014,Soper2000,Skinner2013,Ohto2019} it is important that we can show that the main conclusions of this paper - the idea that certain interfacial reactions can proceed via an `In-and-Out' mechanism - are robust with respect to differences in the underlying theory. 
Accordingly, we generated new MACE models at two differing levels of theory to test the proposed mechanism. 
First, we developed a new DFT model at the \textbf{BLYP-D3} level. 
This was done by recalculating all force and energy labels of our 8000-structure dataset (with otherwise identical DFT parameters) and then training a new model from scratch (energy RMSE: 1.3 meV/atom; force RMSE: 67.8 meV/\AA{}). 
In addition, we used transfer learning to obtain a MACE model at the \textbf{RPA} level.
We implemented this training using a 700-structure subset of the main structural dataset consisting of 500 bulk structures (that is, \ch{CO2}, bicarbonate, carbonic acid, and transition-state-like species in bulk water) and 200 gaseous structures (energy RMSE: 1.2 meV/atom; force RMSE: 18.8 meV/\AA{}). 
The structural predictions of these models - specifically, the bulk density $\rho$ and the interfacial tension $\gamma$ - are shown in Table \ref{table:models}. 
Density estimates were obtained through 500 ps $NPT$ simulations at 300 K and 1 bar pressure; $\gamma$ estimates were obtained using 500 ps $NVT$ simulations of 180-water molecule systems ($15 \times 15 \times 100$ \AA{}). 
For $\gamma$, we used the Kirkwood-Buff approach for calculating interfacial tensions \cite{Kirkwood1951}.
Good agreement is found between the predictions of our models and those of prior \textit{ab initio} studies, with densities replicated to within 1 \% and $\gamma$ predictions consistent to within their error bars.

To test our proposed interfacial mechanism, we first obtained a series of density profiles for each model using free MD simulations. 
In each case, we extracted the water density, the \ch{CO2} density, and the carbonic acid density. 
These profiles are shown in Figure \ref{fig:dens_plot_models}, where the carbon density (primary axis) and water density (secondary axis) are plotted as a function of the distance from the instantaneous interface. 
Looking at these profiles, we see very similar adsorption propensities for each model: 
\ch{CO2} adsorbs on top of the interface at around 1 \AA{}, whilst carbonic acid adsorbs at/just below the interface around the zero. 
Where \ch{CO2} and carbonic acid represent the start- and endpoints of our interfacial reactions, these profiles are consistent with an `In-an-Out'-style reaction mechanism, one in which the reactive species ultimately returns to the interface following reaction.

Having confirmed the interfacial adsorption propensities for \ch{CO2} and carbonic acid, we can now more directly evaluate the reaction mechanism through restrained MD simulations at the air-water interface. 
In these simulations, we gradually switched the CO coordination $s_\mathrm{CO}$ from 2.0 (\ch{CO2}) to 3.0 (carbonic acid, CA) over a 40 ps period, after which the restraint was fully switched off. 
For each model, three trajectories were initialized from different starting configurations and velocities (9 simulations in total), and the distance of the reacting species from the instantaneous interface, $d_\mathrm{int}$, was monitored. 
The results of these simulations are shown in Figure \ref{fig:inter_traj_models}. 
In each panel, we plot $d_\mathrm{int}$ and $s_\mathrm{CO}$ (primary and secondary axes, respectively) as a function of simulation time. 
Across the 9 different trajectories, we observe similar trends in $d_\mathrm{int}$: as $s_\mathrm{CO}$ is switched from 2.0 to 3.0, $d_\mathrm{int}$ decreases from an originally positive value to a negative value; following the removal of the restraint, $d_\mathrm{int}$ starts to increase again and tends towards the instantaneous interface 0. 
This behavior in $d_\mathrm{int}$ is entirely representative of an `In-and-Out' mechanism.
Of course, these types of restrained simulations are prone to lots of variation, and we maintain that the most convincing evidence of this mechanism can be found in the main text from our statistical analysis of the metadynamics trajectories. 
However, that we can reproduce this `In-and-Out' behavior from simple restrained MD runs is reassuring and provides further, 
evidence of the existence of this type of reaction mechanism.

\subsection*{Bicarbonate conversion}
In Figure 2 of the main text, our minimum energy profiles for the bulk and interfacial reactions show a shallow minimum for the bicarbonate species. 
To confirm this part of the profile as an actual minimum, we perform additional free energy additional analysis on the conversion between bicarbonate and carbonic acid. 
Umbrella sampling simulations were conducted to obtain the free energy profiles corresponding to the loss (gain) of a proton from carbonic acid (bicarbonate),
\begin{equation}
    \mathrm{HCO_3^- + H_3O^+ \rightleftharpoons H_2CO_3 + H_2O,}
\end{equation}
under both bulk and interfacial conditions. 
Profiles are shown in Figure \ref{fig:umbrella} as a function of a simplified aqueous protonation state $s_\mathrm{(OH)_{aq}}$, in which we set $\sigma_{\mathrm{i}}$ equal to 1 by specifically assigning carbon-bound (Oc) and -unbound (Ow) waters. 
Under both bulk and interfacial conditions, there is a clear minimum in the free energy profiles around $s_\mathrm{(OH)_{aq}}$ = 1, corresponding to bicarbonate. 
Differences between these profiles and those presented in Figure 3 of the main text can be ascribed to differences in the methodology as well as the collective variable used.

\subsection*{Structural analysis}
In Figure 4 of the main text, we present a series of profiles relating to the density, solvent interactions, and degree of hydrogen bonding for each of the main species involved in the \ch{CO2} hydration reaction. 
To extract the relevant data for analysis, we specify specific regions within CV space relating to each species around its equilibrium structure. 
These regions are shown Figure \ref{fig:selection}.
Structures encountered during the metadynamics runs with $s_\mathrm{(CO)}$ and $s_\mathrm{(OH)}$ within these thresholds are extracted for profiling.

To understand how the density profiles of Figure 4 compare with those from free MD simulations, we plot the latter for each species in Figure \ref{fig:profiling}.
These simulations were performed using the same system setups as used in the metadynamics runs. 
Simulations were performed under the $NVT$ ensemble, with a total run time of 2.5 ns for each system. 
It is interesting to note the resemblance these profiles show with those extracted from the metadynamics runs. 
This is indicative of a convergence in the metadynamics runs and suggests that we are able to recover equilibrium positions from our enhanced sampling runs through the isolation of the equilibrium structures.

In addition to the species discussed above, in Figure \ref{fig:hydronium} we also show the density profile obtained for the hydronium ion (\ch{H3O+}) obtained from free MD. 
The setup for this simulation is similar to that discussed above; the only difference comes with the exchange of the main carbon molecule for an additional \ch{H+}.
Similar to previous simulations, we find that the \ch{H3O+} ion adsorbs at the air-water interface. 
Using $\Delta F = -RT \mathrm{ln} (\rho / {\rho}_0)$, we find that this profile relates to an adsorption energy of 1.3 kcal/mol.
This is in exact agreement with previous measurements obtained from experimental SFG \cite{Das2020}, attesting to the quality of this potential for treating ions at aqueous interfaces.

\begin{figure}[p]
    \centering
    \includegraphics[scale=0.7]{SI_water_rdfs.pdf}
    \caption{\label{fig:rdf_h2o}
    Radial distribution (g(r)) for Ow-Ow pairs obtained from \textit{ab initio} MD and MACE-MD and plotted against distance r. 
    AIMD simulations were run for 25 ps. MACE simulations were performed for 1 ns. 
    System and simulation details are shown above. 
    }
\end{figure}

\begin{figure}[p]
    \centering
    \includegraphics[scale=0.7]{SI_co2_rdfs.pdf}
    \caption{\label{fig:rdf_co2}
    Radial distribution (g(r)) for C-Ow pairs (\ch{CO2}-water) obtained from \textit{ab initio} MD and MACE-MD and plotted against distance r. 
    AIMD simulations were run for 45 ps. MACE simulations were performed for 1 ns.
    System and simulation details are shown above. 
    }
\end{figure}

\begin{figure}[p!]
    \centering
    \includegraphics[scale=0.7]{SI_ca_rdfs.pdf}
    \caption{\label{fig:rdf_ca}
    Radial distribution (g(r)) for C-Ow pairs (carbonic acid - water) obtained from \textit{ab initio} MD and MACE-MD and plotted against distance r. 
    AIMD simulations were run for 20 ps. MACE simulations were performed for 1 ns.
    System and simulation details are shown above. 
    }
\end{figure}

\begin{figure}[b]
    \centering
    \includegraphics[scale=0.95]{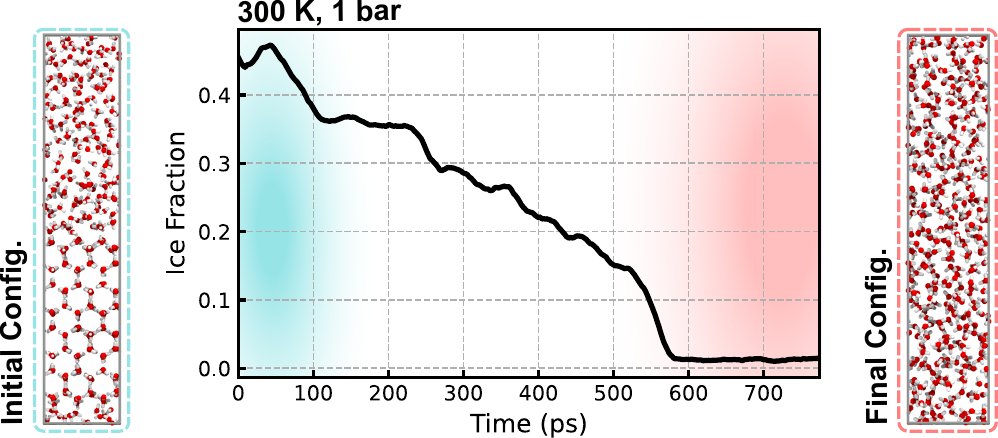} 
    \caption{\label{fig:melting} 
    %
    \textbf{(Left)} The biphasic setup used for this analysis. 
    Ice and liquid water phases were constructed in a 1:1 ratio (192 molecule each). 
    $NPT$ simulations were performed over 1 ns, during which the locally averaged Steinhardt $\overline{q}_6$ order parameter ($L$=6, $w_l$) was monitored as a function of time.
    $\overline{q}_6$ values greater than 0.42 were ascribed to ice-like geometries; values less than 0.42 were ascribed to water-like geometries.
    \textbf{(Middle)} Plot of the fraction of ice-like environments at each stage of the simulation. 
    The disappearance of the ice phase occurs at around 550 ps, coinciding with a sharp drop in $\overline{q}_6$.
    \textbf{(Right)} Snapshot of the final system configuration following the disappearance of the ice phase.}  
\end{figure}

\begin{figure}[p]
    \centering
    \includegraphics[scale=0.90]{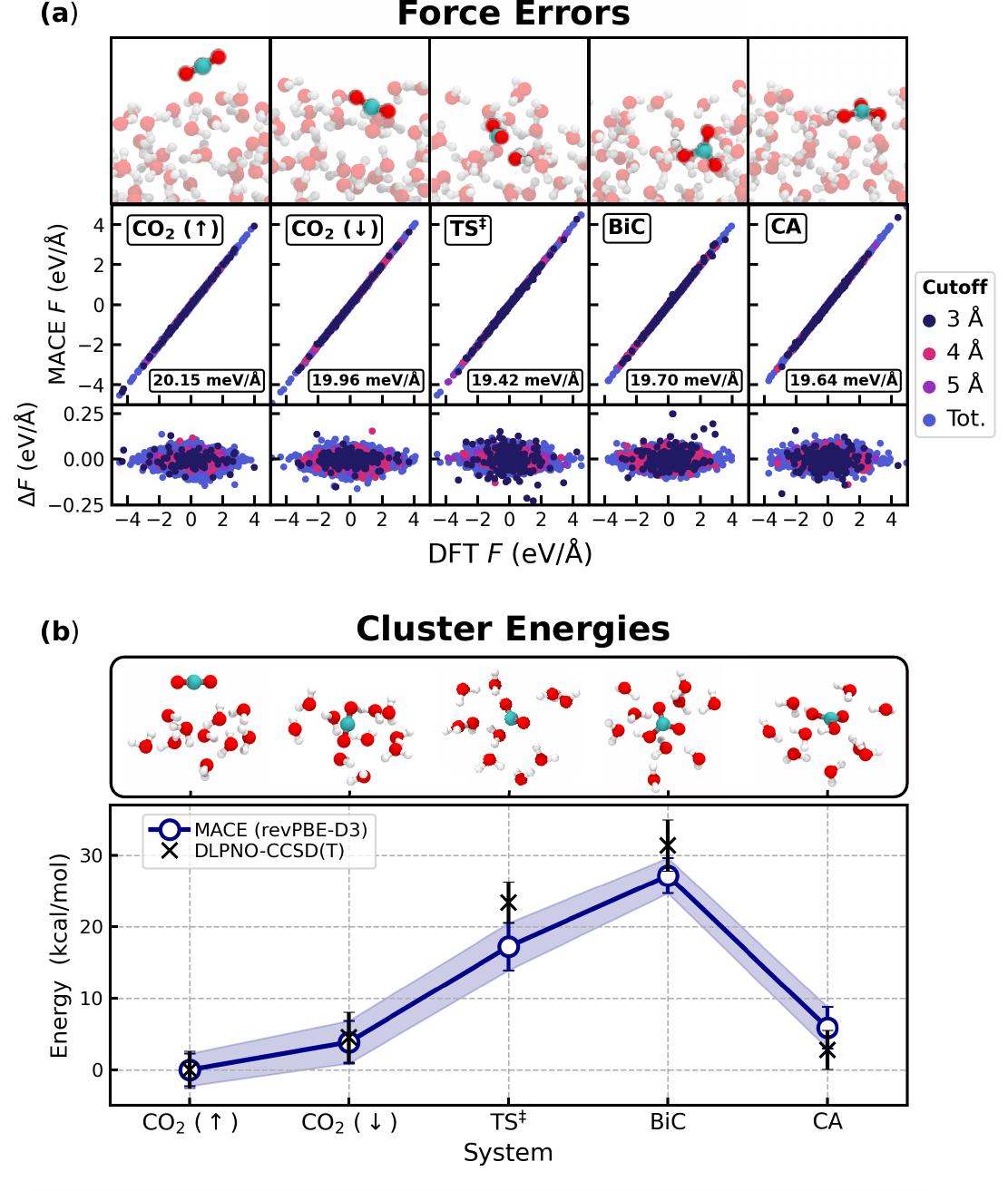}
    \caption{\label{fig:inter_states} 
    Validating interfacial reactivity.
    \textbf{(a)} Comparison of our model's force predictions against those of revPBE-D3 for the five identified interfacial states. 
    For each state, 10 representative structures are extracted from our dataset for analysis. 
    Forces are calculated within differing radial cutoffs of the reactive carbon species and compared on an absolute scale (middle panel) and relative scale ($\Delta F$, bottom panel).
    \textbf{(b)} A comparison of potential energies predicted by our MACE model and from CCSD(T). 
    For each interfacial state, clusters containing the closest 9/10 waters to the central carbon molecule are extracted. 
    DLPNO-CCSD(T) calculations are run using ORCA \cite{ORCA,ORCA5} and using the def2/TZVPP basis set. 
    The standard error in the energies across each state are plotted as error bars on the figure.}
\end{figure}

\begin{figure}[p]
    \centering
    \includegraphics[scale=0.95]{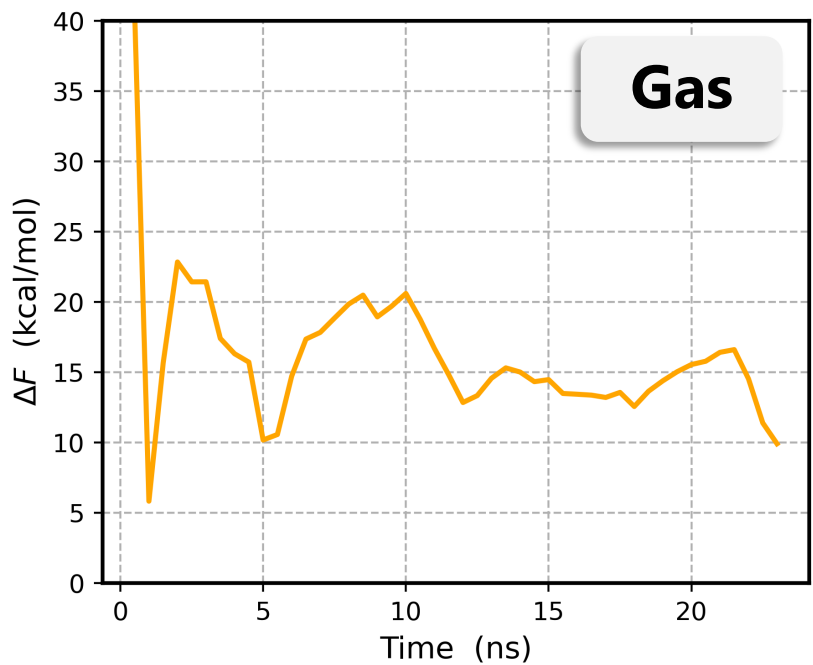}
    \caption{\label{fig:conv_gas}
     $\Delta F$ of carbonic acid in gas phase plotted as a function of the (individual) walker time. 
     Cumulative walker time amounts to over 60 ns for the gas-phase reaction.}
\end{figure}

\begin{figure}[p]
    \centering
    \includegraphics[scale=0.85]{SI_conv_bulk.pdf}
    \caption{\label{fig:conv_bulk}
     $\Delta F$ of carbonic acid at the bulk nanolayer plotted as a function of the (individual) walker time. 
     Free energy differences between bicarbonate and carbonic acid are plotted below.
     Cumulative walker time amounts to over 50 ns for the bulk reaction.}
\end{figure}

\begin{figure}[p]
    \centering
    \includegraphics[scale=0.75]{SI_conv_inter.pdf}
    \caption{\label{fig:conv_inter}
     $\Delta F$ of bicarboante carbonic acid at the interfacial nanolayer plotted as a function of the (individual) walker time. 
     Free energy differences between bicarbonate and carbonic acid are plotted below.
     Cumulative walker time amounts to over 50 ns for the interfacial reaction.}
\end{figure}

\FloatBarrier

\begin{figure}[p]
    \centering
    \includegraphics[scale=0.90]{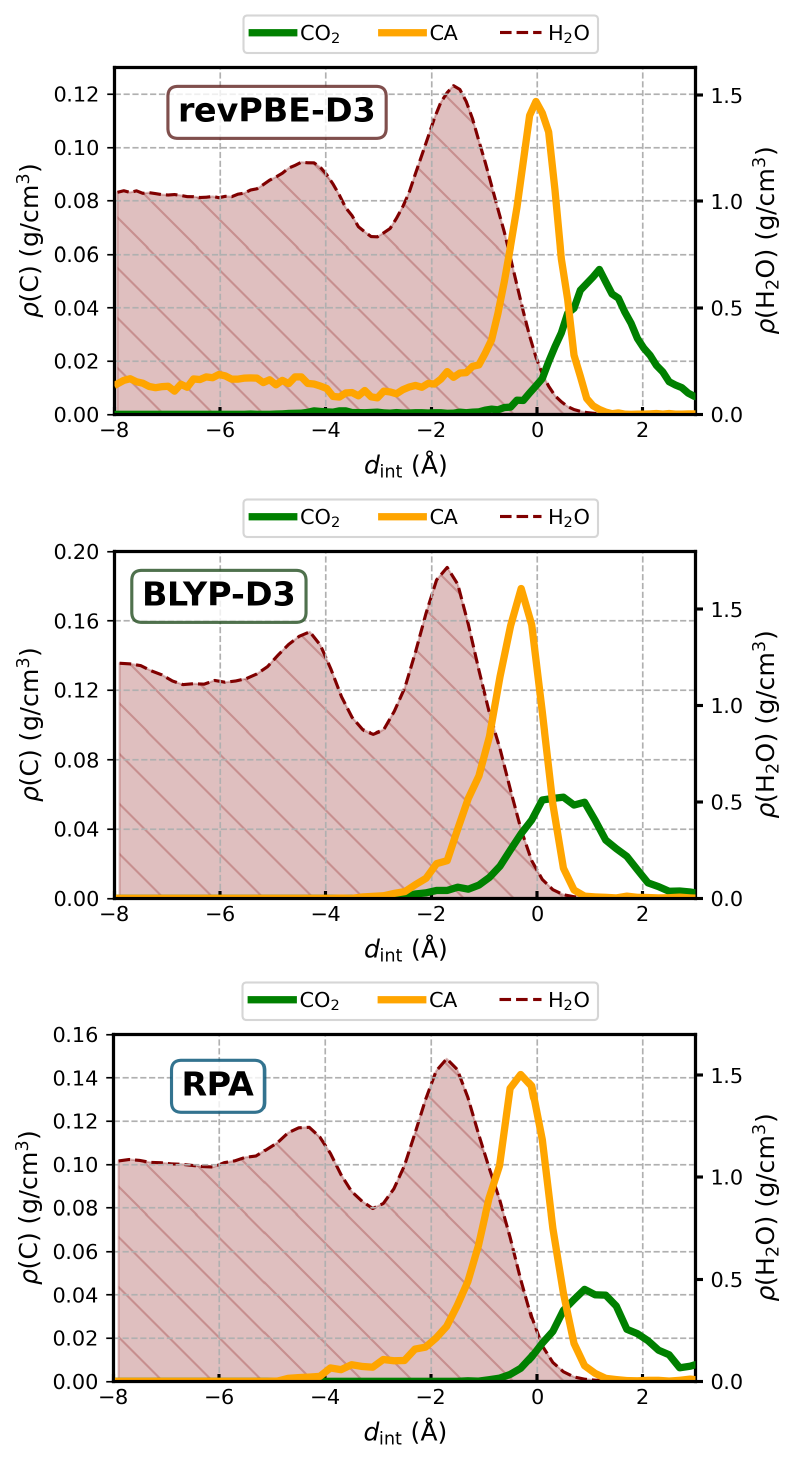}
    \caption{\label{fig:dens_plot_models} 
    Adsorption propensities of \ch{CO2} and carbonic acid at the air-water interface.
    Profiles are shown for our main revPBE-D3 model as well as the BLYP-D3 and RPA models. 
    In each panel, time-averaged densities are plotted as a function of distance from the instantaneous interface, with the carbon density given by the left axis and the water density given by the right axis.}
\end{figure}

\begin{figure}[p]
    \centering
    \includegraphics[scale=0.85]{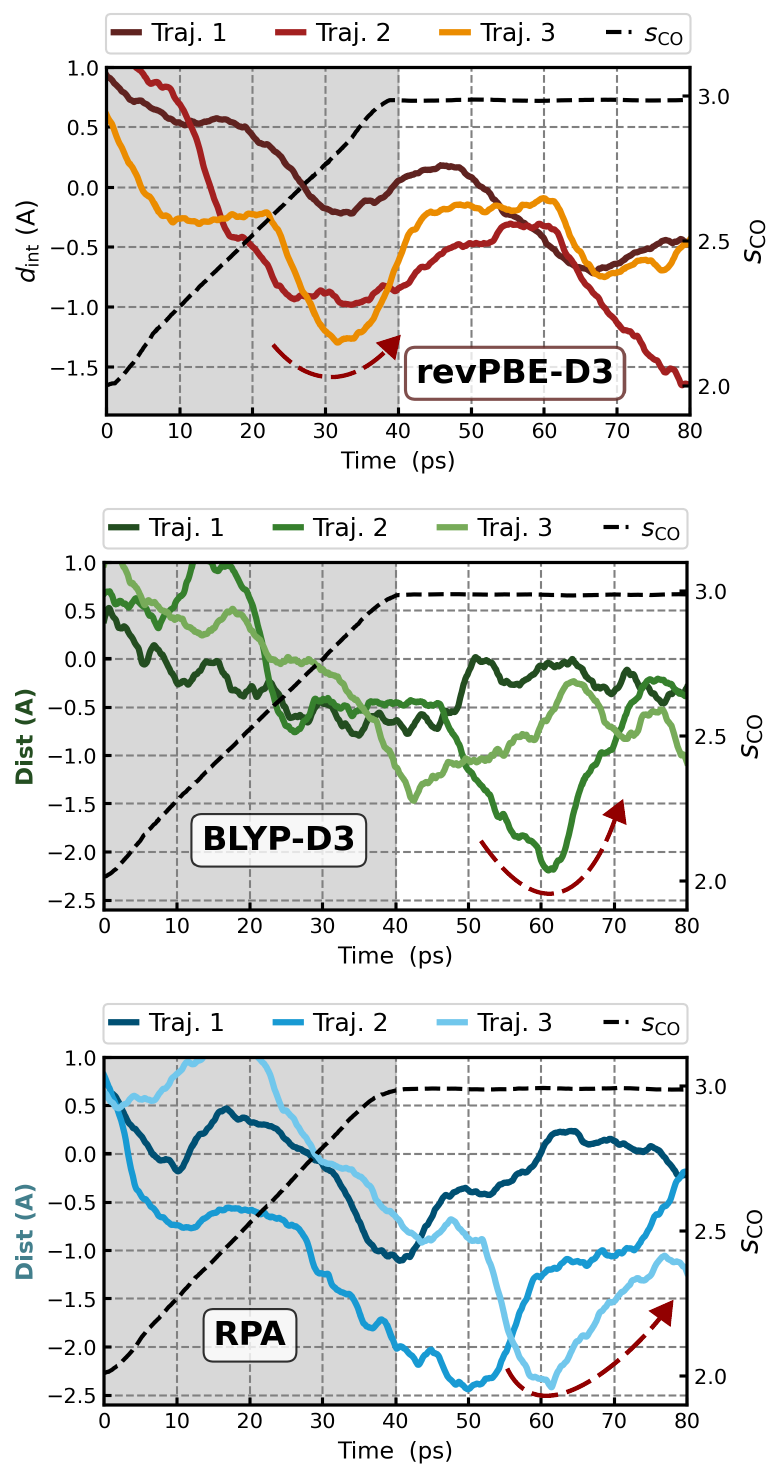}
    \caption{\label{fig:inter_traj_models} 
    The 'In-and-Out' mechanism observed using restrained MD. 
    For each model (revPBE-D3, BLYP-D3, and RPA), three restrained MD runs are performed in which the C-O coordination number, $s_\mathrm{CO}$, is gradually switched from 2.0 (\ch{CO2}) to 3.0 (carbonic acid). 
    The resulting change in the distance of the reacting molecule to the instantaneous interface, $d_\mathrm{int}$, is plotted against simulation time (colored, solid lines - one for each trajectory).
    The shaded portion of the figure represents the period over which $s_\mathrm{CO}$ (dashed line) is switched from 2.0 to 3.0.}
\end{figure}

\begin{figure}[p]
    \centering
    \includegraphics[scale=0.8]{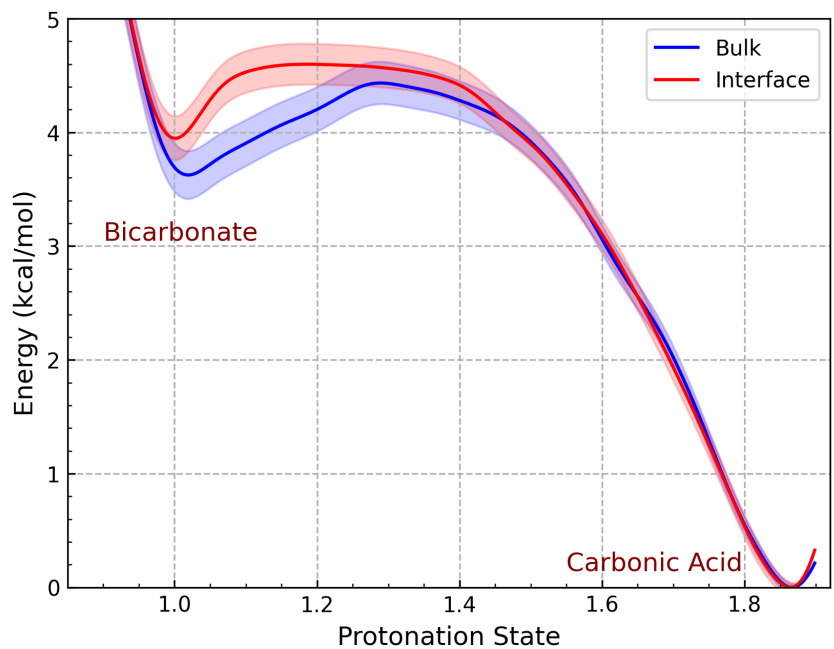}
    \caption{\label{fig:umbrella}
     Free energy profiles tracking the deprotonation of carbonic acid (right-hand side) to form bicarbonate (left-hand side) in bulk and at the air-water interface. 
     Free energies were obtained from umbrella sampling simulations and are plotted as a function of the protonation state of the carbon species. 
     Errors obtained from this integration are plotted as shaded regions. 
     Umbrella sampling simulations were run for 100 ps per umbrella under the $NVT$ ensemble using the same bulk and interfacial system setups employed for metadynamics simulations.}
\end{figure}

\FloatBarrier

     %
     %

\FloatBarrier

\begin{figure}[p]
    \centering
    \includegraphics[scale=0.7]{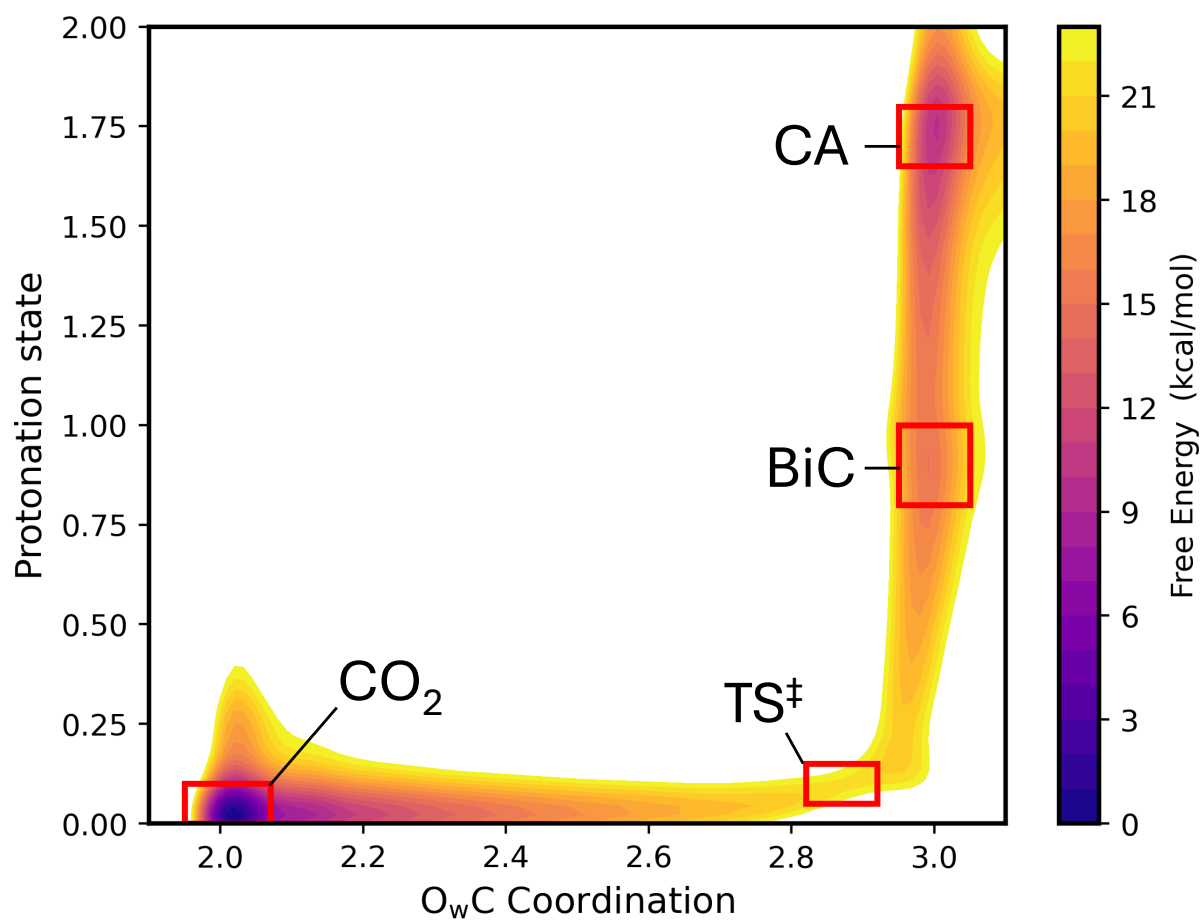}
    \caption{\label{fig:selection}
    Interfacial free energy profile of reaction, labeled with the state boundaries used to isolate \ch{CO2}, bicarbonate, carbonic acid, and TS$\mathrm{\ddag}$ structures for analysis.  
    }
\end{figure}

\begin{figure}[p]
    \centering
    \includegraphics[scale=1]{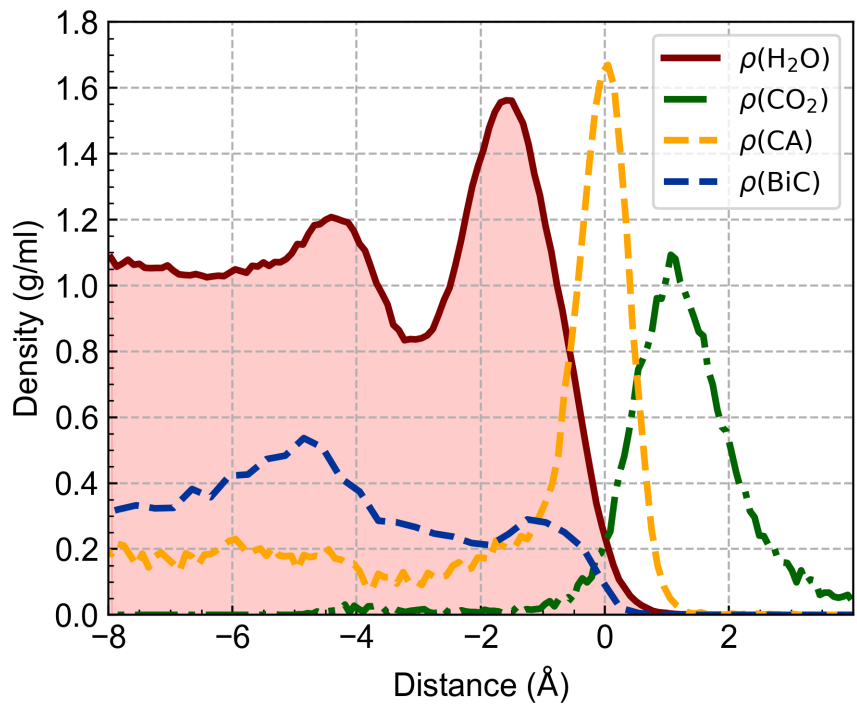}
    \caption{\label{fig:profiling}
    Density profiles obtained for \ch{CO2}, bicarbonate, and carbonic acid from free MD simulations. 
    Simulations were performed under the $NVT$ ensemble for a duration of 2.5 ns and using the same system setups as with our metadynamics interfacial run. 
    Densities are plotted as a function of the distance from the instantaneous interface (water density given by y axis, carbon densities on arbitrary scale).
    }
\end{figure}

\newpage

\begin{figure}[p]
    \centering
    \includegraphics[scale=1]{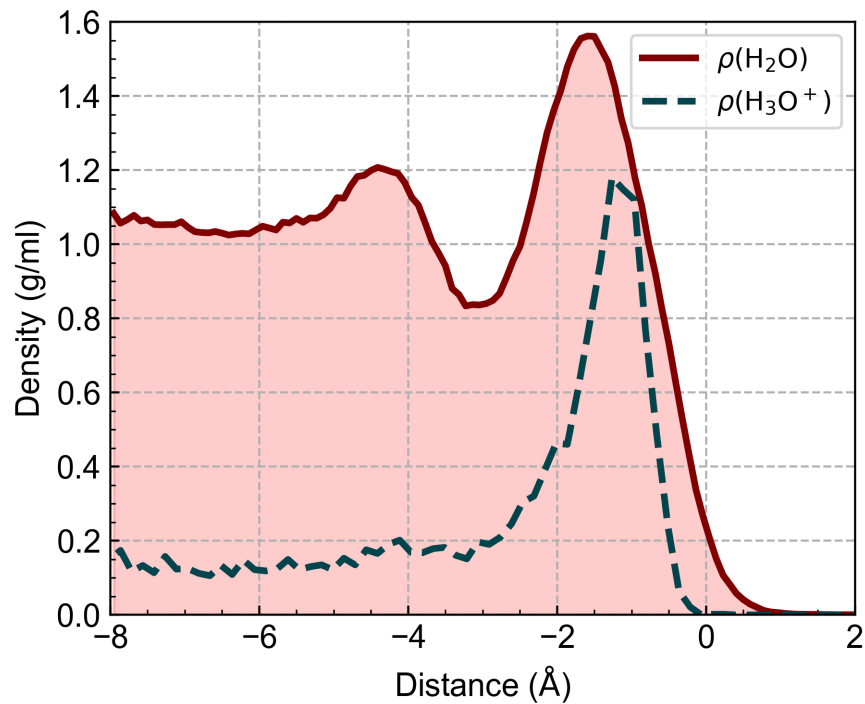}
    \caption{\label{fig:hydronium}
    Density profile obtained for hydronium ion from free MD simulations. 
    Simulations were run under the $NVT$ ensemble for 2.5 ns using the same interfacial setup as with the metadynamics runs. 
    Densities are plotted as a function of the distance from the instantaneous interface (water density given by y axis, hydronium density on arbitrary scale).
    Using $\Delta F = RT \mathrm{ln}(\rho)$, we determine a free energy of stabilization of $-$ 1.3 kcal/mol for hydronium at the interface. 
    }
\end{figure}

\begin{table}[p]
\caption{Key properties and predictions for the revPBE-D3, BLYP-D3, and RPA MACE models.  
        For each model, the training errors - taken as the RMSE between the MACE-predicted forces and those from the reference method - are reported alongside the predicted density ($\rho$) and IFT ($\gamma$) values.
        }
\begin{center}
\begin{tabular}{ >{\centering\arraybackslash\color{black}}m{3cm} >{\centering\arraybackslash\color{black}}m{2cm} >{\centering\arraybackslash\color{black}}m{2cm} >{\centering\arraybackslash\color{black}}m{2cm} }
 \textbf{Reference Theory} & \textbf{Train Er.\ (meV/\AA{})} & \textbf{$\rho$ (g/ml)} & \textbf{$\gamma$ (mN/m)} \\ 
 \hline
  & & & \\
 \underline{revPBE-D3} & 28.8 & 0.991 $\pm$ 0.002 & 84 $\pm$ 1 \\  
  Ref.\cite{Galib2017,Nagata2016} &  & 0.96 $\pm$ 0.03 & 83 $\pm$ 28 \\
  & & & \\
  BLYP-D3 & 67.8 & 1.110  0.002 & 106  $\pm$ 4 \\
  Ref.\cite{DelBen2013,Nagata2016} &  & 1.07 $\pm$ 0.02 & 92 $\pm$ 25 \\
  & & & \\
  RPA & 18.8 & 1.019 $\pm$ 0.002 & 81 $\pm$ 3 \\
  Ref.\cite{DelBen2015} &  & 0.99 $\pm$ 0.02 & / \\
  & & & \\
  Experiment & / & 0.997 & 72.8 \\
  & & & \\
   \hline
\end{tabular}
\end{center}
\label{table:models}
\end{table}

\FloatBarrier






%